\documentclass[11pt,a4paper]{article}
\usepackage{mathtools}

\DeclarePairedDelimiterX\braket[2]{\langle}{\rangle}{#1\,\delimsize\vert\,\mathopen{}#2}

\usepackage{geometry}
\usepackage{booktabs}
\usepackage{jheppub}
\usepackage{appendix}
\usepackage{graphicx}

\usepackage{capt-of}
\usepackage{siunitx}
\usepackage{amssymb}
\usepackage{enumitem}   
\usepackage[english]{babel}
\usepackage{url}
\usepackage{mathtools}
\usepackage{bbold}
\usepackage{slashed}
\usepackage{multirow}
\usepackage[usenames,dvipsnames,svgnames]{xcolor}

\usepackage{slashed}
\usepackage{multirow}
\newcommand{\SMadd}{{\rm SM_{\rm add}} }
\newcommand{\SMmult}{{\rm SM_{\rm mult}} }

\newcommand{\be}{\begin{equation}}
\newcommand{\ee}{\end{equation}}
\newcommand{\bea}{\begin{eqnarray}}
\newcommand{\eea}{\end{eqnarray}}
\newcommand{\Lint}{{\mathcal L}_{\rm int.}}

\newcommand{\MzSM}{\mathcal{M}^0_{\rm SM}}
\newcommand{\MoNP}{\mathcal{M}^1_{\rm NP}}
\newcommand{\MoNPunren}{\widehat{{\mathcal{M}}^1_{\rm NP}}}
\newcommand{\MoNPCT}{\mathcal{M}^1_{\rm NP,\, CT}}

\newcommand{\SigmaLO}{\sigma_{\rm LO}}
\newcommand{\SigmaNLO}{\sigma_{\rm NLO}}

\newcommand{\SigmaLOQCD}{\sigma_{\rm LO_{\rm QCD}}}
\newcommand{\SigmaNLOQCD}{\sigma_{\rm NLO_{\rm QCD}}}
\newcommand{\SigmaNLOEW}{\sigma_{\rm NLO_{\rm EW}}}

\newcommand{\SigmaNNLOQCD}{\sigma_{\rm NNLO_{\rm QCD}}}

\newcommand{\SigmaNP}{\sigma_{\rm NP}}
\newcommand{\SigmaLOQCDNP}{\sigma_{\rm LO_{QCD}+NP}}
\newcommand{\SigmaLOQCDNPH}{\sigma_{\rm LO_{QCD}+NP}^H}

\newcommand{\SigmaNPH}{\sigma_{\rm NP}^{H}}

\newcommand{\SigmaSM}{\sigma_{\rm SM}}

\newcommand{\SigmaSMadd}{\sigma^{\rm add.}_{\rm SM}}
\newcommand{\SigmaSMmult}{\sigma^{\rm mult.}_{\rm SM}}

\newcommand{\SigmaSMNPadd}{\sigma^{\rm add.}_{\rm SM+NP}}
\newcommand{\SigmaSMNPHadd}{\sigma^{{\rm add.}, H }_{\rm SM+NP}}

\newcommand{\SigmaSMNPmult}{\sigma^{\rm mult.}_{\rm SM+NP}}
\newcommand{\SigmaSMNPHmult}{\sigma^{{\rm mult.}, H }_{\rm SM+NP}}

\newcommand{\KNLOQCD}{K_{\rm QCD}^{\rm NLO}}
\newcommand{\KNNLOQCD}{K_{\rm QCD}^{\rm NNLO}}

\def\alphas{\alpha_s}

\newcommand{\MAD}{{\sc\small Mad\-Graph5\_\-aMC@NLO }}
\newcommand{\ct}{c_t}
\newcommand{\ctt}{\tilde{c}_t}
\newcommand{\Ct}{C_t}
\newcommand{\cts}{\ct}
\newcommand{\ctp}{\ctt}
\newcommand{\kt}{\kappa_t}
\newcommand{\ktt}{\tilde{\kappa}_t}
\newcommand{\Kt}{K_t}
\newcommand{\ytSM}{y_{t}^{\rm SM}}
\newcommand{\yt}{y_{t}}
\newcommand{\ytt}{\tilde y_{t}}
\newcommand{\LHNP}{\mathcal{L}_{H,\,{\rm NP}}}
\newcommand\UFO{{\sc\small UFO}}

\newcommand{\Sigmact}{\bar\sigma_{\ct}}
\newcommand{\Sigmactt}{\bar\sigma_{\ctt}}
\newcommand{\Sigmactctt}{\bar\sigma_{\ct,\ctt}}

\newcommand{\Sigmakt}{\bar\sigma_{\kt}}
\newcommand{\Sigmaktt}{\bar\sigma_{\ktt}}

\newcommand{\Sigmatheo}{\vec{\sigma}_{\rm th.}}
\newcommand{\Sigmaexp}{\vec{\sigma}_\mathrm{obs.}}

\newcommand{\mtt}{m(t \bar t)}
\newcommand{\Dytt}{\Delta y (t, \bar t)}

\newcommand{\gev}{{\rm GeV}}

\newcommand{\Yt}{y(t)}
\newcommand{\ptt}{p_T(t)}

\newcommand{\phiC}{\phi}
\newcommand{\phiK}{\phi}

\begin{document}
\title{Top-quark pair production as a probe of light top-philic scalars and anomalous Higgs interactions}

\author[a,b,c]{Fabio Maltoni,}
\author[c]{Davide Pagani,}
\author[a]{Simone Tentori}

\affiliation[a]{Centre for Cosmology, Particle Physics and Phenomenology (CP3),
Universit\'{e} Catholique de Louvain, B-1348 Louvain-la-Neuve, Belgium}
\affiliation[b]{Dipartimento di Fisica e Astronomia, Universit\`{a} di Bologna, Via Irnerio 46, 40126 Bologna, Italy}
\affiliation[c]{INFN, Sezione di Bologna, Via Irnerio 46, 40126 Bologna, Italy}

\emailAdd{fabio.maltoni@unibo.it}
\emailAdd{davide.pagani@bo.infn.it}
\emailAdd{simone.tentori@uclouvain.be}

 \abstract{
 We compute the effects due to the virtual exchange (or the soft emission)  of a scalar particle with generic couplings to the top quark in $t\bar t$ pair production at the LHC.  We apply the results to two cases of interest, extending and completing previous studies. First, we consider the indirect search for light ($m_S<2 m_t$) top-philic scalars with CP-even and/or CP-odd interactions. Second, we investigate how to set constraints on anomalous Yukawa couplings of the Higgs boson to the top quark.  Our results show that the current precision of experimental data together with the accuracy of the SM predictions make such indirect determinations a powerful probe for new physics.}

\preprint{\begin{flushright}
IRMP-CP3-24-17
\end{flushright}
}

 \maketitle
\section{Introduction}
\label{sec:introduction}

At the conclusion of Run III, the Large Hadron Collider (LHC) will have produced approximately one billion top-quark pairs. Even folding in trigger and reconstruction efficiencies, the striking signature of these pairs presents unprecedented opportunities that the high-energy physics community has yet to fully explore. Many analyses, particularly those focused on events within the bulk of distributions, currently confront limitations imposed by systematic effects.  As data volumes increase, the domains in phase space where systematic uncertainties dominate will expand, encompassing progressively higher scales. In addition, this abundance of events promises to enhance our comprehension of both detector performance and fundamental physics, thereby mitigating systematic uncertainties. Anticipating an order of magnitude increase in statistics by the conclusion of the High-Luminosity phase of the LHC, the prospect of conducting more refined, exclusive studies, and probing rare processes becomes increasingly feasible.

On the theoretical side, substantial advancements have been made over the past fifteen years. Achieving next-to-next-to-leading order (NNLO) precision in Quantum Chromodynamics (QCD) \cite{Baernreuther:2012ws,Czakon:2012zr,Czakon:2012pz,Czakon:2013goa,Czakon:2015owf,Czakon:2016dgf, Catani:2019iny, Catani:2019hip, Catani:2020tko} alongside the calculation and the combination with the electroweak (EW) corrections ~\cite{Hollik:2011ps, Kuhn:2011ri, Bernreuther:2012sx, Pagani:2016caq, Czakon:2017wor, Czakon:2017lgo, Gutschow:2018tuk, Frederix:2018nkq, Czakon:2019txp} has been a significant milestone, with various independent collaborations attaining this level of accuracy at a fully differential level. Moreover, the achievement of fully exclusive predictions, which can be seamlessly integrated into simulations, has pushed the boundaries to NNLO precision coupled with parton shower (PS) accuracy \cite{Mazzitelli:2020jio, Mazzitelli:2021mmm}. Advancements have been made also in the calculation of final state signatures. In certain regions of phase space, processes that do not strictly feature one or two on-shell top quarks have been computed with next-to-leading order (NLO) precision in QCD and EW \cite{Denner:2016jyo, Denner:2017kzu, Denner:2023grl} and in some cases even coupled with parton showers \cite{Jezo:2016ujg, Jezo:2023rht}, broadening the scope of the analyses. Higher order corrections, up to NNLO in QCD,  have also been calculated for spin-correlations \cite{Czakon:2020qbd, Frederix:2021zsh}. Resummation effects, crucial particularly at threshold, have also been included up to next-to-next-to-leading logarithmic (NNLL) accuracy \cite{Ahrens:2010zv, Ferroglia:2012ku, Pecjak:2016nee,  Czakon:2019txp}. Looking ahead, efforts towards achieving next-to-next-to-next-to-leading order ($\rm N^3 LO$) QCD accuracy have started, with the calculation for the top-quark decay that has recently appeared in the literature \cite{Chen:2023osm}. With the extended runtime of the High-Luminosity LHC (HL-LHC) in mind,  such a precision appears increasingly attainable.

Given what we have already at disposal and what is expected in the near and medium term, it is therefore mandatory to think how to best leverage such a phenomenal combination of theoretical predictions and experimental data. Fortunately, the top quark, occupying a unique position within the Standard Model as both the sole naturally occurring quark with a mass on the order of the electroweak scale, offers an ideal platform for probing novel concepts aimed at unravelling the mysteries surrounding electroweak symmetry breaking and investigating the potential existence of new phenomena.

At the moment, top-quark pair data in general show quite a good agreement with SM theory predictions, and this agreement would have not been possible without all the advancements in the knowledge of higher-order corrections that we have mentioned before. However, a tension is actually observed in the threshold region \cite{ATLAS:2023fsd,CMS:2024hgo}, which may be solved via further advancements in higher-order corrections, such as the inclusion of, {\it e.g.}, toponium effects in Monte Carlo simulators \cite{Fuks:2021xje,Maltoni:2024tul}. On the other hand,  BSM physics could also be the origin of such effects.

One of the possible BSM dynamics having an effect on  the threshold region in top-quark production is the presence of an anomalous top-quark Yukawa coupling ($\yt$), which affects the predictions for the cross section via EW loops. In particular, already within the SM,  the exchange of the Higgs boson between the two top quarks  leads to a  Sommerfeld enhancement at the threshold, {\it i.e.}, in the non-relativistic regime. An anomalous value of $\yt$ would have therefore an effect, especially at the threshold, on $t \bar t$ distributions and may be experimentally detected.  This idea has already been  investigated in the literature \cite{Schmidt:1992et, Kuhn:2013zoa} and the effects of an anomalous CP-even top-quark Yukawa coupling can be calculated via the public code {\sc \small Hator} \cite{Aliev:2010zk}. Indeed, the CMS collaboration has already exploited this strategy in order to set constraints on $\yt$ \cite{CMS:2019art,CMS:2020djy}. Unlike the case of direct on-shell production, the $t \bar t H$ final state, this strategy is not sensitive on the Higgs-boson decay width, and  in Ref.~\cite{Martini:2021uey} this idea has been further pushed taking into account CP-odd top-quark Yukawa interactions. It is therefore mandatory to have reliable predictions for the indirect detection, via $t \bar t $ signatures, of effects from anomalous $\yt$ originating at one loop.

In this work, not only  we revisit the calculation performed in Ref.~\cite{Martini:2021uey}, but we also further extend it to the case in which the scalar exchanged between the two top quarks can be not only the Higgs but also a generic light top-philic scalar $S$, setting constraints on its interactions with the top quark. For both scenarios, a Feynman diagram is particularly relevant, namely the one with the $s$-channel Higgs boson (or $S$ scalar) stemming from a top-quark loop in the $gg \to t \bar t$ process (see Fig.~\ref{fig:diagrams}(a)). 

In the case of the Higgs boson, to the best of our knowledge, this diagram was not taken into account in Ref.~\cite{Martini:2021uey}: while for CP-even anomalous $\yt$ interactions we find that its contribution is negligible, for the CP-odd ones it is actually the opposite; the $s$-channel diagram cannot be neglected. In the case of a general scalar $S$   we limit ourselves to the mass range $m_S< 2 m_t$, otherwise the best signature would be  a loop-induced $S$ production subsequently decaying into top quarks (see {\it e.g.}~Refs.~\cite{Dicus:1994bm, Bernreuther:2015fts, Hespel:2016qaf, Banfi:2023udd}), which corresponds precisely to the diagram we are speaking of. 

In the case of the scalar $S$ we observe that, if only the CP-odd interaction with the top-quark is present, virtual corrections are not sensitive to $m_S$ in the limit $m_S\to0$ and therefore they are infrared finite. While for the CP-even case, or a mixed one, virtual corrections have to be combined together with real emissions of $S$ in order to achieve IR-finite predictions. We provide in this work all the technical details of the calculation that we have performed for the scalar $S$ and, as we will describe better within the text, these technical details complement the calculation presented in Ref.~\cite{Blasi:2023hvb}, where a top-philic axion-like-particle (ALP) was considered. 

For both the scenarios, the one for the top-philic scalar $S$ and the one for the Higgs boson, we calculate the impact of virtual effects on differential distributions and, taking into account both QCD (up to NNLO) and EW effects (up to NLO) in the SM, we show how top-quark data can be sensitive to them and thus can constraints the strengths of the interaction of the top quark with either the Higgs boson or the top-philic scalar $S$. 

The paper is organised as follows. Sections \ref{sec:theoframe}--\ref{sec:Ssensitivity} concern the case of the light top-philic scalar $S$, while Sections \ref{sec:theoframeH}--\ref{sec:sensitivityH} the case of the Higgs boson. 
In Sec.~\ref{sec:theoframe} we  present the theoretical framework and we give the details of the calculation of virtual effects from a top-philic scalar on the $t \bar t$ cross section. We also show how to combine these effects with precise SM predictions. In Sec.~\ref{sec:distrct} we discuss the different patterns that can arise depending on how is the interaction between the top-quark and the scalar $S$, namely, if it is purely CP-even, CP-odd or a mixture of the two. We then show in Sec.~\ref{sec:Ssensitivity} the sensitivity that we can already achieve with current data on such interactions.
In Sec.~\ref{sec:theoframeH} we show how the calculation for the scalar $S$ presented in Sec.~\ref{sec:theoframe} can be recycled for the case of the Higgs boson and we also discuss the connection with the SMEFT framework. We then show the impact of the virtual effects from the Higgs on the $t \bar t$ cross section and we discuss in details the differences with the calculation in Ref.~\cite{Martini:2021uey}.
In Sec.~\ref{sec:sensitivityH} we discuss the bounds that can be set on CP-even and CP-odd Yukawa interactions of the top-quark. Finally,  we give our conclusions in Sec.~\ref{sec:conclusions}. In the Appendices we collect useful analytical formulas which we used for cross-checking our numerical simulations and additional information entering the statistical analysis.

\section{Scalar $S$: Theoretical framework}
\label{sec:theoframe}

As mentioned in Sec.~\ref{sec:introduction}, our goal is to calculate and study, in top-quark pair production, the effects  that originate from virtual corrections induced by a generic scalar $S$ that couples to the top quark with both CP-even and CP-odd interactions.  In this section we consider the theoretical framework where $S$ is an additional scalar on top of the SM particle content. The case where $S$ is the Higgs boson itself, allowing for its anomalous and/or CP-odd interactions with the top quark, is postponed to Sec.~\ref{sec:theoframeH}.

We start  by introducing in Sec.~\ref{sec:Lagrangians} the notation and defining the Lagrangian used for performing our calculation. Then, in Sec.~\ref{sec:loopcalculation} we present the calculation of the NP one-loop effects induced by $S$ to top-quark pair distributions. Additional explicit results are reported in Appendices \ref{app:UVcounterterms} and \ref{app:qqttanal}. Finally, in Sec.~\ref{sec:SMcombination} we detail how we combine the virtual NP effects  with SM predictions. 
\subsection{Lagrangian and notation}
\label{sec:Lagrangians}

The presence of an additional scalar $S$  on top of the SM particle content can be described by a Lagrangian of the form
\bea
\mathcal{L}_{{\rm SM}+S}\equiv \mathcal{L}_{{\rm SM}} +  \mathcal{L}_{S} + \Lint\,, \label{eq:LangS}
\eea
where $\mathcal{L}_{{\rm SM}}$ is the SM Lagrangian and 
\bea
 \mathcal{L}_{S}&\equiv& \frac{1}{2}\partial_\mu S\partial^\mu S-\frac{1}{2}m_S^2 S^2 -V(S)\label{eq:LdiS}\,,\\
  \Lint&\equiv& -\overline \psi_t (\cts+i\ctp\gamma_5)\psi_t S \label{eq:intL}\,.
\eea
The term $\mathcal{L}_{S}$ represents the Lagrangian of a generic free scalar with mass $m_S$. In principle it includes an unspecified potential $V(S)$, which as we will show does not enter our  calculations. The term $\Lint$ is the interacting Lagrangian of the scalar $S$ with the top-quark field $\psi_t$, where  $\ct$ and $\ctt$ parameterise the CP-even and CP-odd components of the interaction, respectively.

It is also useful to introduce the ``polar'' quantities 
\bea
\Ct &\equiv& \ct+ i \ctt = |\Ct| e^{i\phiC}\,, \label{eq:Ctdef}\\
\phiK&\equiv& \arctan\frac{\ctt}{\ct}\,,\label{eq:phiCtdef}
\eea
where $|\Ct|$ is parametrising the ``total'' strength of the interaction between the top quark   and the scalar $S$. Instead, the admixture of the CP-even and CP-odd components is parameterised by the phase $\phi$. 

\medskip

\begin{figure}[!t]
\begin{center}
\hspace*{-1cm}\includegraphics[width=0.75\textwidth]{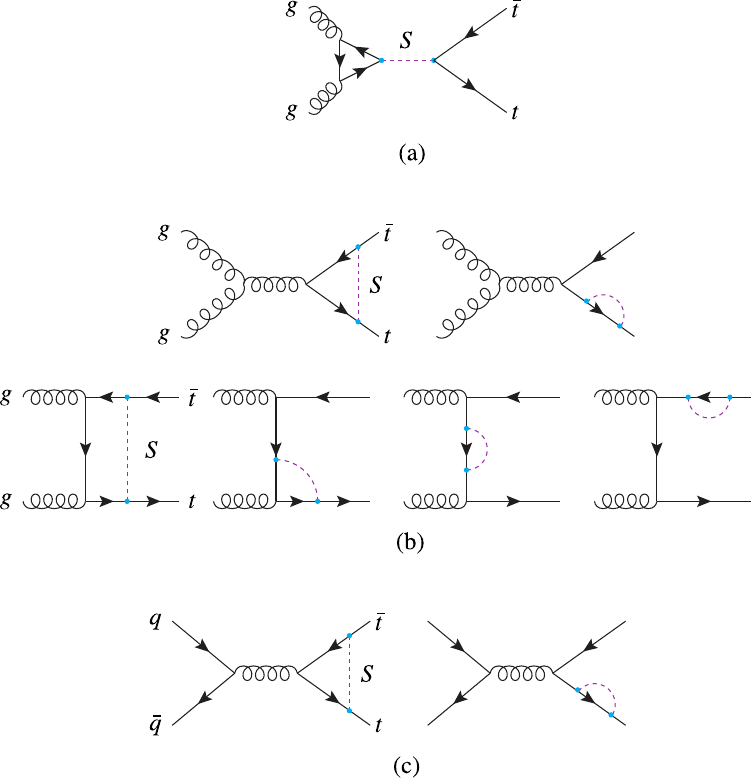}
\caption{Representative Feynman diagrams for the production of a $t\bar t$ pair at a hadron collider including the one-loop corrections due to the virtual exchange of a scalar particle $S$. These are the diagrams leading to the quantity $\MoNP$ discussed in the text. The same diagrams are also possible with the Higgs boson $H$ instead of $S$.}
\label{fig:diagrams}
\end{center}
\end{figure}
\medskip

Since in this work we are interested in NP effects from loop corrections to top-quark pair production induced by a generic scalar $S$, first of all let us look at the Feynman diagrams that are involved  in the calculation.
In Fig.~\ref{fig:diagrams} we show several representative diagrams for the two different $2\to2$ partonic processes leading to the  hadroproduction of top-quark pairs: $q\bar q\to t\bar t$ and $gg\to t\bar t$. We will come back later to the discussion of the different topologies and technical details of the calculation of these diagrams and the renormalisation of those that are UV-divergent. Here we want to stress another point. First of all we notice that none of the virtual diagrams features a self interaction vertex for the scalar  $S$. Thus, the quantity $V(S)$ in Eq.~\eqref{eq:LdiS}, as anticipated, does not enter our one-loop calculation. Then, the same diagrams, with the Higgs boson in the place of $S$, contribute to the NLO EW corrections in the SM itself. This is important since we want to study the possibility of extracting bounds in the $(\ct,\ctt)$ parameter space by comparing SM + NP predictions with data. Therefore, NLO EW corrections of SM origin cannot be neglected, otherwise they would be  fitted/interpreted in data as a BSM effect, especially since both the NLO EW in the SM and the BSM contributions involve diagrams with the same topologies. In the context of what will be discussed in Sec.~\ref{sec:theoframeH} this issue will be even more relevant, as it will be explained in detail.

\subsection{Scalar one-loop corrections to top-quark pair production}
\label{sec:loopcalculation}

\begin{figure}[!t]
    \centering
 
    \includegraphics[width=0.63\textwidth]{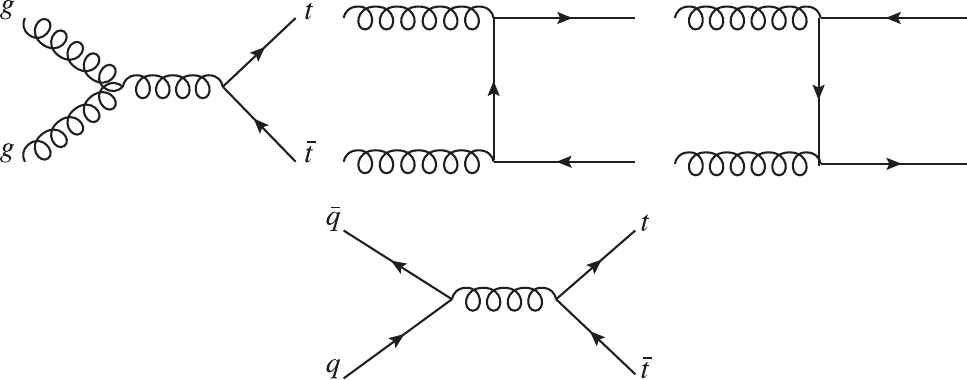}
    \caption{Tree-level QCD diagrams for the  partonic subprocesses $gg \to t \bar t$ (first line) and  $q \bar q \to t \bar t$ (second line). These are the diagrams entering the quantity $\MzSM$ discussed in the text.}
    \label{fig:SMtt}
\end{figure}

We start by describing the calculation of the one-loop corrections induced by a scalar $S$ to the hadroproduction of a top-quark pair. This process stems at the tree level from the $q\bar q$ and $gg$ initial states. The dominant contributions of the Leading Order (LO) prediction is proportional to $\alpha_S^2$ and originates from the squared amplitudes associated to the diagrams depicted in Fig.~\ref{fig:SMtt}. The one-loop corrections originate instead from the diagrams of Fig.~\ref{fig:diagrams}, which are in general UV divergent and they have to be renormalised in order to obtain finite contributions.

Without specifying the initial state, we denote as $\MzSM$ the leading\footnote{In principle also tree-level diagrams featuring EW interactions are possible for the $q\bar q$ initial state, but they lead to smaller contributions. They  are discussed later in Sec.~\ref{sec:SMcombination} and lead to the contributions denoted as $\sigma_{\rm LO_2}$ and  $\sigma_{\rm LO_3}$.} SM tree-level amplitude associated to the diagrams in Fig.~\ref{fig:SMtt} and  $\MoNP$ the {\it renormalised} one-loop amplitude featuring the scalar $S$, defined as
\be
\MoNP=\MoNPunren+\MoNPCT\,,
\ee
with $\MoNPunren$ being the {\it unrenormalised} amplitude and the $\MoNPCT$ the amplitude containing the UV counter terms. 

The following relations hold
\bea
\MzSM&\propto& \alpha_S\,,\\
\MoNPunren&\propto& \alpha_S \, |\Ct|^2 \Longrightarrow \MoNP, \,\MoNPCT\propto \alpha_S \, |\Ct|^2\, , \label{eq:MoNPunren}
\eea
which bear consequences in the calculation and especially in the renormalisation of $\MoNPunren$. 

First, since we do not include EW interactions at the LO, the relevant Lagrangian for this calculation is a subset of $\mathcal{L}_{{\rm SM}+S}$ in Eq.~\eqref{eq:LangS}, namely,
\bea
\mathcal{L}_{{\rm QCD}+S}\equiv \mathcal{L}_{{\rm QCD}} +  \mathcal{L}_{S} + \Lint\, \,, \label{eq:LangSQCD}
\eea
with 
\begin{equation}\label{eq:LQCD}
    \mathcal{L}_{\rm QCD} =-\frac{1}{4}G^{a}_{\mu\nu}G^{a,\mu\nu}+ \overline \psi_t(i \slashed \partial-m_t)\psi_t-g_s \overline\psi_t^{\,i} \gamma^\mu t^{a}_{ij} G^{a}_{\mu} \psi_t^j\, , 
\end{equation}
which is  the QCD Lagrangian with only one fermion, the top quark. This means that the only interaction that can be renormalised is the one between the gluons and the top quark.

Second, according to Eq.~\eqref{eq:MoNPunren}, one-loop corrections can in principle induce effects of order $\ct^2$, $\ctt^2$ and also $\ct \ctt$. This means that also the counterterms can include these same three classes of corrections, such that all UV divergencies are cancelled. 

\begin{figure}[!t]
    \centering
    \includegraphics[width=0.74\textwidth]{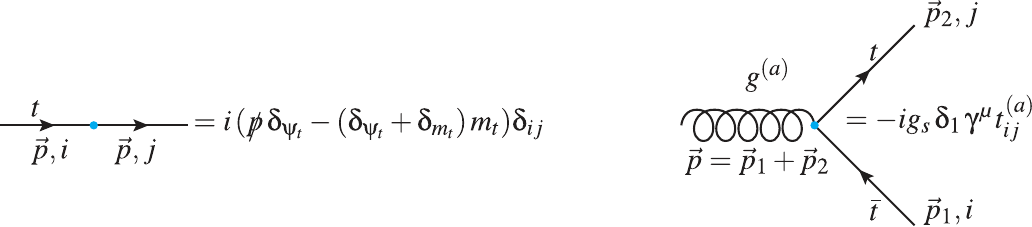}
    \caption{Feynman rules for the UV vertex counterterms entering our calculation, for which  Eq.~\eqref{eq:d1eqdpsit} fully defines $\delta_1$. \label{fig:counterterms} }
\end{figure}

Third, since only QCD interactions are present in $\MzSM$, the quantities that have to be renormalised are only those related to $\mathcal{L}_{\rm QCD}$. In particular, denoting now with a $0$ pedex the bare quantities in $\mathcal{L}_{\rm QCD}$ and rewriting them as
\begin{align}\label{eq:ren}
\psi_{t,0} = \sqrt{Z_{\psi_t}} \psi_t \equiv&\left(1+\frac{1}{2}\delta_{\psi_t}\right)\psi_t\,,\\
G_0 =\sqrt {Z_G} G \equiv&\left(1+\frac{1}{2}\delta_G\right) G\,,\\
g_{s,0} = Z_{g_s} g_s \equiv &\left(1+\delta_{g_s}\right) g_s\,,\\
m_{t,0} = Z_{m_t}m_t \equiv &\left(1+\delta_{m_t}\right) m_t\,,
\end{align}
we obtain
\be
\mathcal{L}_{\rm QCD} \Longrightarrow \mathcal{L}_{\rm QCD} +\mathcal{L}_{\rm CT}\label{eq:LQCDren}\,,
\ee
with the Lagrangian containing the counterterms equal to 
\be
\mathcal{L}_{\rm CT} \equiv i\delta_{\psi_t}\overline \psi_t \slashed \partial\psi_t-(\delta_{\psi_t}+\delta_{m_t})m_t \overline \psi_t\psi_t-g_s\delta_1\overline\psi_t^{\,i} \gamma^\mu t^{a}_{ij} G^{a}_{\mu} \psi_t^j+\cdots\,, \label{eq:LCT}
\ee

\noindent where retaining only the terms proportional to $|\Ct|^2$,

\be
Z_1\equiv 1+\delta_1 = \sqrt Z_G Z_{g_s}Z_{\psi_t} \to  \delta_1=\frac{1}{2}\delta_G+\delta_{g_s}+\delta_{\psi_t}\,.
\ee
In Eq.~\eqref{eq:LCT} we have omitted the part related to the gluon kinetic term, since at one loop the scalar $S$ does not affect the gluon field and therefore neither the wave function normalisation  $Z_G$. Moreover, no $S$ loop is entering the three- or four-gluon vertex and so also $\alpha_s$ cannot receive corrections. Thus, we can directly set to zero both the wave-function counterterms for the gluon field 
and for the QCD gauge coupling, leading to 
\bea
\delta_G&=&0\,,\\
\delta_{g_s}&=&0\,,\\
\delta_1&=&\delta_{\psi_t}\,. \label{eq:d1eqdpsit}
\eea
In conclusion Eq.~\eqref{eq:LCT} can be rewritten as 
\be
\mathcal{L}_{\rm CT} = i\delta_{\psi_t}\overline \psi_t \slashed \partial\psi_t-(\delta_{\psi_t}+\delta_{m_t})m_t \overline \psi_t\psi_t-g_s\delta_{\psi_t}\overline \psi_t^{\,i} \gamma^\mu t^{a}_{ij} G^{a}_{\mu} \psi_t^j\,.\label{eq:LCTok}
\ee
The associated Feynman rules are shown in Fig.~\ref{fig:counterterms}
and we need to calculate only the counterterm for the top-quark wave function ($\delta_{\psi_t}$) and for its mass ($\delta_{m_t}$).
In order to do so, a scheme has to be specified and we have adopted the on-shell scheme, which we briefly describe in the following for our calculation. 

If we  denote the renormalised two-point Green function associated to the top-quark propagator as $G_t( p)$, by taking into account one-loop corrections via Dyson summation, $G_t( p)$ is defined as 
\begin{equation}\label{eq:Greent}
iG_t(p)\equiv\frac{i}{\slashed p-m_t+\Sigma( p)}\, , 
\end{equation}
where $\Sigma( p)$ are the renormalised one-loop corrections induced to the top-quark propagator by $S$, which corresponds to 
 \begin{equation}\label{eq:Sigmat}
\Sigma(p)\equiv \widehat \Sigma^{1}_{\rm NP}( p)+\slashed p\delta_{\psi_t}-(\delta_{\psi_t}+\delta_{m_t})m_t\,,
\end{equation}
with $\widehat \Sigma^{1}_{\rm NP}$ being the UV divergent one-loop corrections (the associated diagram is shown in Fig.~\ref{fig:1looplegt} in Appendix \ref{app:UVcounterterms},  where we also calculate it) while the remaining contributions originate from the associated UV counterterm. 
The on-shell scheme corresponds to requiring
\bea
\Re \left[G^{-1}_t(p)\right] u_t(p)\bigg|_{p^2=m_t^2}&=&0\,,\label{eq:onshell1}\\
\lim_{p^2\to m_t^2} \frac{\slashed p + m_t}{p^2-m_t^2} \Re \left[G^{-1}_t(p)\right] u_t(p) &=&  u_t(p) \label{eq:onshell2} \,,
\eea
where $u_t(p)$ is the polarisation of the top quark.
The quantity $\widehat \Sigma^{1}_{\rm NP}$, as it can be explicitly seen in Appendix \ref{app:loopintegrals},  has the following form
\begin{equation}\label{eq:Sigma1NP}
\widehat \Sigma^{1}_{\rm NP}( p)=\slashed p \Sigma_{V}(p^2) + m_t \Sigma_{S}(p^2) + i\gamma^5 m_t  \Sigma_{P}(p^2)\,,
\end{equation}
and has an important consequence.

The conditions \eqref{eq:onshell1} and \eqref{eq:onshell2} lead to
\bea
\delta_{m_t}&=&\Re[\Sigma_{S}(m_t^2)+\Sigma_{V}(m_t^2)]\,, \label{eq:defdeltamt}\\
\delta_{\psi_t}&=&-\Re[\Sigma_{V}(m_t^2)]-m_t^2 \Re\left[\frac{d}{dp^2}\left(\Sigma_{V}(p^2)+\Sigma_{S}(p^2)\right)\bigg|_{p^2=m_t^2}\right]\label{eq:defdeltaZt}\,.
\eea

Neither in $\delta_{m_t}$ nor in $\delta_{\psi_t}$ the quantity $\Sigma_{P}$ appears due to the presence of the $\Re$ function, and the fact that $\Sigma_{P}$ itself is real. Yet this component in $\widehat \Sigma^{1}_{\rm NP}$ is UV divergent. In order to cancel this UV divergence we should include an additional term $- i \tilde m_t\overline \psi_t  \gamma_5 \psi_t $  in the Lagrangian $\Lint$.~\footnote{In the case $S=H$ the top-Higgs interaction and the top-mass arise from the same operator, as required by $SU(2)$ invariance. In this case, it is necessary to introduce a higher-dimension SMEFT operator in order to consistently modify the top-Higgs interactions without modifying the mass of the top quark. The operator is precisely the one shown later in Eq.~\eqref{eq:SMEFT} of Sec.~\ref{sec:SMEFT}, where the SMEFT point of view is discussed. The $- i \overline \psi_t  \tilde m_t\gamma_5 \psi_t H$ term in $\Lint$ leading to the necessary counterterm, can be obtained after EWSB without imposing in Eq.~\eqref{eq:SMEFT} the $-v^2/2$ term or using the same approach of Ref.~\cite{Maltoni:2018ttu}.} Having said that, for our calculation $\Sigma_{P}(p^2)$ and the associated complications due to the renormalisation can be completely neglected, as we explain in the following.

For both the $q\bar q$ and $gg$ initiated partonic processes we need to calculate, for a generic  observable, the associated cross section $\sigma$. We understand here that $\sigma$ could be also at differential level and we define $\sigma$ at LO as $\SigmaLOQCD$ and including also NP effects as $\SigmaLOQCDNP=\SigmaLOQCD+\SigmaNP$, with
\bea
\SigmaLOQCD&\Longleftarrow&|\MzSM|^2\propto\alpha_S^2\,,\\
\SigmaNP&\Longleftarrow&2\Re\left[\MzSM (\MoNP)^*\right] \propto\alpha_S^2 |\Ct|^2 \, \label{eq:proprtoRE}.
\eea
Because of the optical theorem, $2\to2$ tree-level amplitudes as  $\MzSM$ have to be real and therefore any contribution that has an imaginary component in $\MoNP$ is not entering $\SigmaNP$. In other words, even if $\Sigma_{P}$ is divergent and not renormalised, it is not entering our results, similarly to any other possible imaginary component from loop diagrams in $\MoNP$. We will understand this technical detail in the following, but it is important to remember that for a generic calculation an additional counterterm $\delta_{\tilde m_t}$ is  necessary and cannot be ignored. The presence of the subscript QCD in $\SigmaLOQCD$ and $\SigmaLOQCDNP$ is a reminder that at the tree level additional perturbative orders are possible; this will be manifest in Sec.~\ref{sec:SMcombination}.

\begin{figure}
    \centering
    \includegraphics[width=0.6\textwidth]{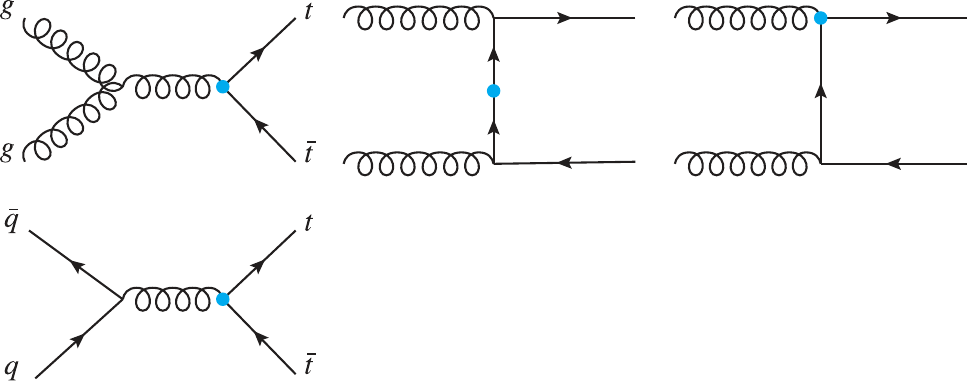}
    \caption{Representative diagrams for $gg\to t \bar t $ (first row) and $q \bar q \to t \bar t $ (second row) entering $\MoNPCT$.}
    \label{fig:countertt}
\end{figure}

Once the quantities $\Sigma_{V}$ and $\Sigma_{S}$ are known, it is possible to calculate $\MoNP$ for both the $q\bar q$ and $gg$ initiated processes. The unrenormalised one-loop amplitude $\MoNPunren$ corresponds to the calculation of the one-loop diagrams in Fig.~\ref{fig:diagrams}, excluding those with a loop on the external top-quark legs. Following the LSZ theorem for scattering amplitudes, in the on-shell scheme the associated contribution is exactly cancelled by the $\sqrt{Z_{\psi_t}}$ normalisation of the external field. The amplitude associated with the UV counter terms $\MoNPCT$ corresponds to the diagrams in Fig.~\ref{fig:countertt}, where the Feynman rules are specified in Fig.~\ref{fig:counterterms} and in turn by Eqs.~\eqref{eq:defdeltamt} and \eqref{eq:defdeltaZt}. 
\medskip

For our study, we have generated a NLO {\UFO}~model \cite{Degrande:2011ua, Darme:2023jdn} via the help of {\sc \small NLOCT} \cite{Degrande:2014vpa} and {\sc \small FeynRules} \cite{Alloul:2013bka} and performed calculations with {\MAD}~\cite{Alwall:2014hca, Frederix:2018nkq}. For the case of the process $q\bar q \to t \bar t$ we have analytically computed the quantity $2\Re\left[\MzSM (\MoNP)^*\right]$ finding perfect agreement with the results from {\MAD}~and thus validating the {\UFO}~model.

We emphasise that such {\UFO}~model is very similar to the one used to perform the study of top-quark pair production of Ref.~\cite{Blasi:2023hvb}, where instead of a generic scalar $S$ a top-philic ALP $a$ has been considered. The main differences are two. First, in the case of the ALP an additional interaction between $a$ and the gluons is present, unlike the case of $S$. The technical details regarding this additional interaction have been described in depth in Ref.~\cite{Blasi:2023hvb}, especially Appendices C and D of that work. Second,  $S$ can feature both CP-odd and CP-even interactions, violating the CP symmetry, while the top-philic  ALP  in the model of Ref.~\cite{Blasi:2023hvb} is itself CP-odd and features only CP conserving interactions, namely $\ct=0$ in the notation of this paper. Excluding the trivial interaction between $a$ and the gluons, the calculation presented in this work is therefore a generalisation of the one presented in Ref.~\cite{Blasi:2023hvb} and we provide in this work, especially in this section and in Appendices \ref{app:UVcounterterms} and \ref{app:qqttanal}, further technical details that were not discussed in Ref.~\cite{Blasi:2023hvb}. For the purely $\ctt$-dependent component and excluding the  interaction between $a$ and the gluons, our calculation is equivalent not only to what has been presented in Ref.~\cite{Blasi:2023hvb}, but also to the results for the $gg\to t \bar t$ process in Ref.~\cite{Phan:2023dqw}.\footnote{In Ref.~\cite{Phan:2023dqw}, however, the $\overline{\rm MS}$ scheme has been used for the renormalisation of the top-quark wave function. This difference has anyway an impact only on Green functions and not on scattering amplitudes, but only the latter are relevant in our study.} 

\medskip

In the generic case for a scalar $S$, $\SigmaNP$ can be written as
\be
\SigmaNP\equiv\Sigmact \ct^2+\Sigmactt \ctt^2+\Sigmactctt \ct \ctt \label{eq:sigmanpincts},
\ee
where the dependence on $\ct$ and $\ctt$ has been made explicit. It is easy to see that the contribution proportional to both $\ct$ and $\ctt$ vanishes, {\it i.e.}
\be
\Sigmactctt =0 \label{eq:zero}\,. 
\ee
This is true not only for the virtual corrections but also in the case of the emission of a  real scalar $S$. This property is also consistent with the fact that $\Sigma_{P}$ is proportional to $\ct\ctt$ and it is divergent. Indeed,  as explained before $\Sigma_{P}$ anyway does not contribute to $\SigmaNP$. 
It is also important to note that the quantities $\Sigmact$ and $\Sigmactt$, which at this stage of the discussion involve only virtual corrections, can also be negative and there can be also large cancellations among them. 

Concerning the CP-violating term $\ct \ctt$, as discussed in the literature \cite{Demartin:2014fia} one can regain sensitivity to it  by considering the decays of the top quarks and building CP-sensitive observables. Since in this work we consider only  observables built from the top momenta before decay, they are not sensitive to the term $\ct \ctt$ and consequently neither to the relative sign between $\ct$ and $\ctt$. 

\medskip

\subsection{SM predictions and combination with NP effects}
\label{sec:SMcombination}

In the previous section we have described the calculation of virtual NP effects induced by a generic scalar $S$  on top of the LO prediction for top-quark pair hadroproduction. The precision reached by experiments nowadays, however, requires that higher-order corrections in the SM have to be taken into account in order to correctly identify these NP effects. In particular, as explained before, NLO EW corrections in the SM contain also the contributions from the Higgs, induced by the same diagrams of Fig.~\ref{fig:diagrams} where the top-Higgs interactions are the SM ones. 

In this section we show how to combine the calculation described in the previous section with predictions at NNLO QCD accuracy and including Complete-NLO corrections from QCD and EW origin. The predictions at this accuracy have been calculated in Ref.~\cite{Czakon:2017wor}, based on the previous results of Refs.~\cite{Czakon:2013goa, Czakon:2016dgf, Pagani:2016caq}. We start by recalling the structure of this calculation.

Using a notation similar to the one adopted in Refs.~\cite{Frixione:2014qaa, Frixione:2015zaa, Pagani:2016caq, Frederix:2016ost, Czakon:2017wor, Frederix:2017wme, Frederix:2018nkq, Broggio:2019ewu, Frederix:2019ubd,Pagani:2020rsg, Pagani:2020mov, Pagani:2021iwa, Pagani:2021vyk}, the (differential) cross section of inclusive top-quark pair production in the SM, $pp\to t \bar t (+X)$ can be written as   
\noindent
\begin{equation}
\SigmaSM (\alpha_s,\alpha) = \sum_{m+n\geq 2} \alpha_s^m \alpha^n \bar \sigma_{m+n,n}\, .
\end{equation}
\noindent
LO predictions correspond to  $m+n=2$, NLO ones to $m+n=3$ and NNLO ones to $m+n=4$
\begin{align}
\sigma_{\rm LO}(\alpha_s,\alpha) &= \alpha_s^2 \, \bar\sigma_{2,0} + \alpha_s \alpha \, \bar\sigma_{2,1} + \alpha^2 \, \bar\sigma_{2,2} \nonumber\\
 &\equiv \sigma_{\rm LO_1} + \sigma_{\rm LO_2} + \sigma_{\rm LO_3}\, ,  \nonumber\\
\sigma_{\rm NLO}(\alpha_s,\alpha) &= \alpha_s^3 \, \bar \sigma_{3,0} + \alpha_s^2 \alpha  \, \bar \sigma_{3,1} + \alpha_s \alpha^2  \, \bar \sigma_{3,2} + \alpha^3  \, \bar\sigma_{3,3} \nonumber\\
 &\equiv \sigma_{\rm NLO_1} + \sigma_{\rm NLO_2} + \sigma_{\rm NLO_3} + \sigma_{\rm NLO_4}\, ,  \nonumber\\
\sigma_{\rm NNLO}(\alpha_s,\alpha) &= \alpha_s^4  \, \bar\sigma_{4,0} + \alpha_s^3 \alpha  \, \bar\sigma_{4,1} + \alpha_s^2 \alpha^2  \, \bar \sigma_{4,2} + \alphas \alpha^3  \, \bar \sigma_{4,3} +  \alpha^4  \, \bar \sigma_{4,4} \nonumber\\
 &\equiv \sigma_{\rm NNLO_1} + \sigma_{\rm NNLO_2} + \sigma_{\rm NNLO_3} + \sigma_{\rm NNLO_4} + \sigma_{\rm NNLO_5} \;.
\label{eq:blobs}
\end{align}
\noindent
In fact, $\sigma_{\rm LO_1}=\SigmaLOQCD$, the quantity introduced already in Sec.~\ref{sec:loopcalculation}, and the so-called NLO QCD and NLO EW corrections are $\sigma_{\rm NLO_1}$ and $\sigma_{\rm NLO_2}$, respectively, denoted also for simplicity as $\SigmaNLOQCD$ and $\SigmaNLOEW$. The Complete-NLO predictions correspond to $\SigmaLO+\SigmaNLO$, while the NNLO QCD corrections are $\sigma_{\rm NNLO_1}$, denoted also for simplicity as $\SigmaNNLOQCD$. The terms  $\sigma_{\rm NNLO_i}$ with $i>1$ are mixed QCD-EW NNLO contributions, which are not available at the moment. 

The current best predictions in the SM are obtained by combining the Complete-NLO predictions with NNLO QCD corrections, which can be achieved in two different ways:
\begin{enumerate}
\item  Additive scheme:
\be
\SigmaSMadd \equiv \SigmaLO+\SigmaNLO+\SigmaNNLOQCD \,.\\
\ee
\item Multiplicative scheme:
\be
\SigmaSMmult \equiv \SigmaSMadd + (\KNLOQCD-1) \, \SigmaNLOEW\,, \label{eq:SigmaSMmult}
\ee
\end{enumerate}

with $\KNLOQCD$ defined as the standard QCD $K$-factor
\be
\KNLOQCD\equiv\frac{\SigmaLOQCD+\SigmaNLOQCD}{\SigmaLOQCD}\,.
\ee

The additive approach, $\SigmaSMadd$, is a rigorous fixed-order calculation deriving from the perturbative expansion in powers of $\alpha_S$ and $\alpha$ (and number of loops). Only known terms in the expansions are kept.  The multiplicative approach relies on the fact that in particular regimes, such as the tails of the distributions in top-quark pair production, QCD and EW dominant corrections factorise and can provide a good approximation of $\sigma_{\rm NNLO_2}$, which can have a non-negligible size. In fact, as we will see in Sec.~\ref{sec:distrct}, the impact of the NP effects considered in this work is large in another region of the phase space, the threshold. The two approaches are somewhat complementary and formally only differ by unknown higher order terms.  

The choice of the additive or multiplicative approach has also an impact on the procedure adopted to obtain predictions including the loop effects induced by the scalar $S$. Similarly to the case of the NLO EW corrections in the SM, one can ``dress'' the NP contributions with NLO QCD corrections, {\it i.e.}, multiplying $\SigmaNP$ by $\KNLOQCD$. In doing so, we assume that this is a good approximation, regardless of the mass of $S$. In this work we consider either both the SM and NP contributions without this rescaling by $\KNLOQCD$, additive approach, or rescaling both of them.
These two approaches correspond to 
\bea
\SigmaSMNPadd \equiv \SigmaSMadd + \SigmaNP\,, \label{eq:addcts}
\eea
in the additive approach and 
\bea
\SigmaSMNPmult \equiv \SigmaSMmult + \KNLOQCD \, \SigmaNP\,, \label{eq:multcts}
\eea
in the multiplicative one.

We have refrained from considering the cases where NLO EW corrections in the SM are not rescaled by $\KNLOQCD$ while the NP ones are instead rescaled and {\it vice versa}. Indeed, both choices are very aggressive in increasing (decreasing) the relative impact of NP effects on the total prediction. In  Ref.~\cite{Czakon:2017wor} also the possibility of rescaling NLO EW corrections not only via NLO QCD corrections but also NNLO QCD ones has been discussed. We have verified that in our work this choice would have a minimal impact and therefore we do not discuss it.

\medskip

In our simulations, for SM predictions we used {\MAD}~for the Complete-NLO predictions, while NNLO QCD corrections have been obtained by interpolating the $K$-factors of NNLO QCD,
\be
\KNNLOQCD\equiv\frac{\SigmaLOQCD+\SigmaNLOQCD+\SigmaNNLOQCD}{\SigmaLOQCD+\SigmaNLOQCD}\,,\label{eq:KNNLO}
\ee
 that can be derived from the  publicly available ancillary files of Ref.~\cite{Czakon:2016dgf}, and rescaling the NLO QCD predictions. To this purpose, we have generated NLO QCD predictions for a generic binning and especially with the same PDF choices of Ref.~\cite{Czakon:2016dgf}: LO, NLO and NNLO five-flavour-scheme replicas from {\sc \small NNPDF3.0} \cite{NNPDF:2014otw}.\footnote{The only subtle point here is that $\KNNLOQCD$ is calculated via the aforementioned ancillary files, where NLO QCD predictions were calculated with NLO PDFs and NNLO predictions with NNLO PDFs. Therefore, in order to obtain NNLO QCD predictions with NNLO QCD PDFs, the NLO predictions to be rescaled via the $K$-factors have to be computed with  NLO PDFs and $m_t=173.3~\gev$.} 
For the same reason, the factorisation and renormalisation scale has been set equal to $H_{T}/4=(E_T(t)+E_T(\bar t))/4$, where $E_T$ is the transverse energy, as done in Ref.~\cite{Czakon:2016dgf}. Having NLO EW corrections, which are evaluated as in Ref.~\cite{Czakon:2017wor} in the so-called $G_\mu$ scheme, the only other remaining free input parameters are 
\be
m_t=172.5~\gev\,,~~~ { m_Z=91.188~\gev\,,}~~~ m_H=125~\gev\,, ~~~ G_F=1.16639~ 10^{-5} \gev^{-2}\,.
\ee
In Appendix \ref{app:table} we explicitly report the individual SM contributions entering $\SMadd$ for the specific binning of the $\mtt$ invariant mass distribution that will be considered in Sec.~\ref{sec:sensitivityH}  and corresponds to the one from the CMS analysis in Ref.~\cite{CMS:2018htd}.

 Last but not least, we anticipate that $\SigmaNP$ is infrared (IR) divergent for $m_S \to 0$ when $\ct\ne 0$ and therefore the contribution of the $t \bar t S$ final state is needed in order obtain reliable predictions for inclusive top-quark pair production. However, we will veto hard radiation of $S$ (as assumed it is experimentally identifiable) and therefore will only include soft contributions in $\SigmaSMNPadd$ or $\SigmaSMNPmult$ of the  fits of Sec.~\ref{sec:Ssensitivity}. We will emphasise it again in Sec.~\ref{sec:Ssensitivity}.

\section{Scalar $S$: Numerical results and phenomenology}
\label{sec:distrct}

\begin{figure}[!t]
    \centering
          \includegraphics[width=0.45\textwidth]{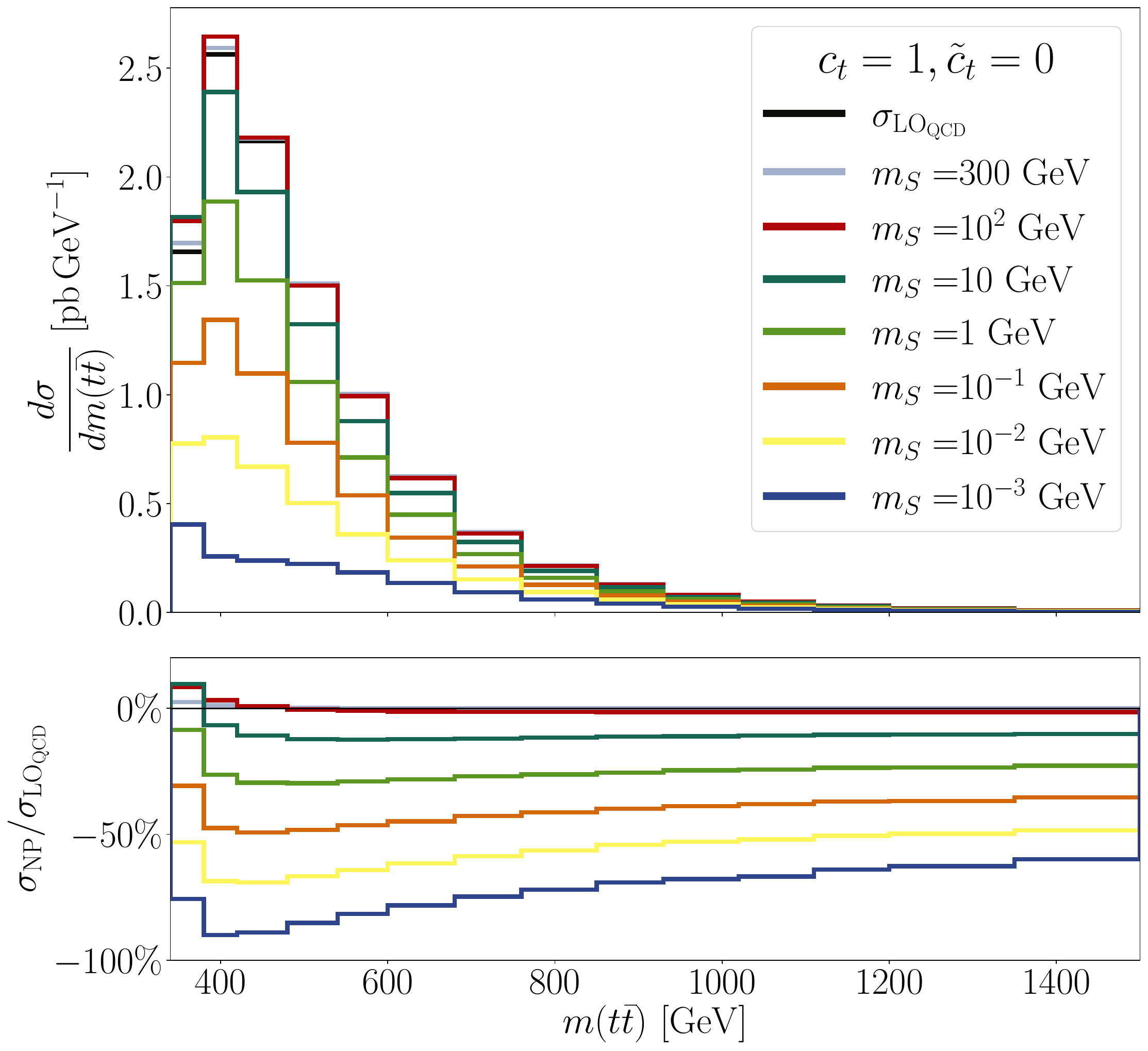}
    \includegraphics[width=0.45\textwidth]{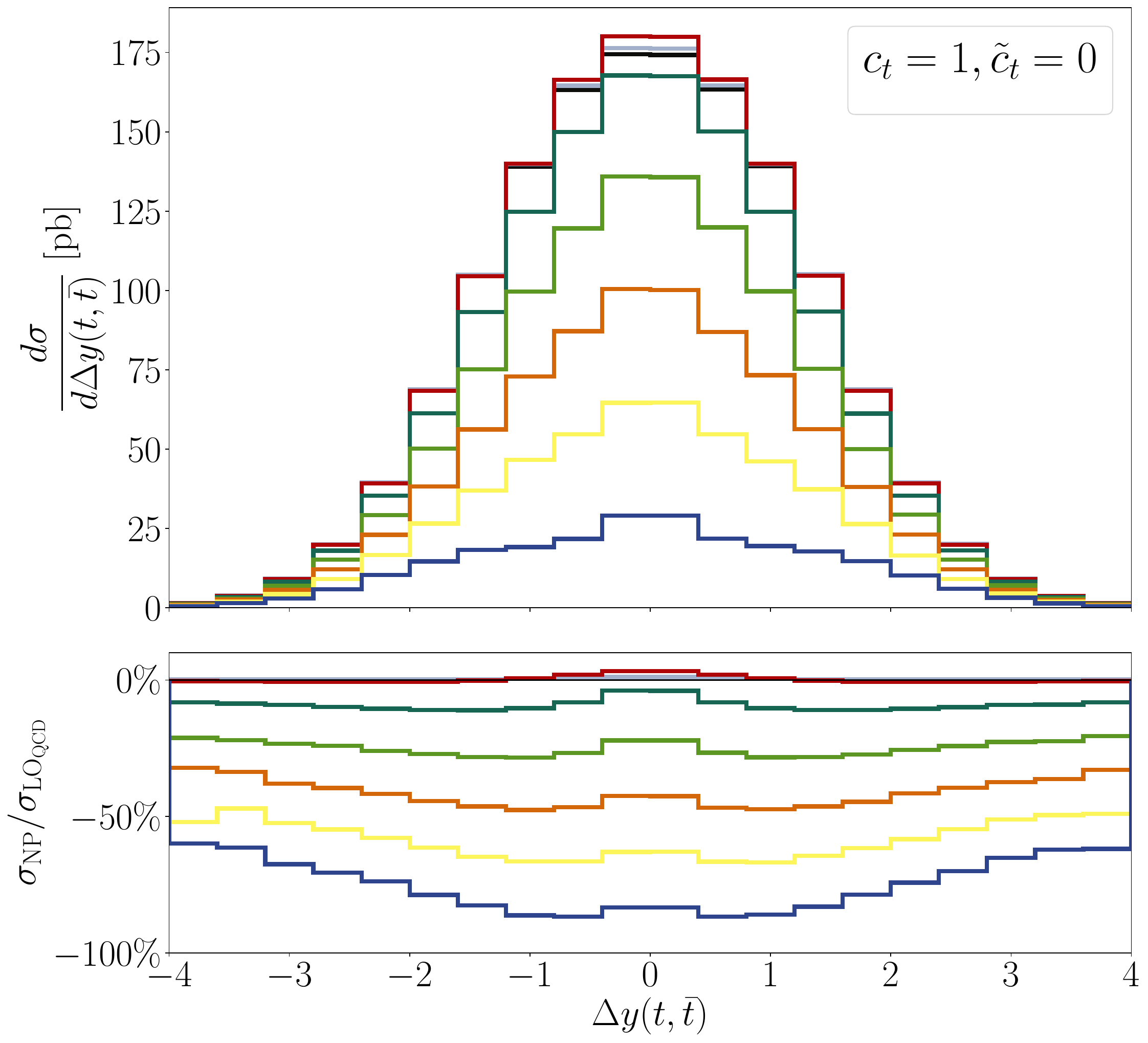}
        \includegraphics[width=0.45\textwidth]{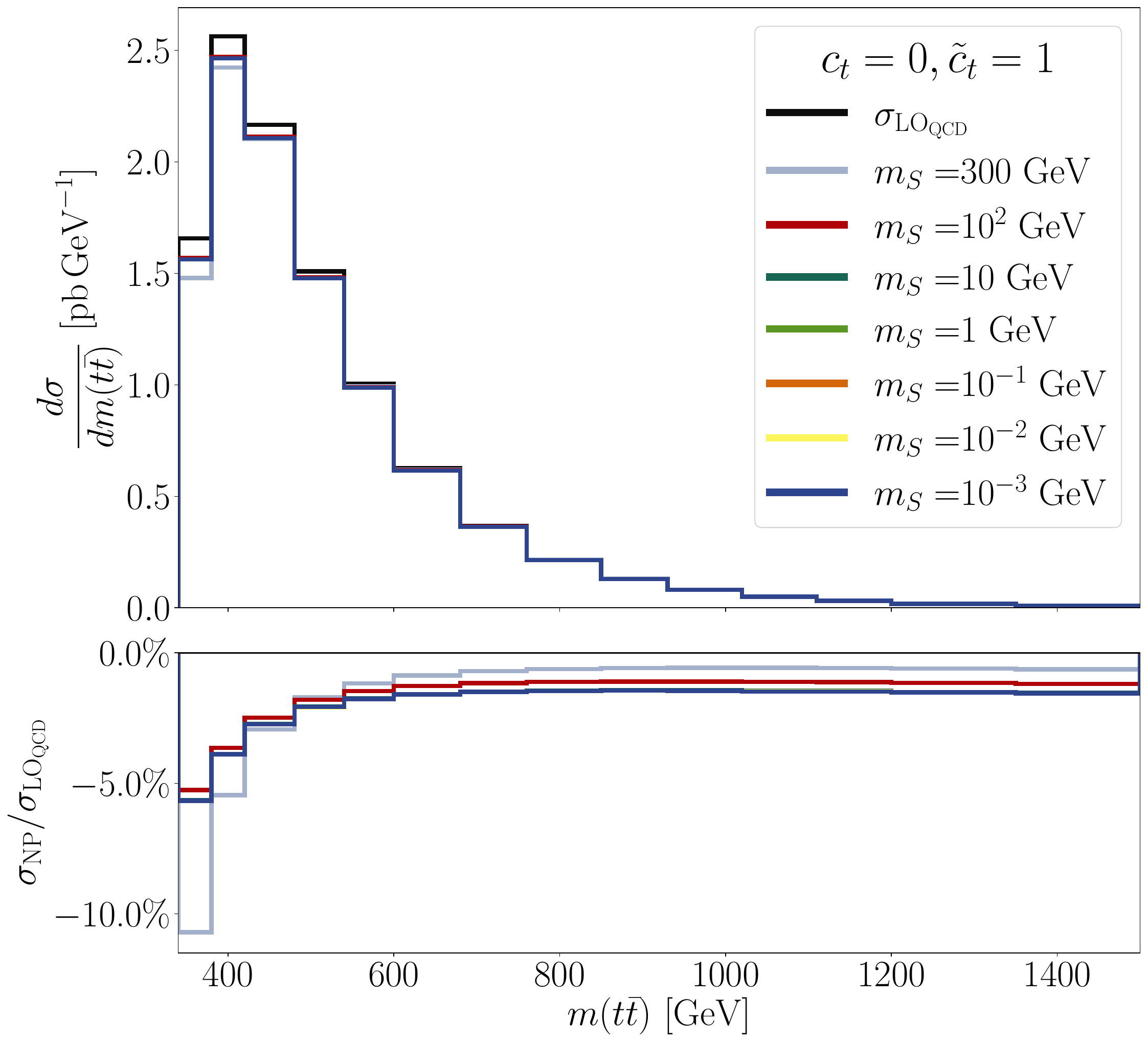}
         \includegraphics[width=0.45\textwidth]{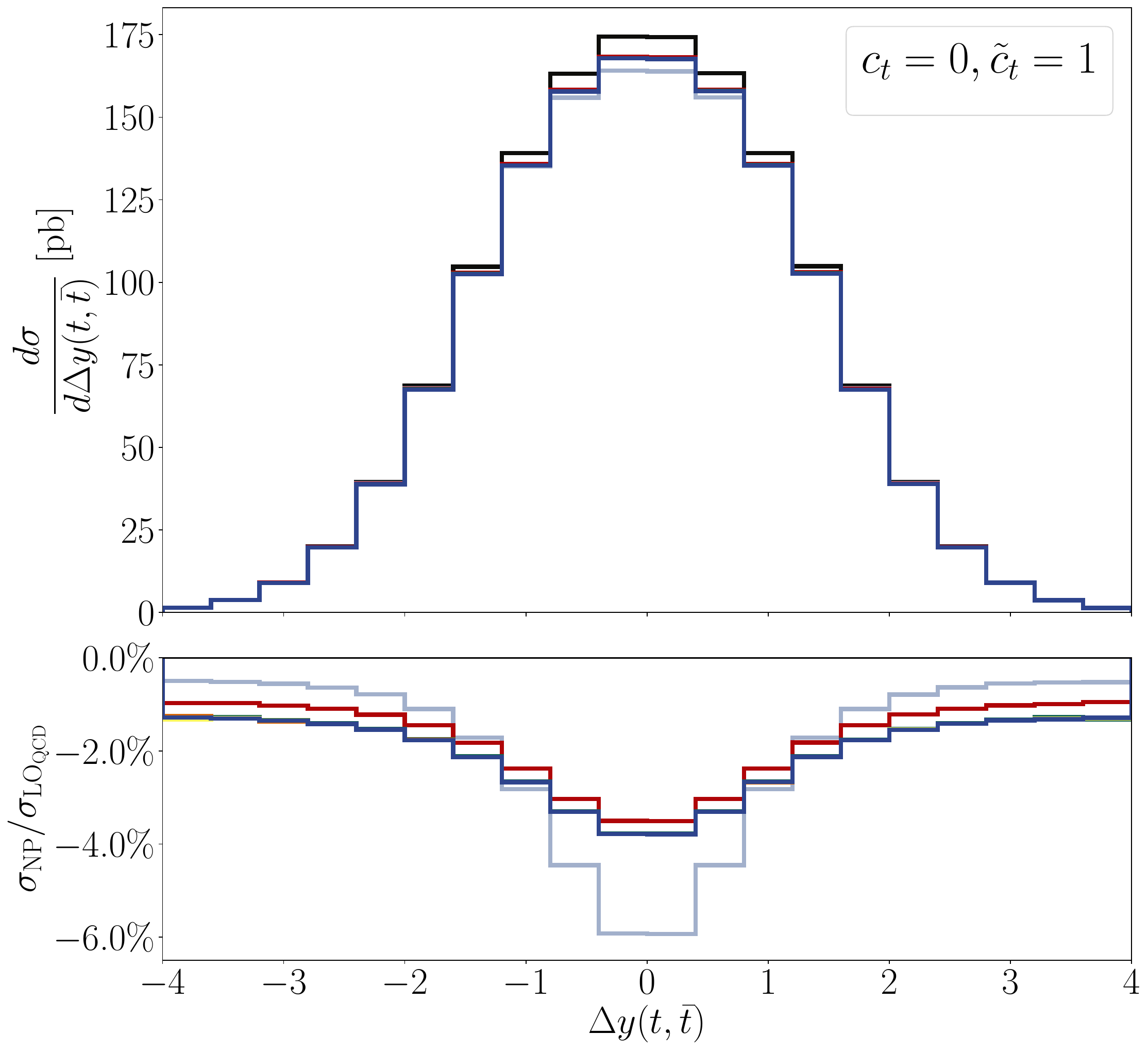}
        \includegraphics[width=0.45\textwidth]{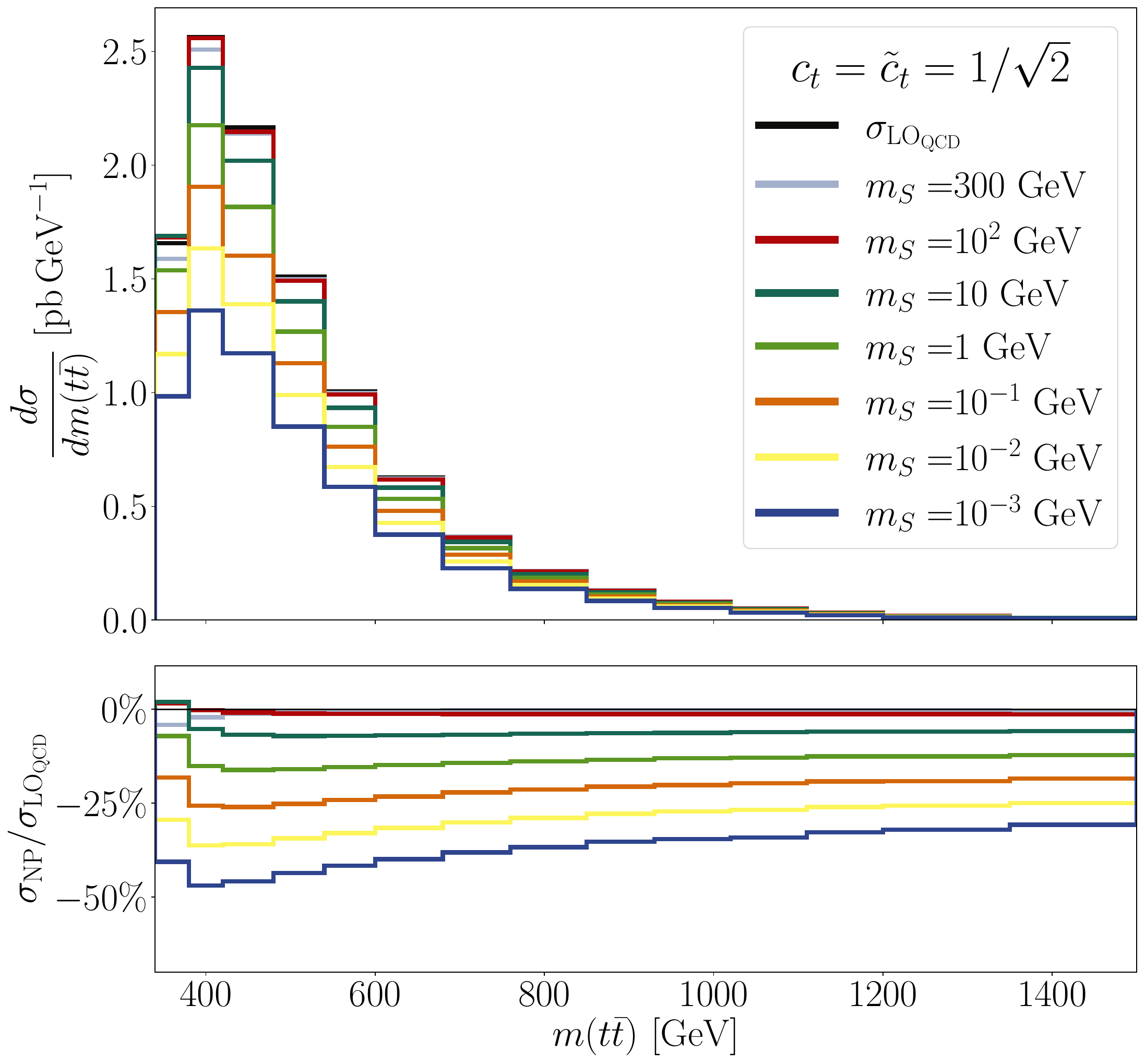}
        \includegraphics[width=0.45\textwidth]{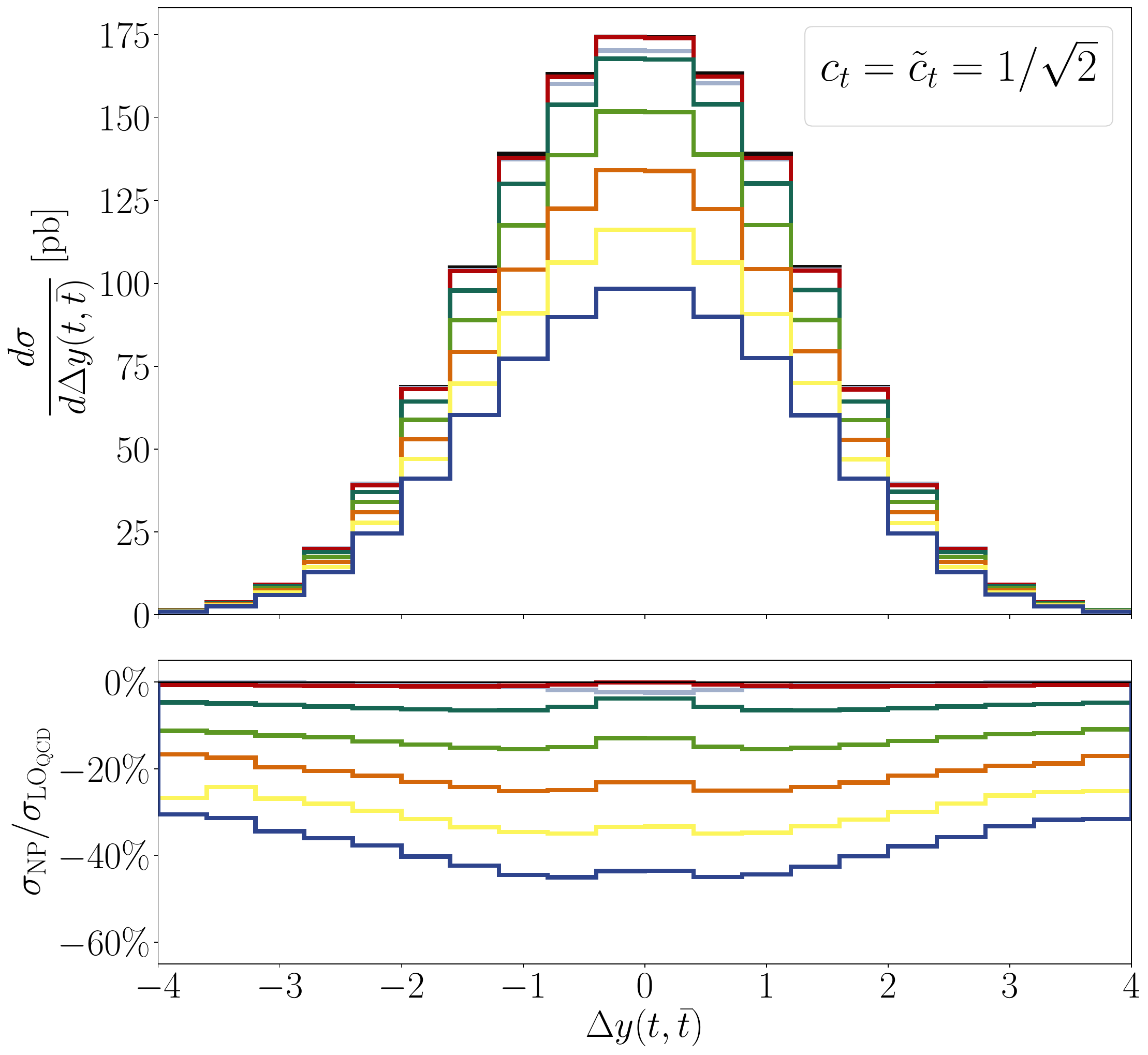}
    \caption{The $\mtt$ (left)  $\Dytt$ (right) distributions for different $m_S$ values considering {\it only virtual contributions}. From top to bottom: purely scalar case, purely pseudoscalar case and maximally mixed case. In the inset of each plot the relative difference w.r.t.~the $\rm LO_{QCD}$ is plotted. As commented in the text, the lines corresponding to the various scenarios are mostly superposed and therefore not distinguishable for the purely pseudoscalar case. }
    \label{fig:differetmass}
\end{figure}

Starting from the Lagrangian in Eq.~\eqref{eq:intL}, which describes the dynamics of the SM with the addition of a scalar $S$ that couples only to the top quark, in Sec.~\ref{sec:loopcalculation} we have calculated the one-loop corrections induced by $S$ to the cross section for the hadroproduction of a top-quark pair, denoted as $\SigmaNP$. This quantity originates from the diagrams in Fig.~\ref{fig:diagrams} and, as shown in Eq.~\eqref{eq:sigmanpincts} depends on  the squared couplings for the CP-even and CP-odd interactions of $S$ with the top quark, $\ct^2$ and $\ctt^2$, and the quantities $\Sigmact$ and $\Sigmactt$ that are fully differential functions of the momenta of the top-quark pair. 

In this section we show and discuss our numerical results for $\SigmaNP$. In order to better understand the underlying dynamics, we do not consider all the higher-order SM corrections introduced in Sec.~\ref{sec:SMcombination} and we just look at the relative size of $\SigmaNP$  w.r.t.~$\SigmaLOQCD$, the LO  cross-section associated to the diagrams of Fig.~\ref{fig:SMtt}. As already said in Sec.~\ref{sec:loopcalculation}, we define $\SigmaLOQCDNP=\SigmaLOQCD+\SigmaNP$.

We start by showing  distributions for $\mtt$, the top-quark pair invariant mass, and $\Dytt\equiv y(t)-y(\bar t)$, the difference between the top quark and antiquark rapidities.~\footnote{These two distributions are the same considered in Ref.~\cite{Martini:2021uey}. For other distributions the behaviour is analogous to that shown here and therefore we do not display them.  For instance, the transverse momentum of the top-quark shows similar features of $\mtt$ and the top-quark rapidity shows similar features of $\Dytt$.} To make the discussion easier to follow, we focus on three benchmarks, all with $|\Ct|^2=\ct^2+\ctt^2=1$, that we list in the following:
\begin{enumerate}
\item Purely Scalar : $(\ct, \ctt)=(1,0)\Longleftrightarrow (|\Ct|,\phiC)=(1,0)$, 
\item Purely Pseudoscalar: $(\ct, \ctt)=(0,1)\Longleftrightarrow  (|\Ct|,\phiC)=(1,\pi/2)$, 
\item Maximally mixed: $(\ct, \ctt)=(1/\sqrt{2},1/\sqrt{2})\Longleftrightarrow  (|\Ct|,\phiC)=(1,\pi/4)$. 
\end{enumerate}
Given the dependence of $\SigmaNP$ on only $\ct^2$ and $\ctt^2$, with no mixed $\ct\ctt$ terms, the first two cases are sufficient for extrapolating the size of $\SigmaNP$ in any configuration; they correspond to $\Sigmact$ and $\Sigmactt$ in Eq.~\eqref{eq:sigmanpincts}, respectively. The third case is useful to assess the presence of possible cancellations, which as we will see, happen at the threshold where $\Sigmact$ is positive while $\Sigmactt$ is negative. 

In Fig.~\ref{fig:differetmass} we show the $\mtt$ (left) and $\Dytt$ (right) differential cross sections for $t \bar t$ production. The three rows correspond to the three aforementioned benchmarks.
In the main panel we show  $\SigmaNP$ for different values of $m_S$, which we have considered in a range from 1 MeV up to values below the $t\bar t$ threshold $2m_t$, {\it i.e.}, up to 300 GeV, avoiding the case of a resonant $S$ in the one-loop $s$-channel diagram in the $gg\to t \bar t$ amplitude.\footnote{Clearly, if $m_S\ge 2m_t$, the one-loop $s$-channel  diagram is dominated by $S$ production and subsequent decay into top-quark pair. In this scenario, the sensitivity would be directly on $S$ production and $t \bar t$ would be the signature in the resonant decay and the strategy also at the experimental level would be completely different.} In the inset we show the $\SigmaNP/\SigmaLOQCD$ ratio, {\it i.e.}, the relative size of the NP effects. 

The most striking feature of Fig.~\ref{fig:differetmass} is the very different qualitative behaviour for the purely CP-odd case (central row) and the other two cases. The CP-even case (upper row) features a large dependence on $m_S$, while the CP-odd does not, with the exception of the two values $m_S=300~\gev$ and $m_S=100~\gev$. The origin of this difference is due to the fact that the quantity $\Sigmact$ is infrared (IR) divergent for $m_S\to 0$, while $\Sigmactt$ is not IR divergent. The different values for $m_S=100~\gev$ and especially  $m_S=300~\gev$ within the purely pseudoscalar case are instead due to the one-loop  $s$-channel diagram in the $gg\to t \bar t$ amplitude; for $m_S$ approaching the value $2m_t$ this diagram approaches the resonant configuration and leads to non negligible contributions. Still, this effect is only at the few percent level for $|\Ct|=1$, while the mass dependence in the purely scalar case is much larger. This also explains while the maximally mixed case (lower row), which is a mixture of the other two, appears in general much more similar to the purely scalar one. Besides the threshold region for the $\mtt$ distribution, consistently with Eq.~\eqref{eq:sigmanpincts}, we can see from the insets  that the relative size of one-loop corrections in the maximally mixed case is half of  the purely scalar one.

The IR sensitivity of the purely scalar case, and therefore $\Sigmact$, can also be better understood by looking at the expressions in Appendix \ref{app:qqttanal}, where it is explicitly shown for the $q \bar q \to t t~$ process that $\Sigmact$, in the $m_S\to 0$ limit, contains terms that are proportional to $\log(m_S^2/Q^2)$, where $Q$ is one of the other scales of the process, as $\sqrt{s}$ or $m_t$. Conversely, the explicit expressions reported in Appendix \ref{app:qqttanal} show that such IR divergences are not present for $\Sigmactt$.\footnote{Actually, although the content of Appendix \ref{app:qqttanal} cannot be directly extended to the $gg\to  t \bar t$ the quantity $2f_1(s)$, with $f_1$ defined in Eq.~\eqref{eq:f1ren}, and be used in the limit $m_S\to 0$, Eq.~\eqref{eq:IRf1explicit}, to correctly quantify in the top-left plot of Fig.~\ref{fig:differetmass} the difference between the lines at different $m_S$ values.} Such IR divergencies are unphysical, in the sense that experimentally the exclusive production of a $t \bar t$ pair and no $S$ emissions {\it at all}, especially for soft $S$ with small $m_S$, is not realistic; the emission of at least soft $S$ on top of $t \bar t$ has to be considered and leads to the cancellations of the aforementioned  $\log(m_S^2/Q^2)$ terms.

\begin{figure}
    \centering
    \includegraphics[width=0.33\textwidth]{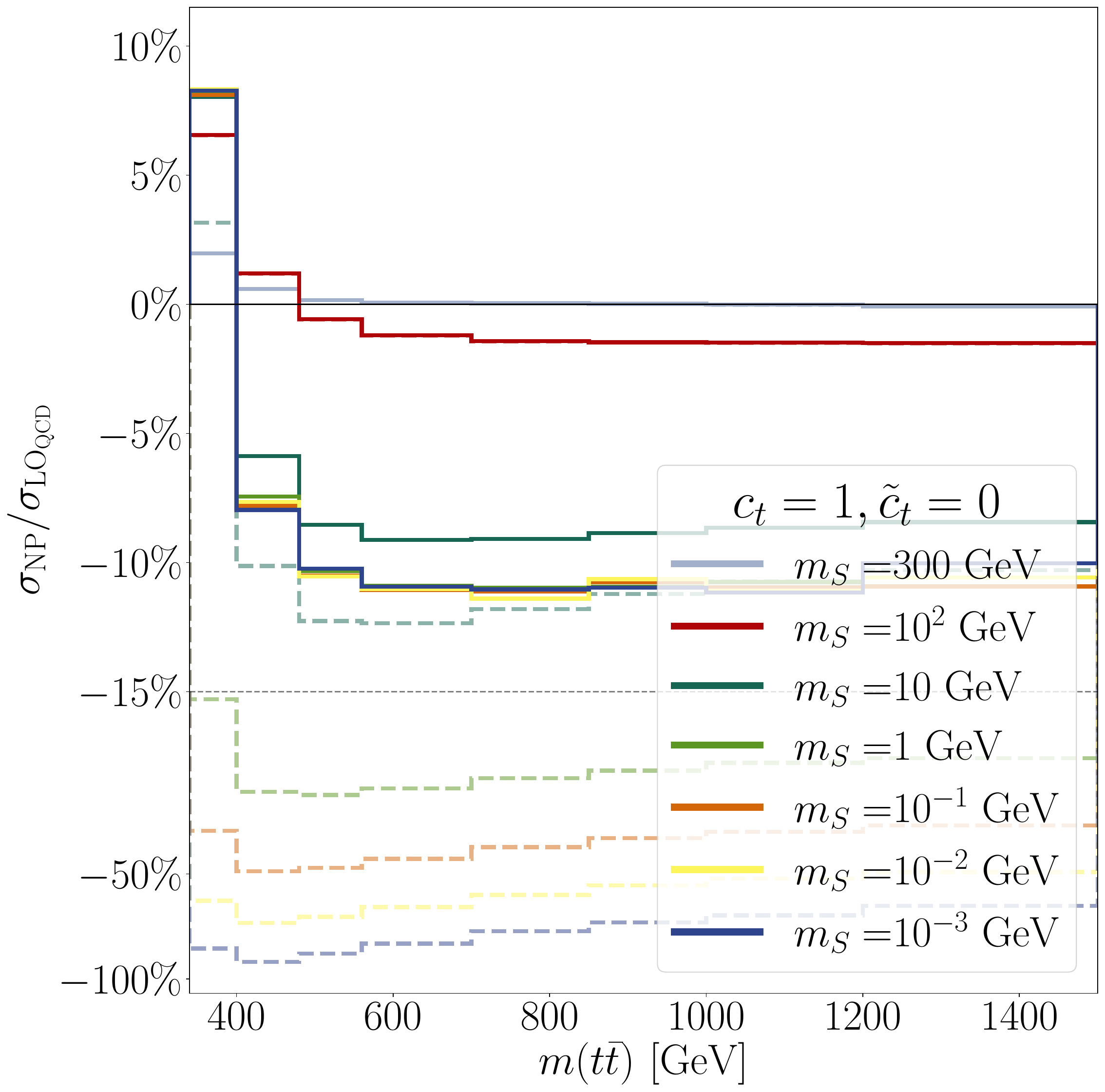}
    \includegraphics[width=0.32\textwidth]{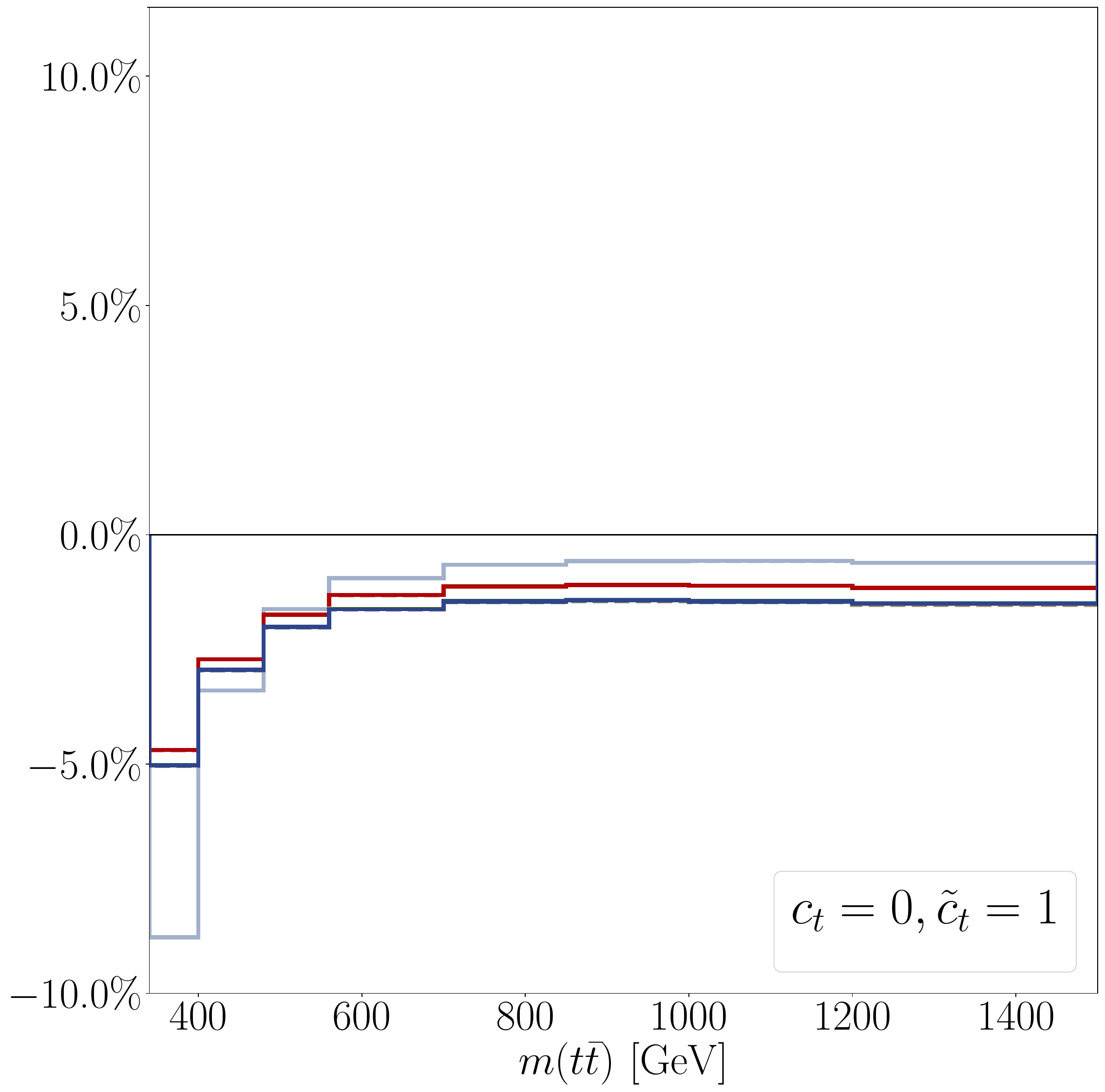}
    \includegraphics[width=0.32\textwidth]{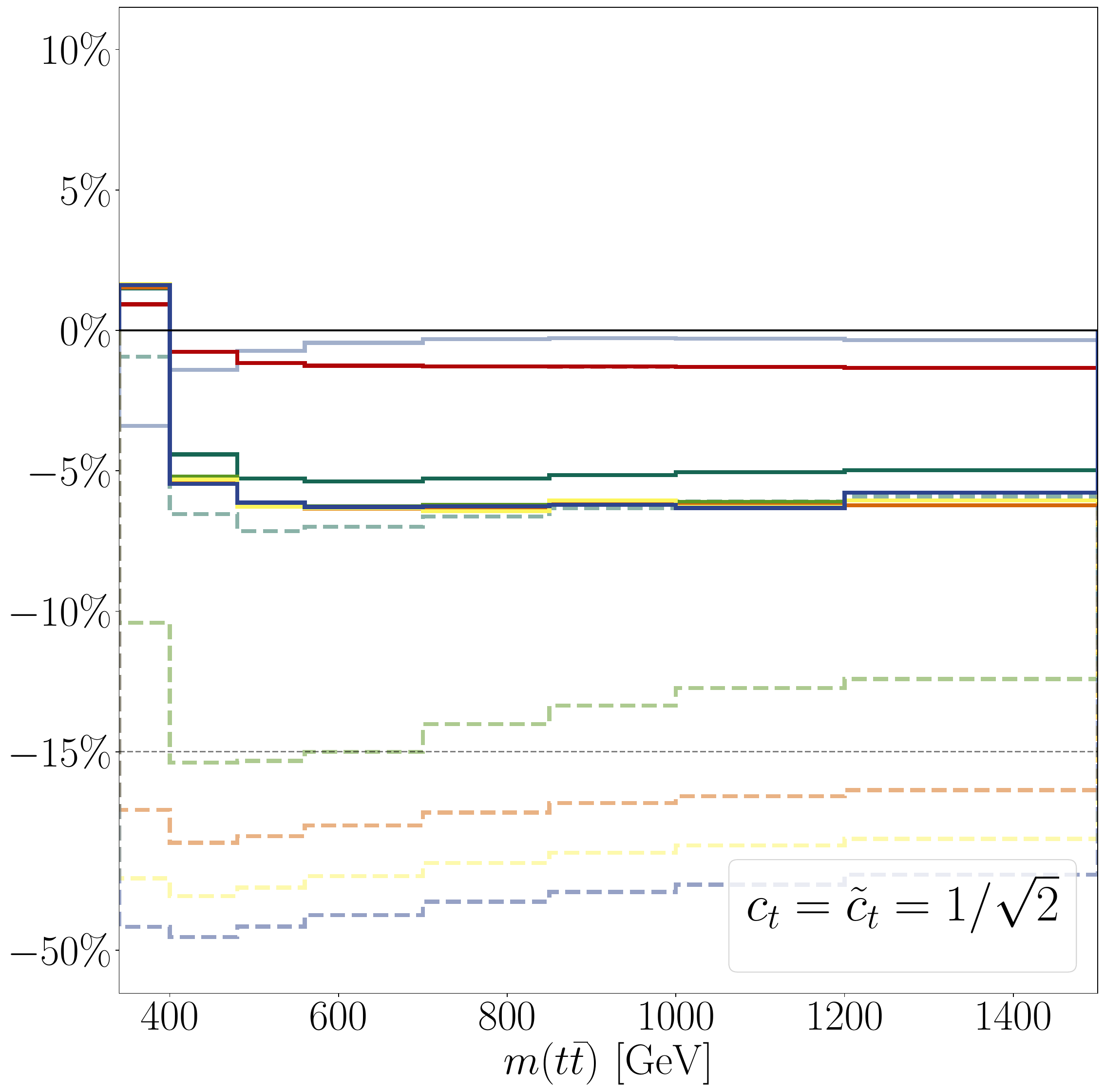}
    \caption{The same information of the insets of the left plots of Fig.~\ref{fig:differetmass}, $\mtt$ distributions, but now solid lines take into account both the virtual corrections and the real emission of the scalar $S$, while the dashed lines only the virtual corrections (as the solid in Fig.~\ref{fig:differetmass}). From left to right: the purely scalar benchmark, the purely pseudoscalar benchmark and the maximally mixed one. The veto \eqref{eq:cutptS} is imposed on $S$ real emissions. Below the value of $15\%$, the horizontal dotted line, the vertical scale switches from linear to logarithmic.}
    \label{fig:realemission}
\end{figure}

We therefore then consider the case in which on top of the one-loop virtual corrections, also the contribution from $pp\to t \bar t S$ production is taken into account in $\SigmaNP$. Clearly, if $S$ is emitted with a large transverse momentum, regardless of the possible signature emerging from $S$ itself, the event will be directly identified through a different signature than just top-quark pair production. On the contrary, if $S$ is soft and undetectable the contribution of  $pp\to t \bar t S$ will ``contaminate'' the signature from inclusive $t \bar t$ production. To this purpose, we apply a cut 
\be
p_T(S)<20~\gev\,, \label{eq:cutptS}
\ee
in order to veto hard emissions of $S$. If the decay products of $S$ were detectable, it would be much more efficient to look directly to  $pp\to t \bar t S$  production, or even just $S$  production,  with the subsequent  $S$ decay. Thus, this scenario is not particularly interesting for the study that we are performing, which primarily targets scenarios where $S$ is undetectable and therefore direct production is not sensitive to them.

In Fig.~\ref{fig:realemission} we show, for the $\mtt$ distributions already considered in Fig.~\ref{fig:differetmass}, the  $\SigmaNP/\SigmaLOQCD$ ratio, {\it i.e.}, the relative size of the NP effects. Solid lines include the contributions from real emissions of $S$, with the cut in \eqref{eq:cutptS} applied. Dashed lines originate from virtual contributions only and therefore are the same lines of the insets of Fig.~\ref{fig:differetmass}, where they were instead drawn as solid.
In principle one could also apply a different cut than the one in \eqref{eq:cutptS}, which is quite stringent from an experimental point of view, but it will be manifest in Sec.~\ref{sec:Ssensitivity} that relaxing this cut solid results are basically unchanged. Thus, the following discussion still holds true also for a less stringent veto on the $S$ radiation. 

In Fig.~\ref{fig:realemission} we see a different picture w.r.t.~Fig.~\ref{fig:differetmass}. For the purely pseudoscalar case (central plot) the real emission is negligible, regardless of the masses considered; solid and dashed lines overlap. In the purely scalar case (left plot), the IR sensitivity to $m_S$ is removed by including the soft emission of $S$, so solid and dashed lines are very different for light masses, with the difference between dashed and solid growing for smaller masses and becoming indistinguishable from $m_S=100~\gev$. On the contrary, solid lines almost overlap for $m_S<1~\gev$, indeed for these values the IR sensitivity has been removed and any possible power corrections of the form $m_S^2/Q^2$, where $Q$ is one of the scales of the process, is negligible. For $m_S>1~\gev$ such power corrections are instead non-negligible. The maximally mixed case (right plot) shows distributions that correspond to the average of the analogous ones in the left and central plot, consistently with  Eq.~\eqref{eq:sigmanpincts}. 

The shapes and especially the sign of the $\SigmaNP/\SigmaLOQCD$ ratio are very different for the $\mtt$ distributions of the CP-even and CP-odd cases.
In the purely scalar case, corrections are negative and of the order $ -\ct^2 \times 10\%$ for large invariant masses for $m_S<10~\gev$. For larger values of $m_S$, corrections are still negative but smaller and smaller in absolute value by increasing $m_S$. Going towards the threshold region instead, they reach values a bit smaller than  
$ +\ct^2 \times 10\%$ for $m_S<10~\gev$ and instead smaller for lager $m_S$ values. In the purely pseudoscalar case, on the contrary, corrections are always negative. For $m_S<10~\gev$, corrections are of the order of $ -\ct^2 \times 2\%$ for large invariant masses, growing in absolute value up to $ -\ct^2 \times 5\%$ at the threshold. Increasing $m_S$ the absolute value of the corrections decreases for large $\mtt$ (halved for $m_S=300~\gev$) while it grows at the threshold (doubled for $m_S=300~\gev$). As already said, the maximally mixed case (right plot) shows distributions that correspond to the average of the analogous ones in the left and central plot. Clearly,  by varying the angle $\phiC$ that parameterises the mixing of the CP-even and CP-odd component, very different shapes can be obtained. 

\begin{figure}
    \centering
    \includegraphics[width=0.33\textwidth]{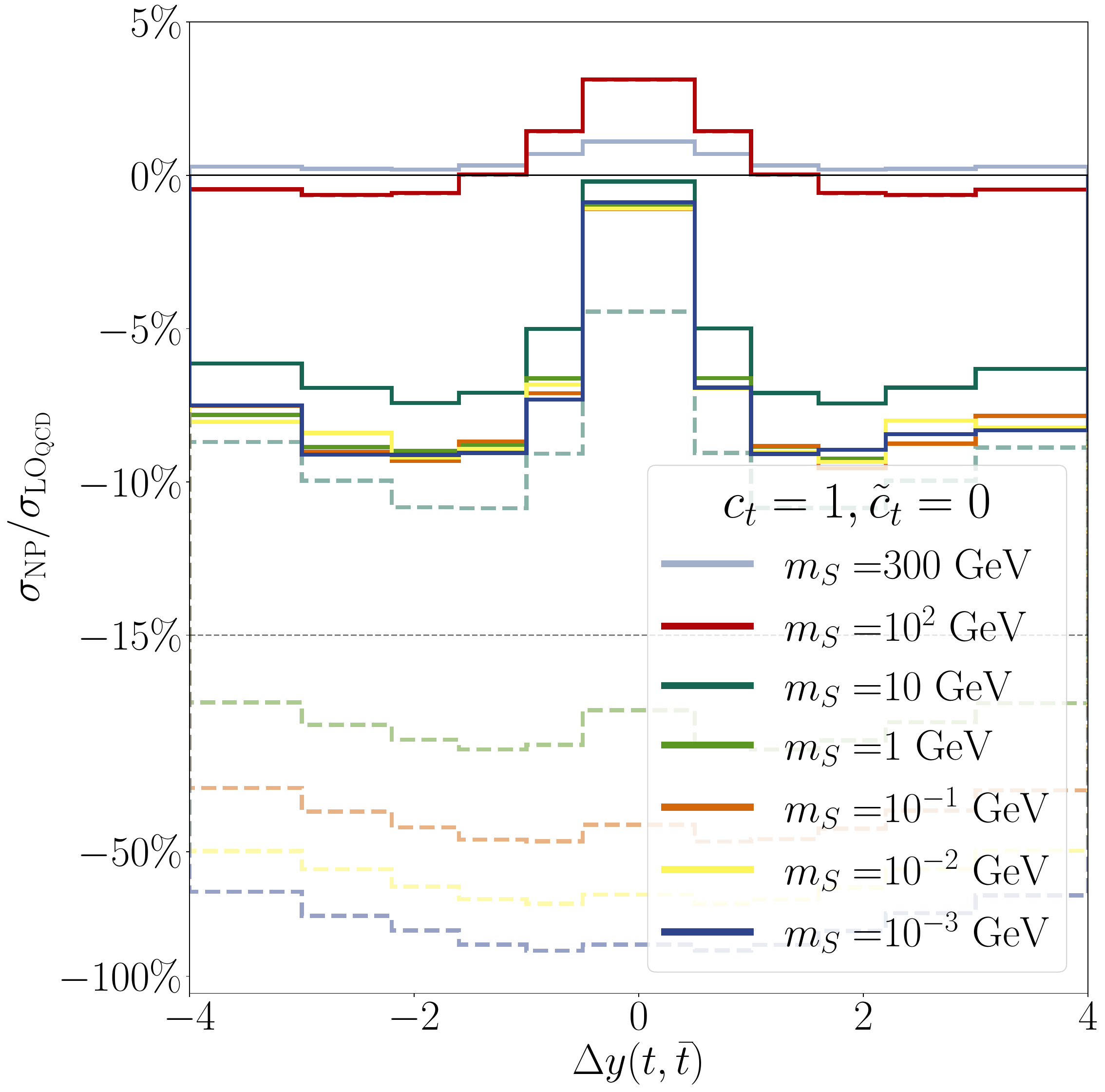}
    \includegraphics[width=0.32\textwidth]{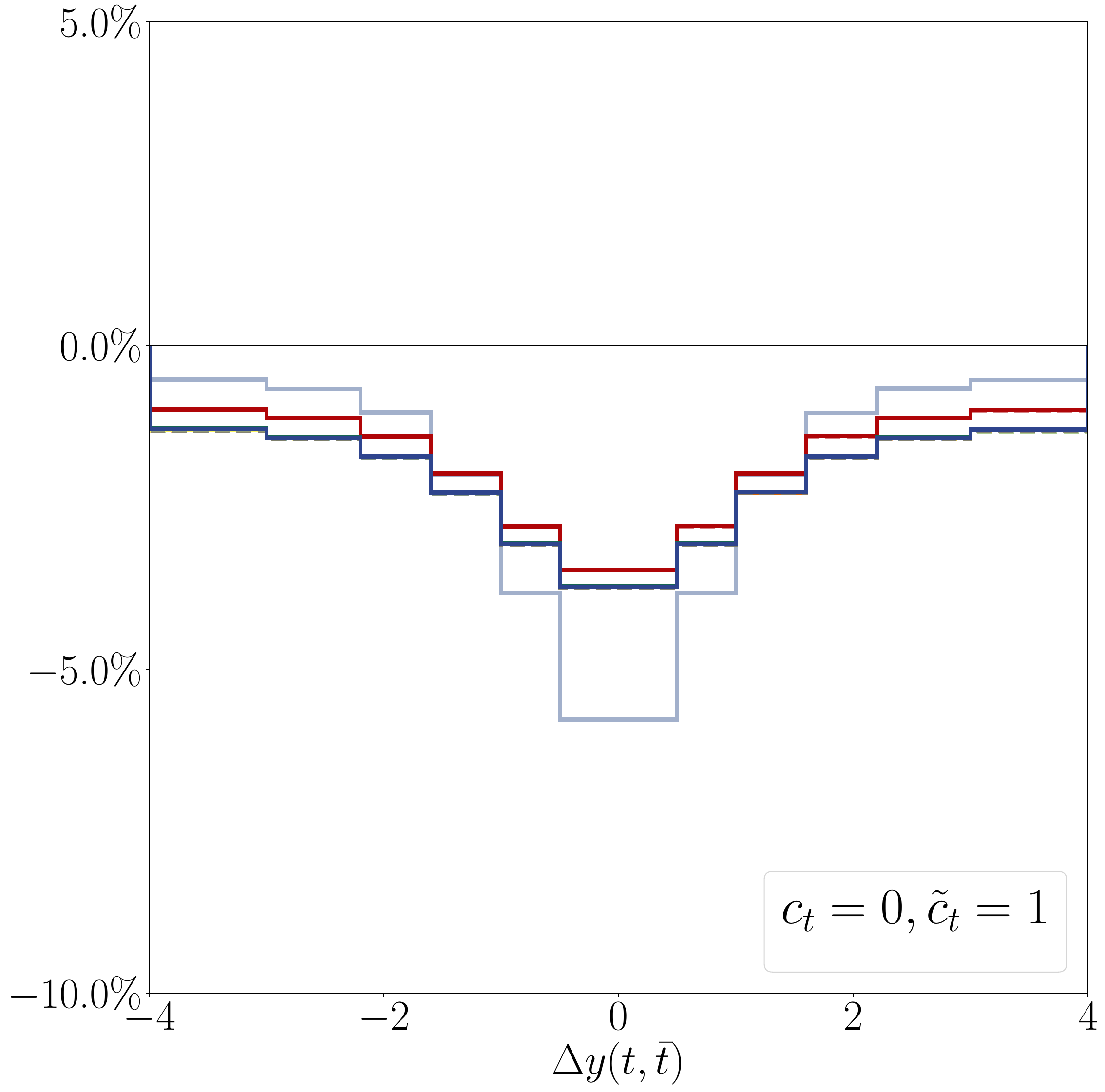}
    \includegraphics[width=0.32\textwidth]{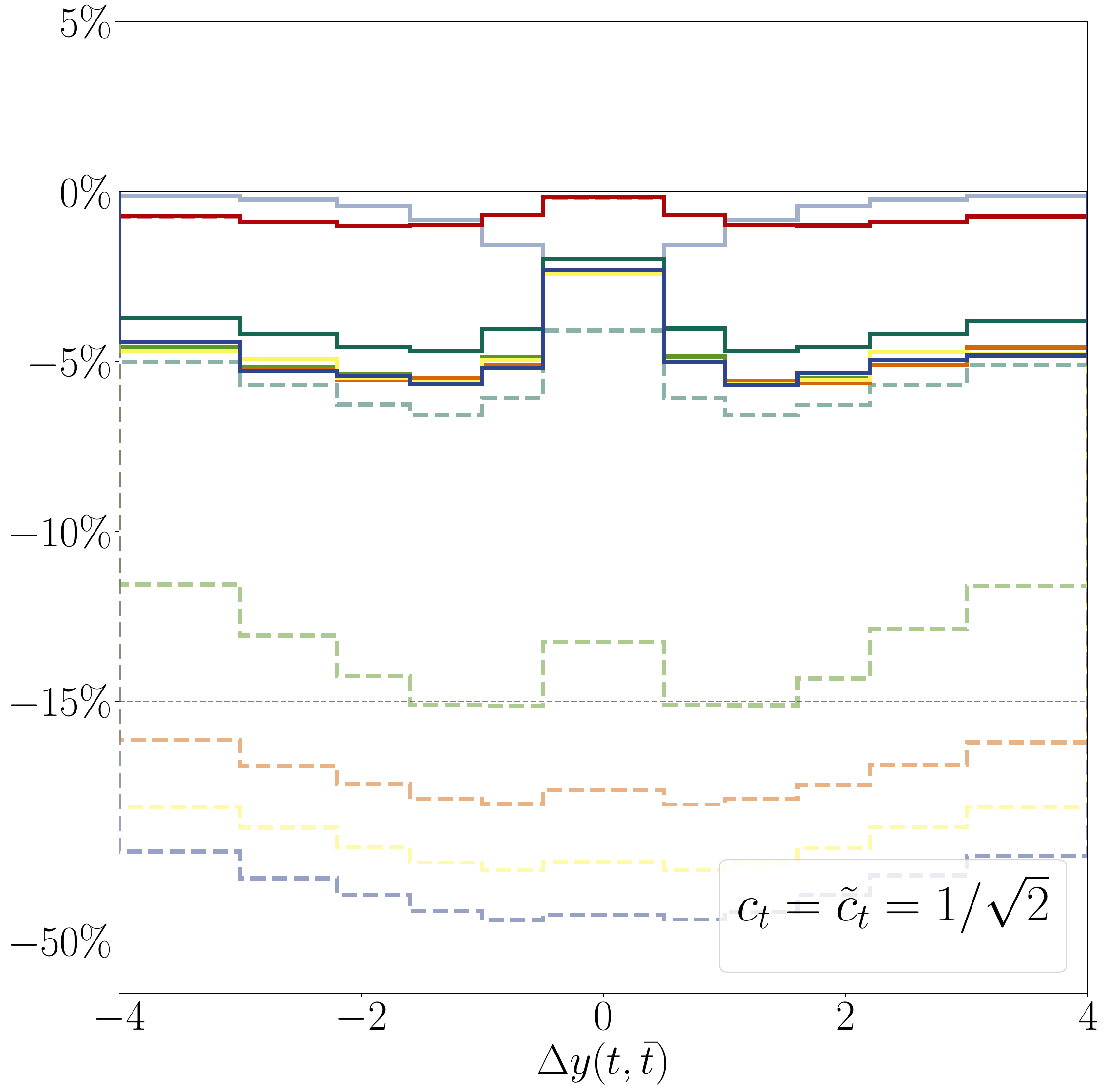}
    \caption{Same as Fig.~\ref{fig:realemission}, but for the $\Dytt$ distributions appearing also in the right plots of Fig.~\ref{fig:differetmass}.}
    \label{fig:realemissiondrap}
\end{figure}

In Fig.~\ref{fig:realemissiondrap}, we display the analogous information of Fig.~\ref{fig:realemission} for the $\Dytt$ distribution. We clearly see a strong correlation between the central region in Fig.~\ref{fig:realemissiondrap}, small $\Dytt$, and the threshold region in  Fig.~\ref{fig:realemission}, small $\mtt$. We also clearly observe in Fig.~\ref{fig:realemissiondrap} how the IR sensitivity on $m_S$ disappears once the soft $S$ emission is included. Two aspects are also even more evident in this figure. First, we can see how, for the size and the shapes of the corrections  are different for the purely scalar (left plot) and pseudoscalar (central plot) case. This leads to a flattening of the corrections for the maximally mixed case (right plot). Second, for $\Dytt\simeq0$, the configuration $m_S=300~\gev$, and in a smaller extent also the $m_S=100~\gev$ one, the $\SigmaNP/\SigmaLOQCD$ ratio is much smaller in the purely scalar case w.r.t.~the purely pseudoscalar one.  The origin of this feature is the same that will be discussed in Sec.~\ref{sec:distrkt}, a Higgs boson $H$ with CP-even and CP-odd couplings, for which this effect is very relevant.

\section{Scalar $S$: Sensitivity study}
\label{sec:Ssensitivity}

In this section we present an explorative study for the potential bounds that can be obtained, via the measurement of top-quark pair production at the LHC, on the $\ct$ (CP-even) and $\ctt$ (CP-odd) couplings describing the interaction of a scalar with mass $m_S<300~\gev$ with the top-quark, see also Eq.~\eqref{eq:LangS}.

In Sec.~\ref{sec:statistics} we will first introduce the statistical method used to extract bounds, which will be used also in Sec.~\ref{sec:sensitivityH}, and specify the data and theory predictions. In Sec.~\ref{sec:boundscts} we present the bounds we find for different values of $m_S$ in the range $1~{\rm MeV}<m_S<300~\gev$.

 \subsection{Statistical method, theory predictions and data}
 \label{sec:statistics}
 
Our goal is to present an exploratory study for the potential bounds that can be obtained at the LHC on the $\ct$ (CP-even) and $\ctt$ (CP-odd) couplings. We neither aim to fully exploit the potential of differential distributions that can be already now measured with high accuracy at the LHC, nor to  quantify the projections for the full data-set of the High-Luminosity program; these aspects are left for future studies. Here, we want simply to show how the measurements of such process can lead to constraints for $\ct$ (CP-even) and $\ctt$ (CP-odd) couplings of a top-philic additional scalar and the estimate the corresponding bounds.
 
This strategy is being pursued by experimental collaborations, in particular CMS  \cite{CMS:2019art,CMS:2020djy}, for the case where $S$ is in fact the Higgs boson $H$, which is precisely what we will discuss in Secs.~\ref{sec:theoframeH}-\ref{sec:sensitivityH}. The experimental analyses are far from trivial, due to both statistical and experimental reasons, and based on doubly differential distributions in  $\mtt$ and $\Dytt$, the same variables that have been discussed in Sec.~\ref{sec:distrct}. We describe in the following how, based on the results of such analyses, we have devised a simplified framework where to mimic the accuracy already achieved at the experimental level with an integrated luminosity of  $35.8 \,\rm fb^{-1}$. 

The statistical analysis  performed for estimating the precision that can be achieved in the determination of $\ct$ and $\ctt$ relies on the minimisation of a $\chi^2$ function built on binned experimental data ($\Sigmaexp$) and corresponding theory predictions ($\Sigmatheo$). The arrow represents a vector of values corresponding to the specific binning considered. The $\chi^2$ function can be written as
\begin{align}\label{eq:chi}
    \chi^2(c_t,\tilde c_t,m_{\rm S}) =  \left(\Sigmatheo -\Sigmaexp \right)\cdot\mathbf{V}^{-1}\cdot \left(\Sigmatheo -\Sigmaexp \right),
\end{align}
where   $\mathbf{V}$ is the experimental covariance matrix.  We now specify $\Sigmatheo$, $\Sigmaexp$ and $\mathbf{V}$.

The theory prediction $\Sigmatheo$ is given by $\SigmaSMNPmult$, see the expression in Eq.~\eqref{eq:multcts}, evaluated for the same binning of the data considered, $\Sigmaexp$. In fact, for our exploratory study we use pseudo-data for $\Sigmaexp$. First, we have considered several top-quark pair analyses \cite{ATLAS:2015lsn,ATLAS:2016pal,ATLAS:2019hxz,CMS:2016oae,CMS:2018adi,CMS:2018htd} in order to select a binning that is not only useful for our fit but also realistic. Given features discussed in Sec.~\ref{sec:distrct},  we have identified the CMS measurement in the lepton+jets channel of Ref.~\cite{CMS:2018htd} for the $\mtt$ distribution as the most  useful, since it has a large number of bins in the low $m(t\bar t$) region. Then, we have generated pseudo data assuming the SM only\footnote{In this work, the only exception to this procedure is what is done in Sec.~\ref{sec:pseudofit}, as explained therein.}: we have calculated for the binning of $\mtt$ of Ref.~\cite{CMS:2018htd} $\SigmaSMNPmult$ with $\ct=\ctt=0$, in other words, the best SM best prediction, the quantity $\SigmaSMmult$ defined in Eq.~\eqref{eq:SigmaSMmult}. This procedure fully defines the quantities $\Sigmaexp$ and $\Sigmatheo$. Further information entering the fit (binning, values of  $\SigmaSMNPmult$, {\it etc.}) is reported in Appendix~\ref{app:table}.

The last piece of information that has to be specified is the covariance matrix $\mathbf{V}^{-1}$. In Sec.~\ref{sec:sensitivityH} we will perform the same study discussed in this Section with the Higgs boson $H$ in the place of $S$. For consistency, $\Sigmaexp$ and $\mathbf{V}^{-1}$ will be the same used here and in particular it will be more clear the choice we have taken for $\mathbf{V}^{-1}$ and we explain in the following. Starting from the value of $\mathbf{V}^{-1}$ in Ref.~\cite{CMS:2018htd}, we have rescaled the entries of the matrix by a $(2.5)^2$ factor,\footnote{This rescaling is equivalent to reduce by a factor 2.5 the experimental errors from Ref.~\cite{CMS:2018htd}. In practice, we have tuned this number in order to obtain what is described in the following sentence in the main text.} such that in the case of the Higgs boson discussed in Sec.~\ref{sec:sensitivityH} we find for the CP-even component of top-Higgs interaction the same uncertainty reported in the aforecited Refs.~\cite{CMS:2019art,CMS:2020djy}. In this way we can mimic, in a simplified statistical framework, the precision already achieved in the experimental analyses based on an integrated luminosity of  $35.8\, \rm fb^{-1}$.

In the results presented in Sec.~\ref{sec:boundscts}, when a two-parameter fit is performed ($\ct$ and $\ctt$), the $(1\sigma, 2\sigma, 3\sigma)$ confidence level intervals are obtained via the condition 
\begin{align}\label{eq:chi2_2par}
    \Delta\chi^2(c_t,\tilde c_t,m_{\rm S})\equiv\chi^2(c_t,\tilde c_t,m_{\rm S}) - \min(\chi^2) \leq (2.30, 6.18, 11.83)\,,
\end{align}
while when only one parameter $c$ is considered, which can be the CP-even coupling, the CP-odd or a specific linear combination of them for fixed $\phiC$,   the condition reads 
\begin{align}
    \Delta\chi^2(c,m_{\rm S})\equiv\chi^2(c,m_{\rm S}) - \min(\chi^2)\leq (1, 4, 9)\,.
\end{align}

As already mentioned, the dependence on $\ct$ and $\ctt$ of $\SigmaSMNPmult$ is symmetric under the independent transformations $\ct\to-\ct$ and $\ctt\to-\ctt$.  In our setup, this symmetry will be manifest in the contour plot in the $(\ct,\ctt)$ parameter space; following the notation in Eqs.~\eqref{eq:Ctdef} and \eqref{eq:phiCtdef}, the information in the full parameter space can be extrapolated from the region $0\le \phiC\le \pi/2$. Since we use pseudodata corresponding to $\ct=\ctt=0$, the minimisation of the  $\chi^2$ function is by definition at $(\ct,\ctt)=(0,0)$. In general,  four  minima can be present.

\subsection{Bounds on $\ct$ and $\ctt$}
\label{sec:boundscts}

\begin{figure}[!t]
    \centering
    \includegraphics[width=\textwidth]{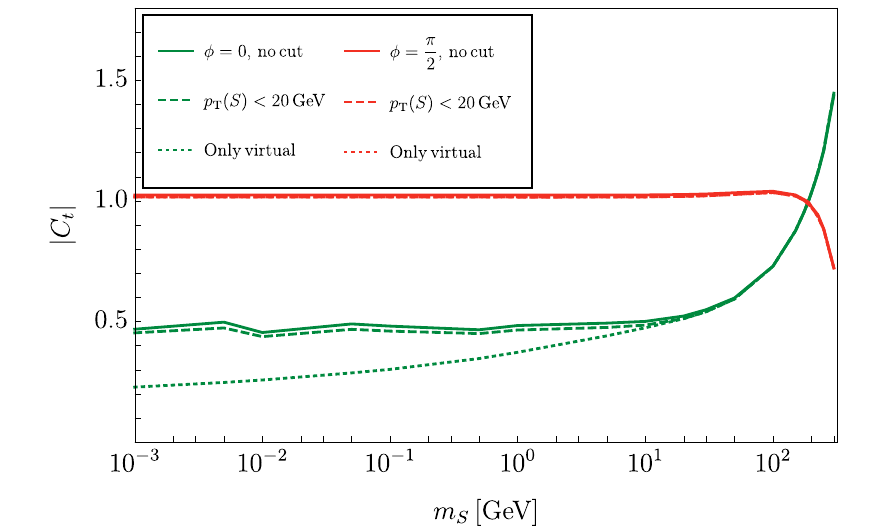}
    \caption{Bound at $2\sigma$ level on $\ct$ (green) and $\ctt$ (red) as a function of  $m_S$.  Solid lines:  virtual corrections and real $S$ emission. Dashed lines: same as solid lines but with the $S$ veto \eqref{eq:cutptS} applied. Dotted lines: only virtual corrections. }
    \label{fig:boundsandreal}
\end{figure}

We present in this section the bounds that can be set in the $(\ct, \ctt)$, or equivalently $(|\Ct|,\phiC)$ plane, via the measurement of top-quark distributions, following the statistical analysis explained in Sec.~\ref{sec:statistics}.

We start by considering the purely scalar ($\ctt=0\Leftrightarrow\phiC=0$) and purely pseudoscalar case ($\ct=0\Leftrightarrow\phiC=\pi/2$) and discuss the impact of the inclusion of real emission contributions from $t \bar t S$ production. In Fig.~\ref{fig:boundsandreal} we show the bounds that can be set on $|\Ct|$, as a function of $m_S$, for the purely scalar (green lines) and purely pseudoscalar (red lines) cases. The solid lines include the contribution from the real radiation of $S$, with no cuts at all applied on $S$, the dashed lines with the cut in \eqref{eq:cutptS} and the dotted line do not include such contribution and therefore corresponds to only the inclusion of the virtuals.

First, we notice again that for the purely pseudoscalar case the bound is completely insensible to the inclusion of the real emission contributions. On the contrary, the bounds for the purely scalar case do depend on it, consistently with what has been discussed in Sec.~\ref{sec:distrct}. However, we also clearly see now that the bounds are not sensible to the inclusion of the radiation with $p_T(S)>20~\gev$; dashed and solid lines are very close to each other. Nevertheless, in order to be as close as possible to what is effectively measured in inclusive $t \bar t $ production, a cut as \eqref{eq:cutptS} has to be applied; if the emission is too hard it is  possible to experimentally reconstruct the real emission  as a different process. Including the radiation, with or without such cut, we see that bounds for both the scalar and the pseudoscalar case are constant for masses below 10 GeV. For the former $|\ct|\lesssim 1$ and for the latter  $|\ct|\lesssim 0.5$.

\begin{figure}[!t]
    \centering
    \includegraphics[width=0.8\textwidth]{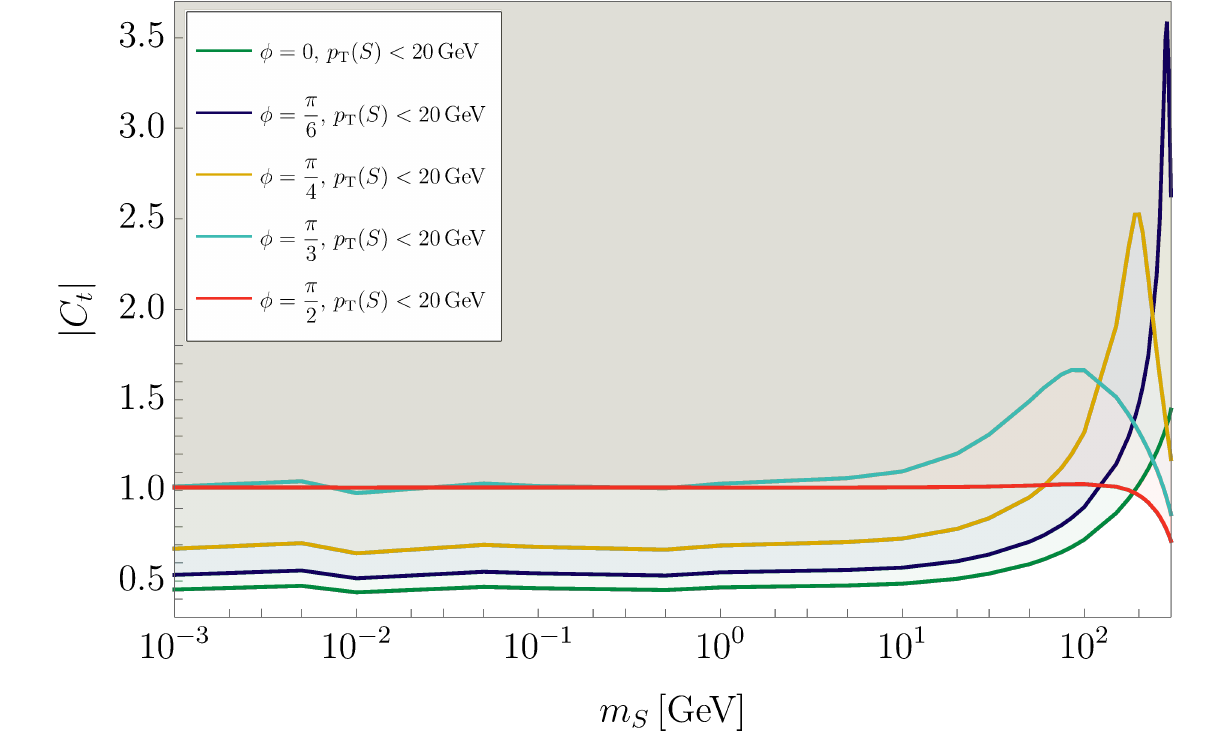}\\
    \vspace{0.5cm}
        \includegraphics[width=0.8\textwidth]{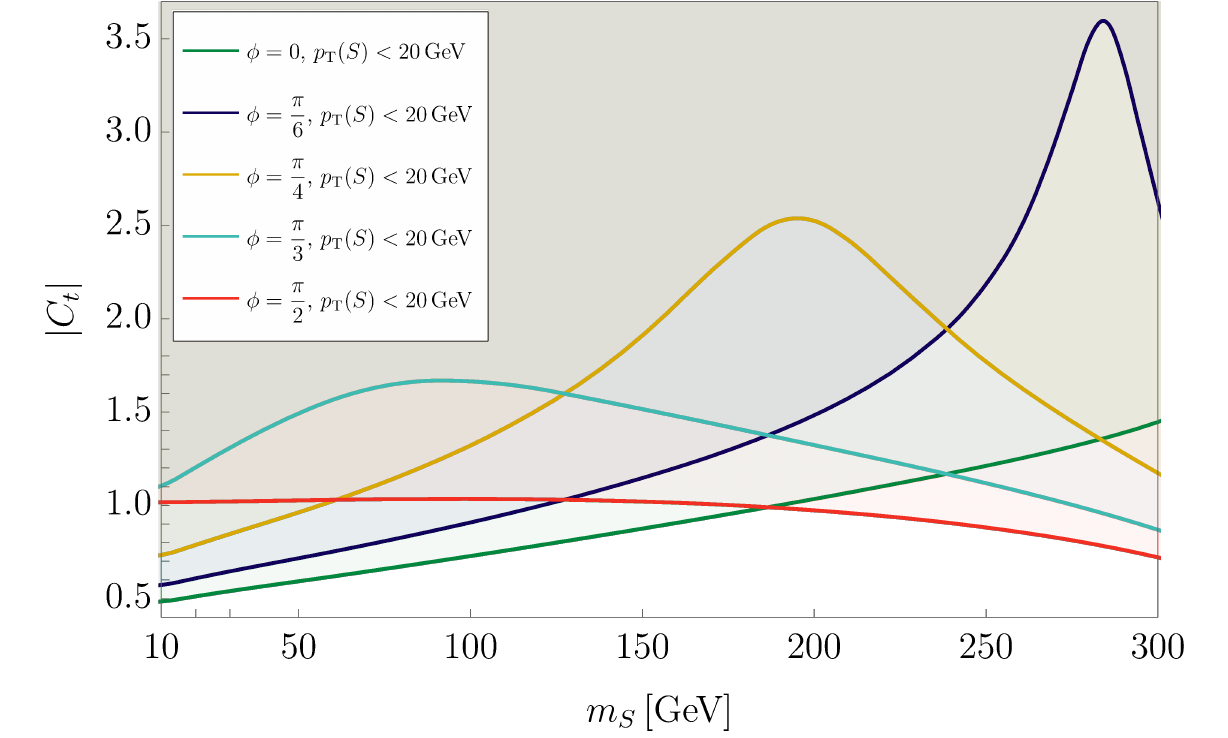}
    \caption{Bound at $2\sigma$ level on $|\Ct|$ as a function of  $m_S$ for different $\phiC $ values.  Virtual corrections and real $S$ emission, with the $S$ veto \eqref{eq:cutptS} applied, are taken into account. The excluded regions correspond to the shaded areas of the plots.}
    \label{fig:boundsphi}
\end{figure}

By including the radiation and applying the veto \eqref{eq:cutptS}, we can inspect the bounds for different values of $\phiC$. In particular, given Eqs.~\eqref{eq:Ctdef}, \eqref{eq:phiCtdef} and \eqref{eq:sigmanpincts} we scan in steps of $ \cos^2\phiC=1/4$, {\it i.e.}, $\phiC=0,\,\pi/6,\,\pi/4,\,\pi/3,\, \pi/2$. In Fig.~\ref{fig:boundsphi}, we show the bounds on $|C_t|$ for the aforementioned set of $\phiC$ choices. The upper plot is similar to Fig.~\ref{fig:boundsandreal}, in fact the cases $\phiC=0$ and $\phiC=\pi/2$ are the same of the dashed lines in that figure. The lower plot is simply showing the same content of the upper plot, but with the horizontal axis in linear scale.

\begin{figure}
    \centering
    \includegraphics[width=0.8\textwidth]{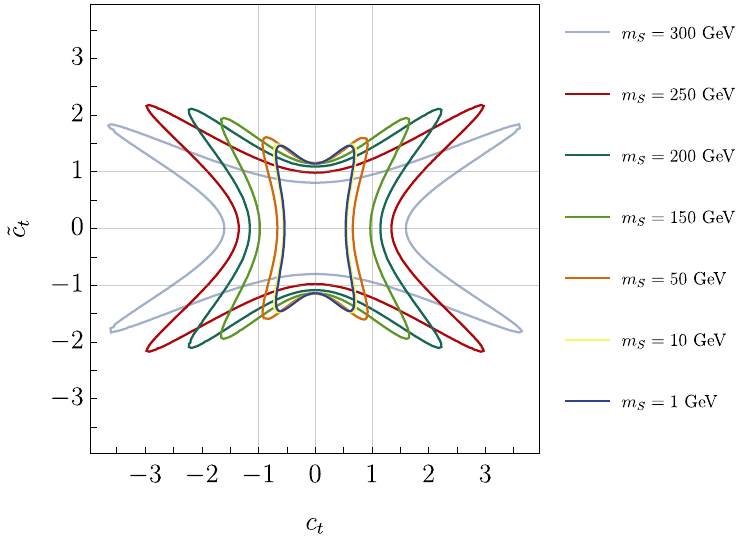}
    \caption{Bound at $2\sigma$ level in the $(\ct,\ctt)$ plane for different $m_S$ values. One should notice that the $m_S$ values considered are very different from those in Figs.~\ref{fig:differetmass}--\ref{fig:realemissiondrap}.}
    \label{fig:ChiDifferentmass}
\end{figure}

There is clearly a non-trivial pattern in the dependence on $m_S$ and $\phiC$, due to the large cancellations that may or may not take place between the $\ct$ and $\ctt$ dependent contributions in Eq.~\eqref{eq:sigmanpincts}, as can also be seen by comparing the left and central plot of Fig.~\ref{fig:realemission}. 
In order to further investigate this pattern, for a few representative cases of $m_S$ in the range $1~\gev<m_S<300~\gev$,\footnote{For values $m_S<1~\gev$, bounds are insensitive to the values of $m_S$ itself.} we have derived the $2\sigma$ intervals in the $(\ct,\ctt)$ plane. Results are shown in Fig.~\ref{fig:ChiDifferentmass}. Consistently with what is shown also in the other figures of this section, for $\ctt\simeq0$ bounds on $\ct$ becomes less stringent at larger masses, while for $\ct\simeq0$ bounds on $\ctt$ are really sensitive to $m_S$ only for large values of $m_S$ and become more stringent. From this plot we observe that the value of $\phiC$ for which the constraint on $|\Ct|$ is the weakest is between $0$ and $\pi/2$. Such value of $\phiC$ decreases by increasing $m_S$ and the weakness of the bound at that value of $\phiC$, compared to the other values at a given $m_S$, increases for large $m_S$.

\section{Higgs boson $H$: Theoretical framework}
\label{sec:theoframeH}

In this section we adapt what has been discussed in Sec.~\ref{sec:theoframe}  for the particular case where $S$ is the Higgs boson itself, $S=H$, allowing for its anomalous and/or CP-odd interactions with the top quark. In this way, we obtain the results of the virtual corrections to top-quark pair production that are induced by anomalous CP-even and CP-odd interactions of the Higgs boson with the top-quark. 

In Sec.~\ref{sec:LagrangiansH} we provide the relevant Lagrangian, the notation used and the main formulas that are analogous to the ones provided in Sec.~\ref{sec:theoframe} for the case of the scalar $S$. In Sec.~\ref{sec:HversusS} we explain in detail how we have derived such expressions, recycling the results of Sec.~\ref{sec:theoframe} for now the Higgs case, highlighting important differences w.r.t.~the scalar $S$ case. In Sec.~\ref{sec:SMEFT} we show how this calculation can be reinterpreted in the SMEFT framework, which further support the consistency of our approach.

\subsection{Lagrangian, notation and relevant formulas}
\label{sec:LagrangiansH}

In analogy with what has already been discussed in  Sec.~\ref{sec:Lagrangians} for the scalar $S$, the case of a BSM Higgs boson that allows for anomalous and/or CP-odd interactions with the top quark can be described by a Lagrangian of the form
\bea
\mathcal{L}_{{\rm SM} + (\kt,\,\ktt)}\equiv \mathcal{L}_{{\rm SM}} +  \LHNP, \label{eq:LangH}
\eea
where $\mathcal{L}_{{\rm SM}}$ is the SM Lagrangian and 
\be
  \LHNP
  \equiv -\overline \psi_t \left [\frac{\left(\yt-\ytSM\right)}{\sqrt{2}}+i\frac{\ytt}{\sqrt{2}}\gamma_5\right]\psi_t H\,
  \equiv -\frac{\ytSM}{\sqrt{2}}~\overline \psi_t \left [\left (\kt-1\right)+i\ktt\gamma_5\right]\psi_t H\, \label{eq:LHNP},
\ee
where $H$ is the Higgs field and 
\be
\ytSM\equiv \frac{\sqrt{2}m_t}{v}\,. \label{eq:ytSMdef}
\ee
In Eq.~\eqref{eq:ytSMdef}, $m_t$ is the top-quark mass and $v$ is the Higgs vacuum expectation value. The parameters $\yt$ and $\ytt$ parameterise the CP-even and CP-odd components of the top-quark Yukawa interaction, respectively.  Adopting the so-called kappa-framework, they can be rewritten in term of $\kt$, $\ktt$ and $\ytSM$, as shown in the r.h.s.~of Eq.~\eqref{eq:LHNP}. One can notice that the parameter choice $(\kt,\,\ktt)=(1,0)$, or equivalently $(\yt,\,\ytt)=(\ytSM,0)$, corresponds to the SM, {\it i.e.},  $\mathcal{L}_{{\rm SM} + (\kt,\,\ktt)}=\mathcal{L}_{{\rm SM}}$.

Analogously to the case of the scalar $S$, as in Eqs.~\eqref{eq:Ctdef} and \eqref{eq:phiCtdef}, we can also introduce the notation
\bea
\Kt &\equiv& \kt+ i \ktt = |\Kt| e^{i\phiK}\,,\\
\phiK&\equiv& \arctan\frac{\ktt}{\kt}\, ,
\eea
with $|\Kt|$ being the actual kappa-modifier of the strength of the SM top-Higgs interaction, and $\phiK$ parameterising the CP-even and CP-odd admixture of the interaction. 

\medskip

With such notation, the renormalised one-loop virtual corrections induced  by the diagrams of  Fig.~\ref{fig:diagrams}, with $H$ in the place of $S$, can be written as 
\be
\SigmaNPH=(\kt^2-1)\, \Sigmakt + \ktt^2 \, \Sigmaktt  \label{eq:sigmanpinkts}\,,
\ee 
with
\bea
\Sigmakt &=&\frac{(\ytSM)^2}{2} \Sigmact\bigg|_{m_S=m_H}\,, \label{eq:Sigmakt}\\
\Sigmaktt &=&\frac{(\ytSM)^2}{2} \Sigmactt\bigg|_{m_S=m_H}\label{eq:Sigmaktt}\,.
\eea 

As can be clearly seen, Eq.~\eqref{eq:sigmanpinkts} is the analogue of Eq.~\eqref{eq:sigmanpincts}, through which the quantities $\Sigmact$ and $\Sigmactt$ appearing in Eqs.~\eqref{eq:Sigmakt} and \eqref{eq:Sigmaktt} are also defined. The quantities $\Sigmakt$ and $\Sigmaktt$ entail the dependence on the kinematic for the purely $(\kt-1)$-dependent and $\ktt$-dependent component of $\SigmaNPH$. We notice here that while in the case Eq.~\eqref{eq:sigmanpincts} only quadratic terms in $\ct$ and $\ctt$ are present, in Eq.~\eqref{eq:sigmanpinkts} also a linear term in $(\kt -1)$ is present since $(\kt^2-1)=(\kt-1)^2+2(\kt-1)$. We will return on this point in detail in Sec.~\eqref{sec:HversusS}. Similarly to  Eq.~\eqref{eq:sigmanpincts}, instead, in Eq.~\eqref{eq:sigmanpinkts} no mixed term of the kind $(\kt-1)\ktt$ is present. Thus, the quantity $\SigmaNPH$ is  symmetric for $\kt\to-\kt$  and $\ktt\to-\ktt$. However, the SM, which corresponds to $(\kt,\ktt)=(1,0)$,  is instead  not symmetric under the $\kt\to-\kt$ transformation. This difference with the case of the scalar $S$, where $\SigmaNP$ is symmetric for $\ct\to-\ct$  and $\ctt\to-\ctt$ and also the SM itself  since it corresponds to $(\ct,\ctt)=(0,0)$, underlies the different qualitative results between Sec.~\ref{sec:Ssensitivity} and Sec.~\ref{sec:sensitivityH}.

As discussed in Sec.~\ref{sec:SMcombination} for the $S$,  two schemes can be used for combining new physics with SM. Eqs.~\eqref{eq:addcts} and \eqref{eq:multcts} can be converted in the Higgs case to
\bea
\SigmaSMNPHadd \equiv \SigmaSMadd + \SigmaNPH\,, \label{eq:addkts}
\eea
and
\bea
\SigmaSMNPHmult \equiv \SigmaSMmult + \KNLOQCD \, \SigmaNPH\,, \label{eq:multkts}
\eea
respectively. Equations \eqref{eq:addkts} and \eqref{eq:multkts} correspond to the  combination in the additive and multiplicative approach, respectively, of the best theory predictions of the SM and of anomalous top-Higgs interactions. Many more details can be found in Sec.~\ref{sec:SMcombination}, where also all the terms entering Eqs~\eqref{eq:addkts} and \eqref{eq:multkts} are rigorously defined. Here, we just want to point out that when an approach is chosen, either multiplicative or additive, it is imperative that it is used both for the SM contribution and as well for the NP one, which depends on $(\kt-1)$ and $\ktt$. Indeed, while in the case of the $S$ scalar an asymmetric approach for the SM and the NP contributions would be possible, in the case of the Higgs boson would lead to contributions not proportional to $\kt^2$ (the CP-even part) in the SM and NP combined prediction.  

Unlike the case of the scalar $S$, where the contribution from the $t \bar t S$ has been taken into account, we will {\it not} include the contribution from the $t \bar t H$ final state in our studies for the sensitivity on $\kt$ and $\ktt$ in Sec.~\ref{sec:sensitivityH} as that corresponds to a different analysis at the LHC. Moreover, the inclusion of the $t \bar t H$ contribution is not necessary for avoiding IR sensitivity, since $m_H$ is large. For the same reason, as manifest in  Figs.    \ref{fig:realemission} and \ref{fig:boundsandreal}, even if we included such contribution, its impact on our results would be totally negligible.

\subsection{Similarities and differences with the scalar $S$ case}
\label{sec:HversusS}

Instead of repeating what has been discussed in detail in Sec.~\ref{sec:loopcalculation} and Sec.~\ref{sec:SMcombination} in order to show how Eqs.~\eqref{eq:sigmanpinkts}--\eqref{eq:Sigmaktt} have been derived, we will limit ourselves to a discussion on how the calculation presented in  Sec.~\ref{sec:loopcalculation} can be applied to the case of the Higgs boson. We also highlight the main differences and the subtleties that have been neglected in the literature so far. 

 We start observing that the relevant diagrams for the calculation in the case of the Higgs boson are the same of Fig.~\ref{fig:diagrams}, with $H$ in the place of $S$. This is the main reason why the calculation for the scalar $S$ can be used for the Higgs. Moreover, the need of taking into account  the contribution from higher-order corrections in the SM, especially the NLO EW ones,  is even stronger in this case. Indeed,  contributions proportional to the anomalous CP-even interactions, $(\kt-1)$,  will lead at the fully differential level to exactly the same corrections induced by the Higgs-boson component of the NLO EW corrections in the SM. 
 
The calculation of virtual corrections to top-quark pair production in the two scenarios, $S$ or $H$, is fully equivalent, with the case of Eq.~\eqref{eq:LangS} being a generalisation of Eq.~\eqref{eq:LangH} for $m_S\ne m_H$. Also in the case of the Higgs boson, none of the virtual diagrams features a self interaction of $H$ and so,   for our calculation, $\mathcal{L}_{{\rm SM}+S}$ and $\mathcal{L}_{{\rm SM} + (\kt,\,\ktt)}$ Lagrangians are completely equivalent if:
\bea
S=H &\Longrightarrow& m_S=m_H\,, \label{eq:link1}\\
\yt\equiv \kt\, \ytSM = \ytSM + \sqrt{2}\ct\,& \Longleftrightarrow& \kt= 1+ \frac{\sqrt{2}\ct}{\,\ytSM}\,,\label{eq:link2}\\
\ytt\equiv \ktt \,\ytSM= \sqrt{2}\ctt &\Longleftrightarrow& \ktt=  \frac{\sqrt{2}\ctt}{\,\ytSM}\,\label{eq:link3}.
\eea

\medskip 

From the pure calculation side the two scenarios are therefore equivalent  and via Eqs.~\eqref{eq:link1}--\eqref{eq:link3} it is possible to obtain $\SigmaNPH$ in Eq.~\eqref{eq:sigmanpinkts} from $\SigmaNP$ in Eq.~\eqref{eq:sigmanpincts}. However, this cannot be achieved by simply applying them directly as we will explain in the next paragraph. As already said, $\SigmaNPH$ corresponds to the renormalised one-loop virtual corrections induced  by the diagrams of  Fig.~\ref{fig:diagrams}, with $H$ in the place of $S$,  to $\SigmaLOQCD$, is the LO cross section from purely QCD interactions (the diagrams in Fig.~\ref{fig:SMtt}). Renormalisation is again understood in the on-shell scheme, as in the scalar S case.

One may be tempted to assume that $\SigmaNPH$ can be directly derived by applying Eqs.~\eqref{eq:link1}--\eqref{eq:link3} to Eq.~\eqref{eq:sigmanpincts}, but the situation is a bit more complex. Indeed, $\SigmaNPH$ parameterises only the contribution due to the anomalous component of the interactions of the Higgs boson with the top-quark. While in the case of scalar $S$ there is not a SM component of the interaction with the top-quark, in the case of the Higgs there is. In other words, while expanding in powers of $\ct$ and $ \ctt$ in the scalar $S$ case all the top-$S$ vertexes in Fig.~\ref{fig:diagrams} must depend on $\ct$ and $ \ctt$, in the case of the Higgs boson, expanding  in power of $(\kt-1)$ and $\ktt$, there can be also contributions with one of the Higgs-top interactions that are the SM ones. The SM corresponds indeed in the case of the scalar $S$ to $(\ct, \ctt)=(0,0)$ while in the case of the Higgs to $(\kt, \ktt)=(1,0)$; expanding in powers of $\ct$ around the SM there is no linear dependence on $\ct$, expanding in powers of $(\kt-1)$ around the SM there is linear dependence on $(\kt-1)$. 
It is important to notice that if we had directly applied  Eqs.~\eqref{eq:link1}--\eqref{eq:link3} to Eq.~\eqref{eq:sigmanpincts} we would have gotten a factor $(\kt-1)^2$ in front of $\Sigmact $ and not  $(\kt^2-1)$ as in 
Eq.~\eqref{eq:sigmanpinkts}, missing the linear contribution in $(\kt-1)$.

All the previous argument becomes more clear if seen from a SMEFT perspective, which is what we are going to discuss in the next section.
Before doing that we want to mention that our calculation is equivalent to the one of Ref.~\cite{Martini:2021uey}, besides 
the fact that the $s$-channel diagram in Fig.~\ref{fig:diagrams}(a), to the best of our knowledge, has not be considered in Ref.~\cite{Martini:2021uey}. We will discuss in detail in Secs.~\ref{sec:schannel} and \ref{sec:schannelfit} the impact of this particular diagram both in the calculation of $\SigmaNPH$ and in the determination of the $\kt$ and $\ktt$ parameters, but it is important to note that this diagram itself is UV finite and indeed no corresponding CT vertex to the $ggS$ vertex is present.

\subsection{The SMEFT perspective}
\label{sec:SMEFT}
First of all, one may wonder if the case $S=H$ and also $\LHNP$ are really  corresponding or  to the case of the Higgs boson, since $H$ is already part of the SM particle content and especially is part of a SU(2) doublet within a theory that is gauge invariant, at variance with the case of the scalar $S$ described by the Lagrangian in Eq.~\eqref{eq:intL}. The Lagrangian $\LHNP$ defined in Eq.~\eqref{eq:LHNP} explicitly breaks SU(2) gauge invariance, similarly as the Eq.~\eqref{eq:intL} for $\Lint$, but the latter can actually be rewritten in a way that allows to preserve SU(2) gauge invariance. Indeed, we notice that  we can recast our calculation to the SMEFT framework, which is SU(2) gauge invariance. As already partially noticed in Ref.~\cite{Martini:2021uey}, for the case of one-loop corrections to top-quark pair hadroproduction,\footnote{We stress that the following statement is not always true  for any process or any perturbative order.}  the Lagrangian in Eq.~\eqref{eq:LangH} is completely equivalent to
\be
\mathcal{L}_{\rm SMEFT,\, top-Higgs}^{\rm dim=6  }\equiv \mathcal{L}_{\rm SM}+\frac{C^{u\Phi}_{tt}}{\Lambda^2}\left(\Phi^\dag\Phi-\frac{v^2}{2}\right)\overline \psi_{Q_{3,L}} \tilde \Phi \psi_{t,R} + {\rm h.c.}\, \label{eq:SMEFT},
\ee
where $\Lambda$ is the NP scale in the EFT expansion, $C^{u\Phi}_{tt}$ is a complex Wilson coefficient, $Q_{3,L}$ is the SU(2) left-handed doublet $(\psi_{t,L},\psi_{b,L})^T$, $\Phi$ is the Higgs doublet before electroweak symmetry breaking, and $\tilde \Phi^a\equiv \epsilon^a_b\Phi^b $ with $\epsilon^a_b$ being the Levi-Civita tensor acting on the  SU(2) components.
For our calculation $\mathcal{L}_{\rm SMEFT,\, top-Higgs}^{\rm dim=6  }$ in Eq.~\eqref{eq:SMEFT} and $\mathcal{L}_{{\rm SM} + (\kt,\,\ktt)}$ in Eq.~\eqref{eq:LangH} are equivalent, provided that
\bea
\kt&=&1-\frac{v^2}{\Lambda^2}\frac{\Re(C^{u\Phi}_{tt})}{\ytSM}\,, \label{eq:SMEFT1}\\
\ktt&=&-\frac{v^2}{\Lambda^2}\frac{\Im(C^{u\Phi}_{tt})}{\ytSM}\,. \label{eq:SMEFT2}
\eea

The presence of a linear term in $(\kt-1)$ in Eq.~\eqref{eq:sigmanpinkts} and all the discussion at the end of Sec.~\ref{sec:HversusS} is manifest in a SMEFT perspective. Considering Eqs.~\eqref{eq:SMEFT1} and \eqref{eq:SMEFT2}, we take into account not only effects of $\mathcal{O}(1/\Lambda^4)$ but also of  $\mathcal{O}(1/\Lambda^2)$, which cannot have a correspondence for a generic scalar $S$ that is not identified with the Higgs itself. The fact that corrections  linear in $\ktt$ are not present is the same leading to $\Sigmactctt=0$ in Eq.~\eqref{eq:sigmanpincts}.

\medskip

The fact that the calculation in the kappa framework for the Higgs can be recasted in a SMEFT framework is crucial for the consistency of the calculation presented in this work and for allowing to recycle the results already obtained for the case of the scalar $S$. We motivate this statement  in the following. 

While NLO EW corrections in the kappa framework, which violates SU(2) gauge invariance, cannot  be in general performed, when calculations can be embedded in a SMEFT description, as in this case,  they are theoretically consistent. Indeed, SMEFT precisely preserves  SU(2) gauge invariance and allows for EW corrections. An analogous situation, {\it e.g.}, is the one of anomalous Higgs self couplings in single and double Higgs production, extensively discussed by some of the authors of this work in Refs.~\cite{Degrassi:2016wml,Maltoni:2017ims,Maltoni:2018ttu, Borowka:2018pxx}.

The consistency of the kappa-framework for such calculation and  especially the possibility of recycling the calculation for $S$ of Sec.~\ref{sec:loopcalculation} also for the Higgs itself and including not only terms proportional to $(\kt -1)^2\simeq \Re^2(C^{u\Phi}_{tt})$ but also to $ \kt -1\simeq \Re(C^{u\Phi}_{tt})$ relies on some subtleties that are typically ignored in the literature. The operators in Eq.~\eqref{eq:SMEFT} generate new interactions not only between the top quark and the physical Higgs field, but also between the top-quark and the neutral(charged) Goldstone boson $G_{0}(G_{\pm})$. The $t\bar t G_0$ and $t b G_{+}$ vertexes are not modified, but the new vertexes $t\bar t G_0 G_0$ and $t \bar t G_{+}G_{-}$, and also $t\bar t HH$,  do appear. In particular, for our calculation, the only quantity that can be affected by these new vertexes  is the two-point function $\Sigma(p)$ in Eq.~\eqref{eq:Sigmat}, since a new topology with a momentum-independent closed loop of scalars induced by $t\bar t HH$ and especially $t\bar t G_0 G_0$ and $t \bar t G_{+}G_{-}$ can appear. Their contributions are not gauge invariant, also when summed together. This is equivalent to what has been discussed in detail in Ref.~\cite{Maltoni:2018ttu} for the case of modified Higgs-self coupling for the Higgs two-point function (see especially Appendices A and B in that work). However, again similarly to the case discussed in Ref.~\cite{Maltoni:2018ttu}, also for our calculation since the contributions of such diagrams are momentum independent, in the on-shell scheme they do not enter $\delta \psi_t$ and are exactly canceled by $\delta m_t$, regardless of the momentum of the top-quark. Gauge invariance is therefore preserved.

\section{Higgs boson $H$: Numerical results for scalar one-loop corrections to $t \bar t$ distributions}
\label{sec:distrkt}

Starting from the Lagrangian in Eq.~\eqref{eq:LangH}, which describes the dynamics of the SM with the Higgs boson $H$ that can have anomalous CP-even and CP-odd interactions with the top-quark, in Sec.~\ref{sec:LagrangiansH} we have presented the one-loop corrections induced by $H$ to the cross section for the hadroproduction of a top-quark pair, denoted as $\SigmaNPH$. This quantity originates from the diagrams in Fig.~\ref{fig:diagrams} and, as shown in Eq.~\eqref{eq:sigmanpinkts} depends only on four quantities: the squared kappa modifiers for the CP-even and CP-odd interactions of $H$ with the top quark, $\kt^2$ and $\ktt^2$, and the quantities $\Sigmakt$ and $\Sigmaktt$ (see Eq.~\eqref{eq:Sigmakt} and \eqref{eq:Sigmaktt}) that are fully differential functions of the momenta of the top-quark pair. As discussed in Sec.~\ref{sec:LagrangiansH}, the calculation of  $\Sigmakt$, $\Sigmaktt$ and more in general $\SigmaNPH$ has been derived from the calculation for the case where instead of $H$ a generic scalar $S$ was considered, discussed in Sec.~\ref{sec:loopcalculation}.

Analogously to Sec.~\ref{sec:distrkt}, where the case of $S$ was considered, in this section we show and discuss our numerical results for $\SigmaNPH$. Also in this case, we do not consider all the higher-order SM corrections introduced in Sec.~\ref{sec:SMcombination} and entering Eqs.~\eqref{eq:addkts} and \eqref{eq:multkts}, we rather focus on  the relative size of $\SigmaNPH$  w.r.t.~$\SigmaLOQCD$, the LO  cross-section associated to the diagrams of Fig.~\ref{fig:SMtt}. We define the sum of them as  $\SigmaLOQCDNPH=\SigmaLOQCD+\SigmaNPH$.

As in Sec.~\ref{sec:distrkt} we consider three benchmarks that we list in the following:
\begin{enumerate}
\item Purely Scalar: $(\kt-1, \ktt)=(1,0)\Longleftrightarrow (|\Kt|,\phiK)=(2,0)$, 
\item Purely Pseudoscalar: $(\kt-1, \ktt)=(0,1)\Longleftrightarrow  (|\Kt|,\phiK)=(1,\pi/2)$, 
\item Mixing aligned: $(\kt-1, \ktt)=(1,1)\Longleftrightarrow  (|\Kt|,\phiK)=(\sqrt{5},\arctan(1/2))$. 
\end{enumerate}
Given the dependence of $\SigmaNPH$ on only $(\kt^2-1)=(\kt-1)^2+2(\kt-1)$ and $\kt^2$, with no mixed $\kt\ktt$ terms, the first two cases are sufficient for extrapolating the size of $\SigmaNPH$ in any configuration; they correspond to $\Sigmakt$ and $\Sigmaktt$ in Eq.~\eqref{eq:sigmanpinkts}, respectively. The third case is useful to see the cancellations that may take place when $\kt-1\simeq\ktt$. Clearly, since the SM corresponds to the configuration $(\kt,\ktt)=(1,0)$, choosing a similar value for the anomalous component, $\kt-1$ and $\kt$, in the different scenarios will lead to very different $|\Kt|$ values.

\begin{figure}[!t]
    \centering
    \includegraphics[width=0.48\textwidth]{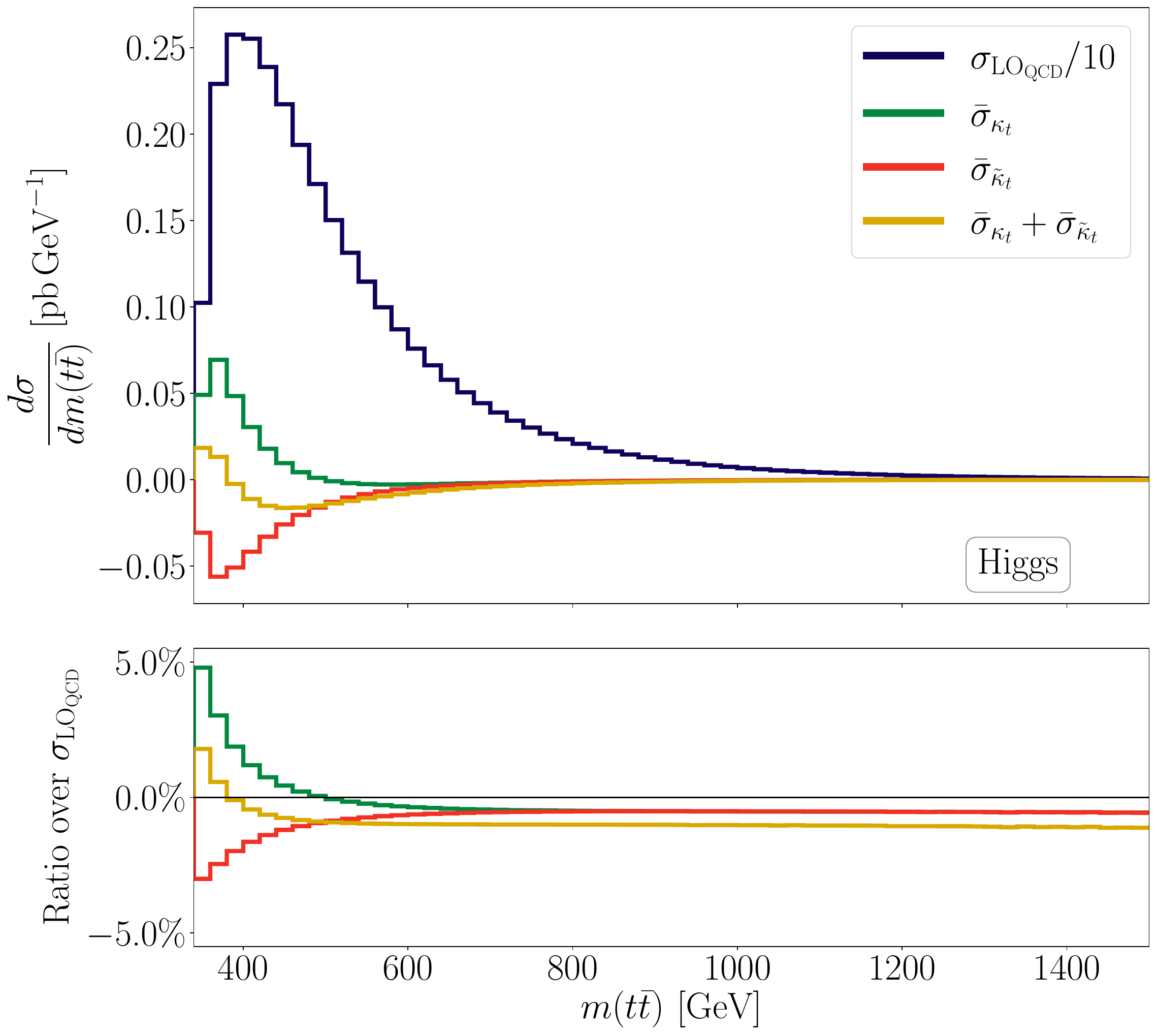}
    \includegraphics[width=0.48\textwidth]{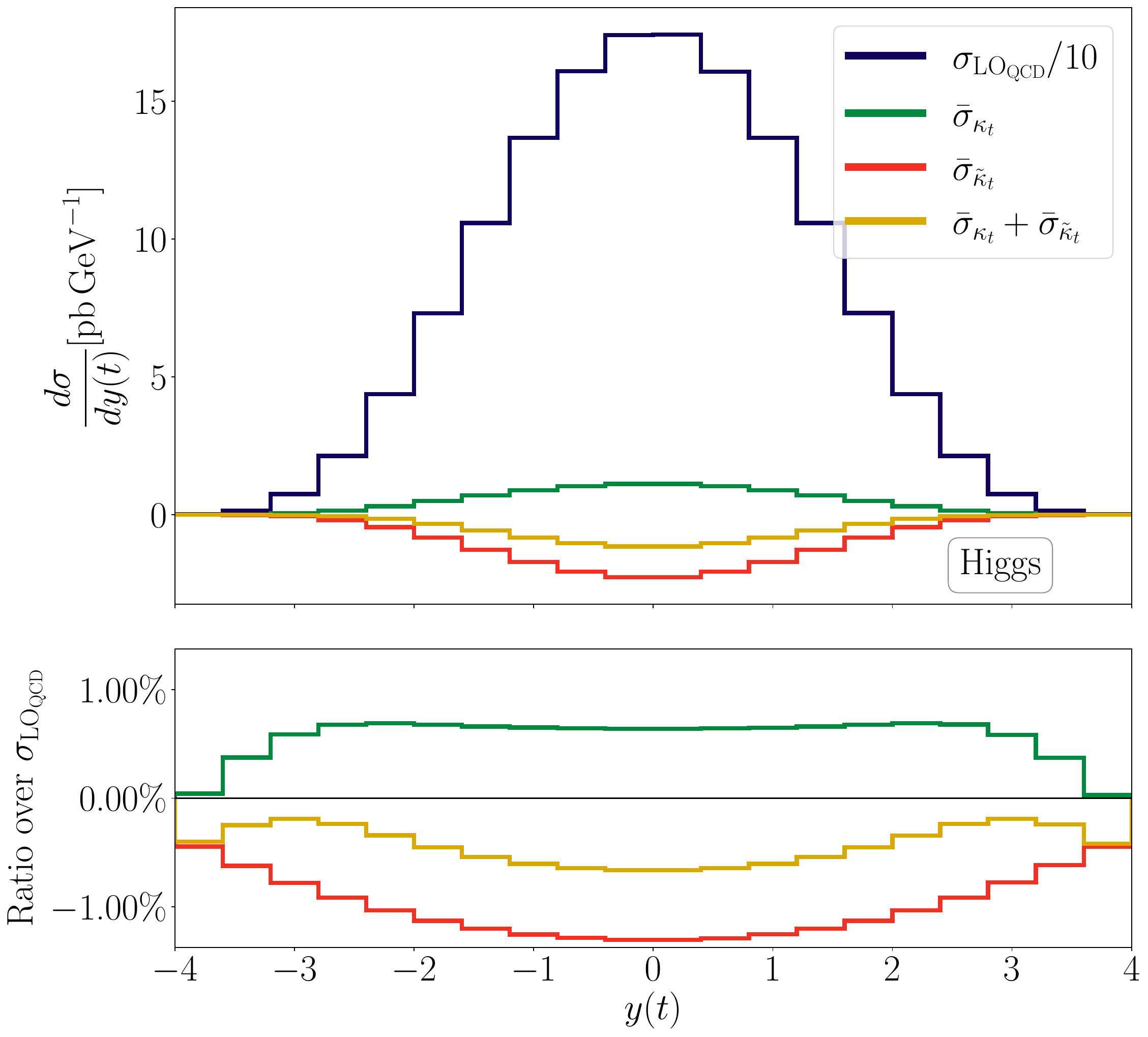}
     \includegraphics[width=0.48\textwidth]{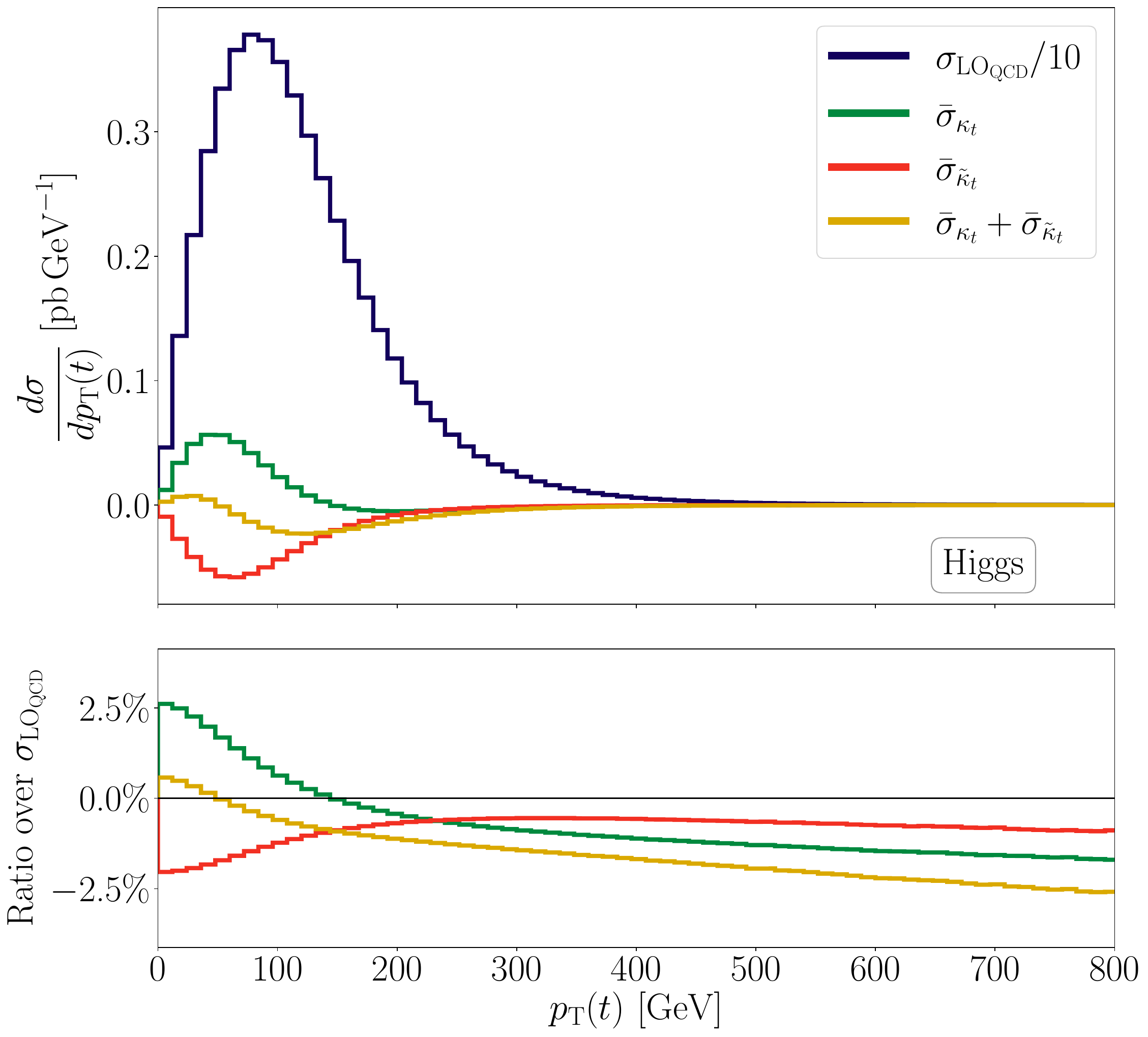}
    \includegraphics[width=0.48\textwidth]{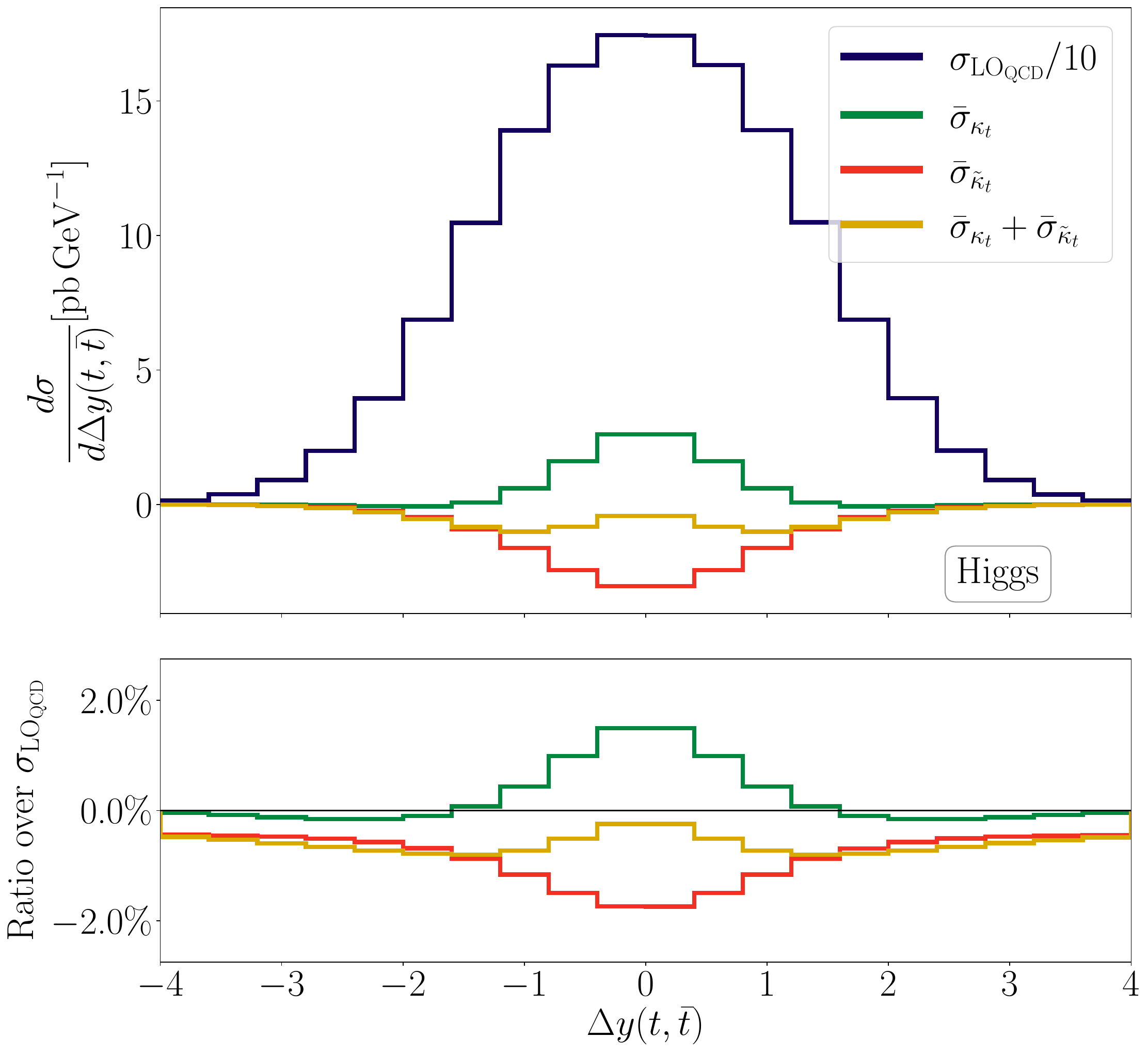}
    \caption{Higgs Boson case. In the main panel of each plot: $\SigmaLOQCD$ divided by ten (blue) and the loop corrections $\SigmaNPH$ evaluated at different $(\kt, \ktt)$ benchmarks: $(1,0)$ green, $(0,1)$  red and $(1,1)$ mustard In the inset of each plot: ratio of $\SigmaNPH$ for the different three benchmarks over $\SigmaLOQCD$. }
    \label{fig:alldistr}
\end{figure}

In Fig.~\ref{fig:alldistr}, we show distributions for $\mtt$ and $\Dytt$, as done in Sec.~\ref{sec:distrct} and Ref.~\cite{Martini:2021uey},  and also for $\Yt$, the rapidity of the top-quark, and $\ptt$, the transverse momentum of the top-quark. For each one of the four plots, related to the aforementioned distributions, we plot in the main panel: $\SigmaLOQCD$ divided by 10 (blue), $\Sigmakt$ (green), $\Sigmaktt$ (red) and $\Sigmakt+\Sigmaktt$ (mustard). The last three curves are equivalent to $\SigmaNPH$ for the three scenarios listed before (purely scalar, purely pseudoscalar and aligned mixing). In the inset we plot the ratio over $\SigmaLOQCD$, {\it i.e.}, the relative corrections to LO predictions.
We notice that at the threshold, in the case of $H$ being a pure scalar, the corrections  are positive while they are negative in the pseudoscalar case.  Thus, in case a discrepancy between theory and data appears, the sign of this discrepancy could be exploited in order to discriminate, within this framework,  a scalar versus pseudoscalar explanation. Also large cancellations can take place as soon as $\phiK\ne0$ and $\phiK\ne\pi/2$, as shown in the aligned mixing case and we expect a lower sensitivity on $|\Kt|-1$ in this configuration.

\begin{figure}
    \centering
     \includegraphics[width=0.48\textwidth]{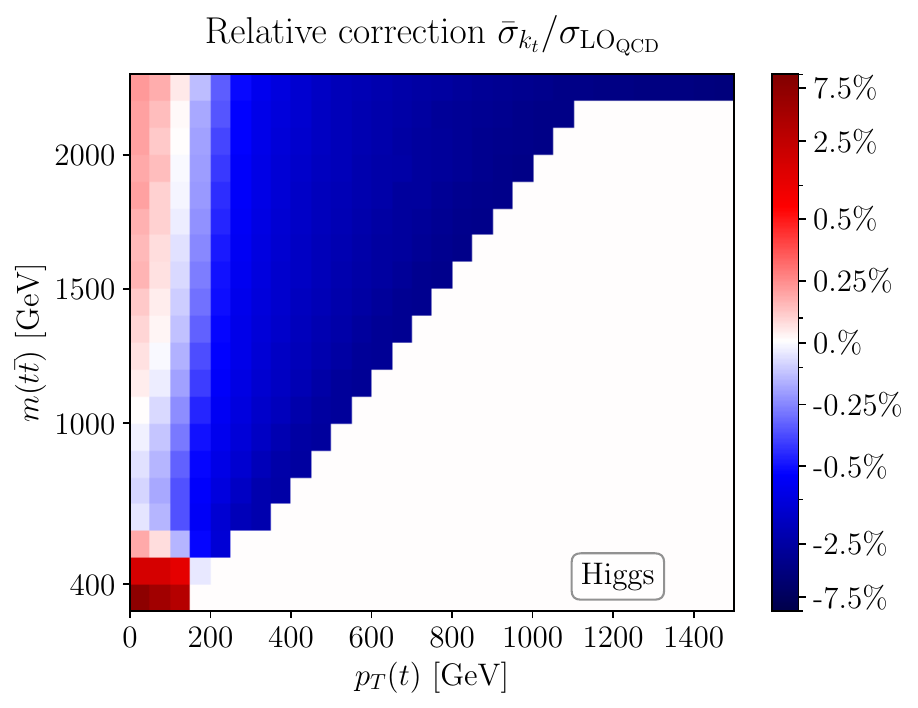}
      \includegraphics[width=0.48\textwidth]{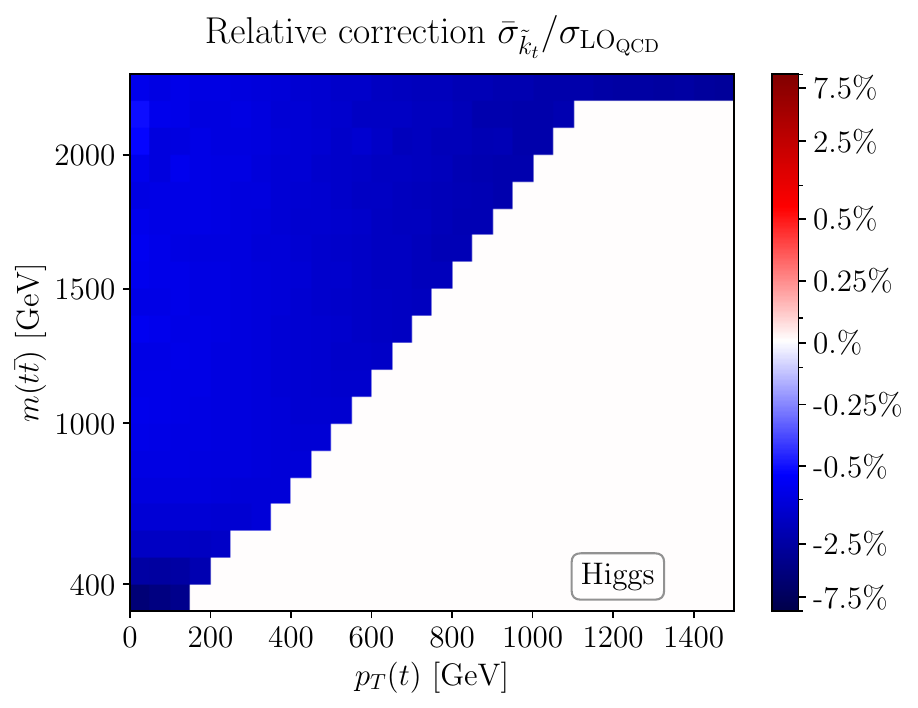}
    \caption{ Higgs Boson. Loop corrections $\SigmaNPH$ normalised to $\SigmaLOQCD$ at the doubly differential level in $\ptt$ and $\mtt$. Left: $(\kt,\ktt)=(1,0)$. Right: $(\kt,\ktt)=(0,1)$. The highest bins in $\ptt$ and $\mtt$ include also the overflow.}
    \label{fig:mttvspt}
\end{figure}

 Considering the other distributions, the behaviour is similar, the differences between the scalar and pseudoscalar corrections are maximal in the area corresponding to the threshold production, namely  $\ptt \simeq0$, $\Yt \simeq 0$ and $\Dytt\simeq 0$. Also, while the corrections from a purely pseudoscalar $H$ remain always negative, in the case of a purely scalar one they change  sign  moving towards the tail of $\mtt$ and $\ptt$ distributions. This is also at the origin of the different behaviour of the scalar for large $\ptt$ w.r.t.~large $\mtt$. In $t \bar t$ production, with no cuts on the phase space, large $\mtt$ is dominated by  $\ptt\ll\mtt/2$ due to the $t/u$-channel diagrams in the $gg\to t \bar t$ process, on the contrary large $\ptt$ is correlated with $\mtt\simeq 2\ptt$. Thus, in single-differential distributions in $\mtt$, for large $\mtt$ values there are cancellations between negative and positive contributions from different $\ptt$ regions, while this dynamics is not present at large values of $\ptt$, and therefore corrections are larger in absolute value.  These features are manifest in Fig.~\ref{fig:mttvspt}, where we show at doubly differential level in $\mtt$ and $\ptt$ the value of the ratio over $\SigmaNP/\SigmaLOQCD$. The left plot refers to the purely scalar case, while the latter to the purely pseudoscalar one.
 
 The top plots of Fig.~\ref{fig:alldistr} can be compared with results presented in Ref.~\cite{Martini:2021uey} and it can be noticed that while the purely scalar case is in agreement with results therein, the purely pseudoscalar case is not. To the best of our knowledge, in Ref.~\cite{Martini:2021uey} the contribution of the $s$-channel diagram in Fig.~\ref{fig:diagrams} (topology (a)) has been neglect. In the following we discuss in detail the impact of such diagram, and why it is especially relevant for the pseudoscalar case.
  
\subsection{Impact of the $s$-channel diagram}\label{sec:schannel}

The $s$-channel diagram in Fig.~\ref{fig:diagrams} is UV-finite and can be removed without violating gauge invariance. Therefore we have analysed the impact of removing such diagram for  the observable that we have considered in Fig.~\ref{fig:alldistr}.

\begin{figure}[!t]
    \centering
    \includegraphics[width=0.33\textwidth]{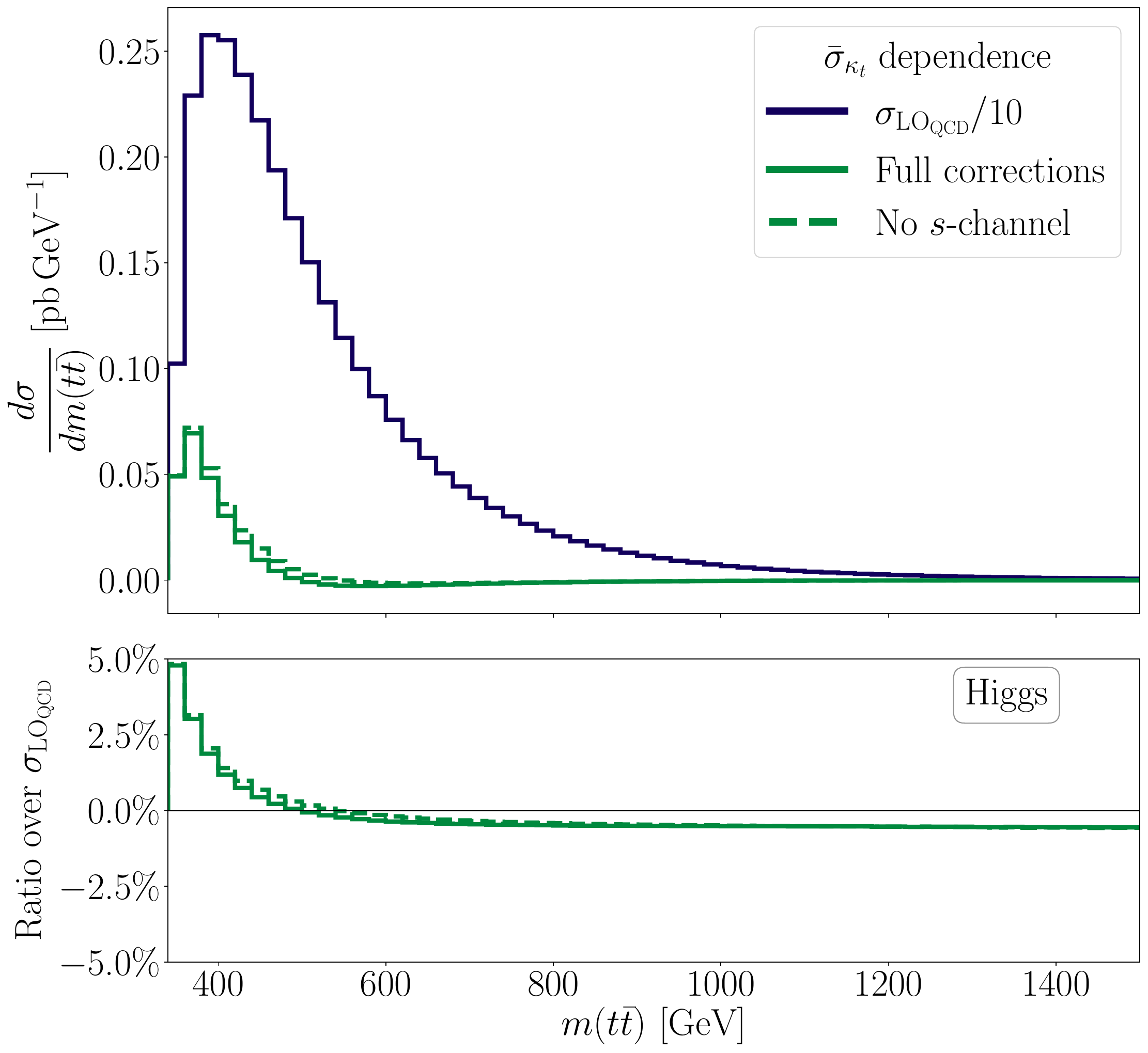} \includegraphics[width=0.32\textwidth]{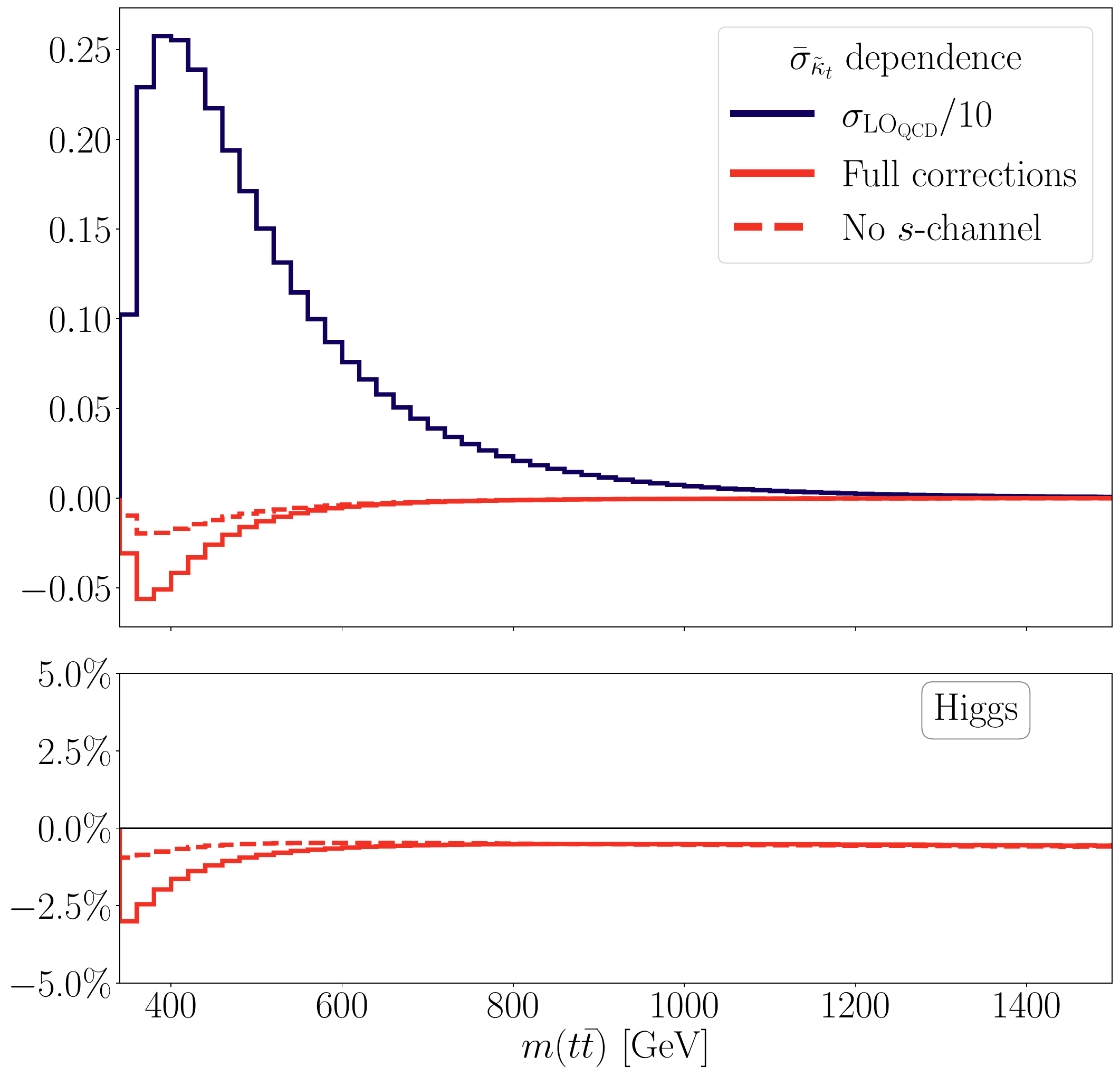} \includegraphics[width=0.32\textwidth]{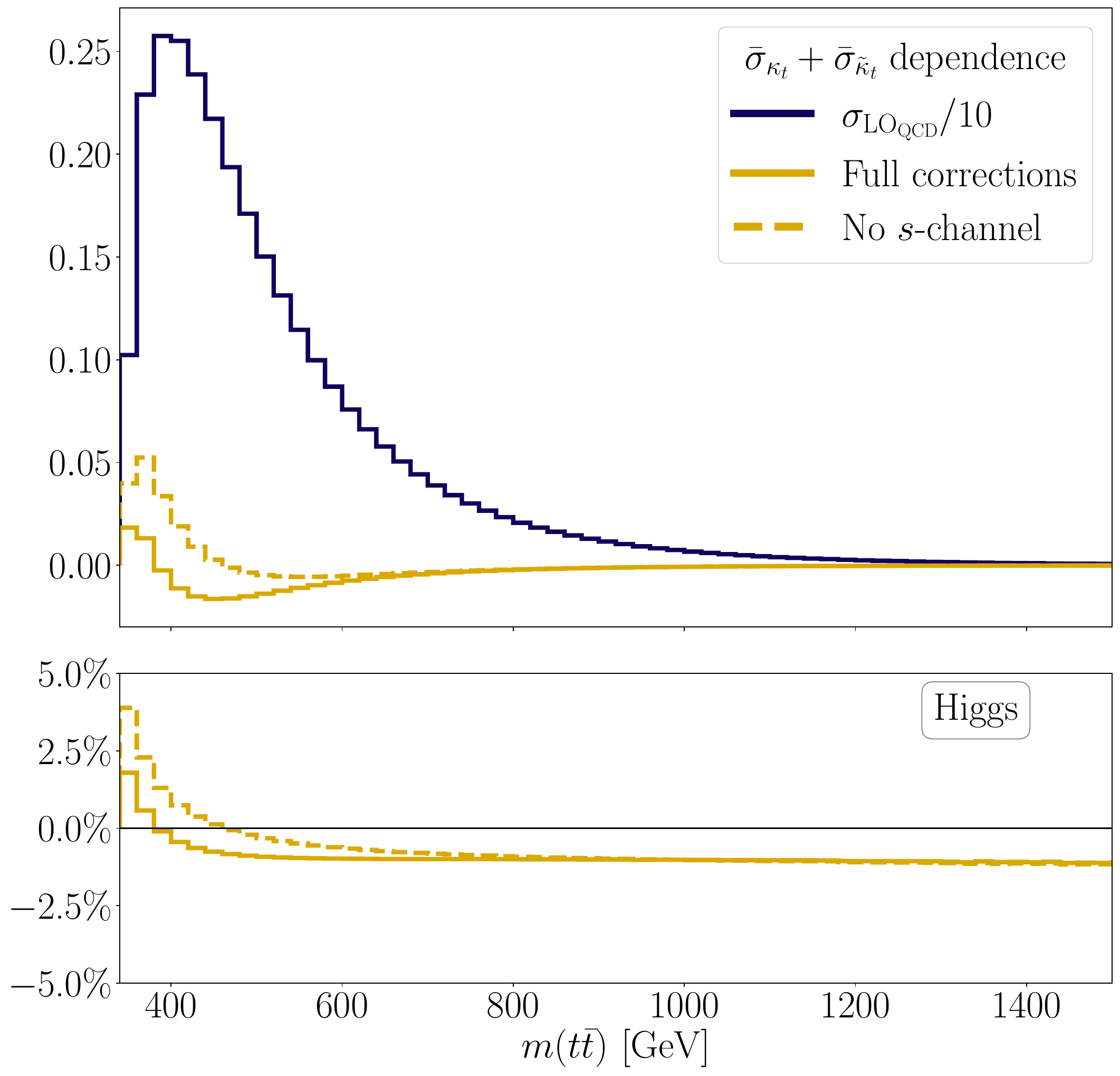}
     \includegraphics[width=0.33\textwidth]{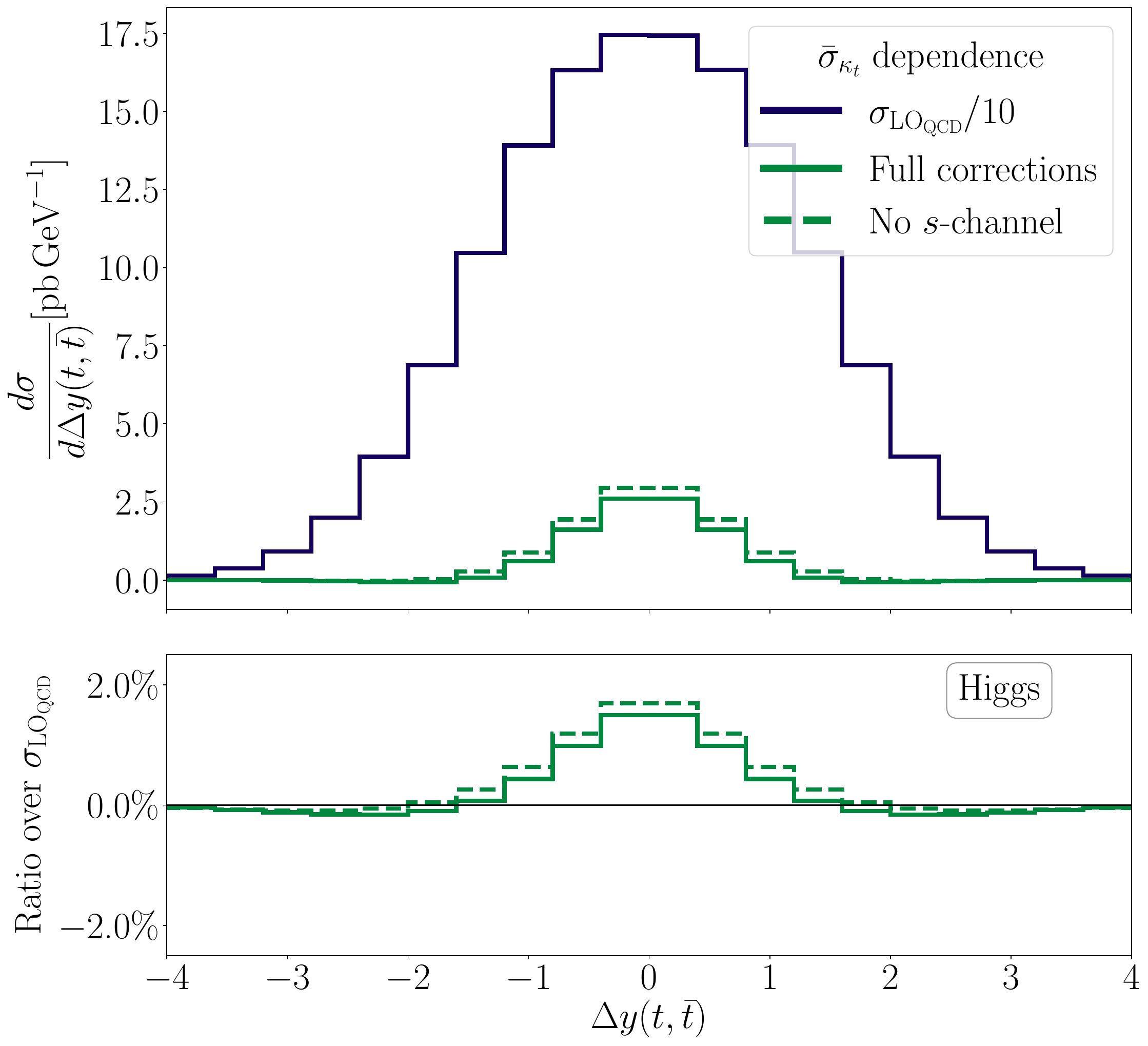} \includegraphics[width=0.32\textwidth]{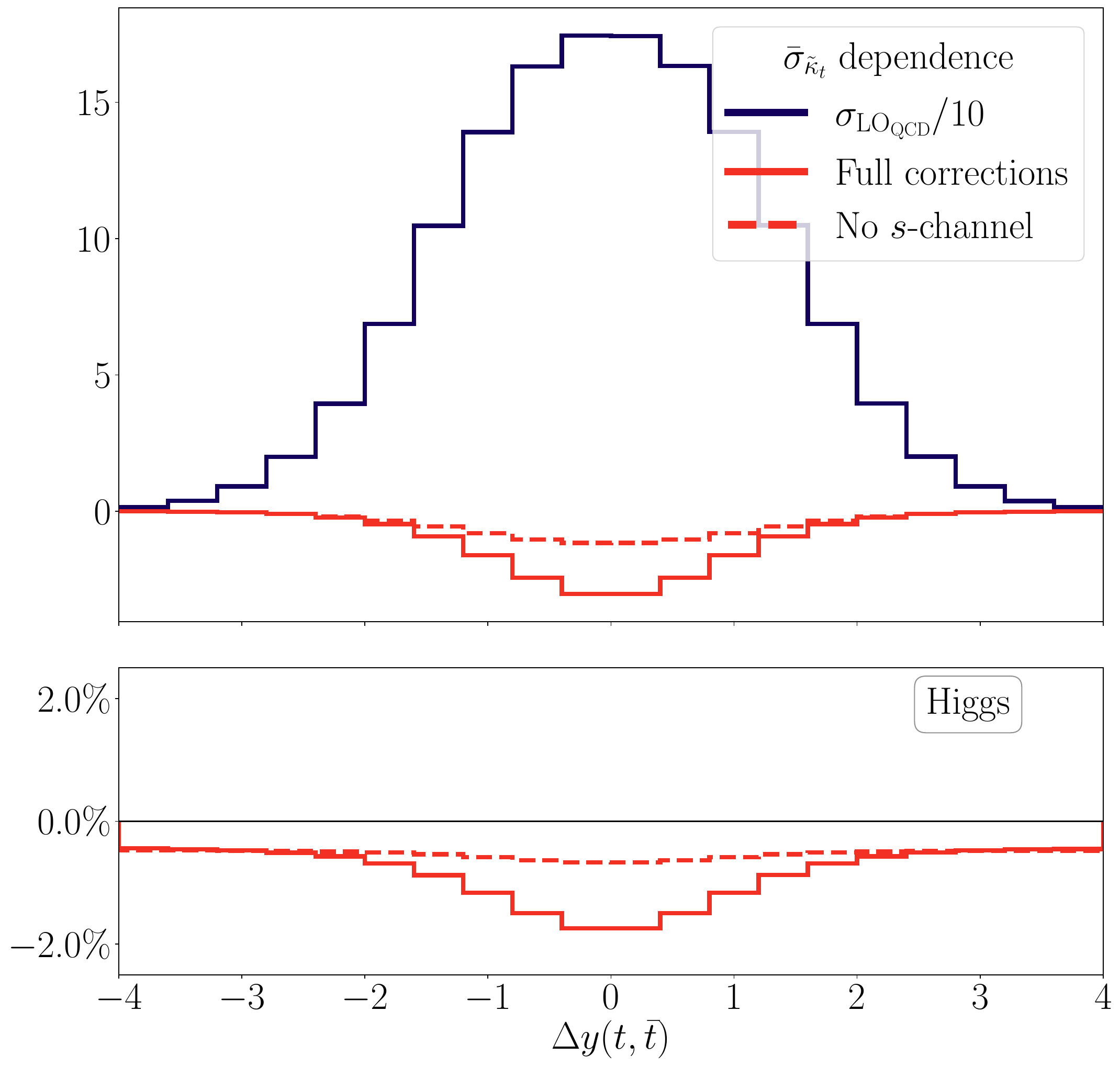} \includegraphics[width=0.32\textwidth]{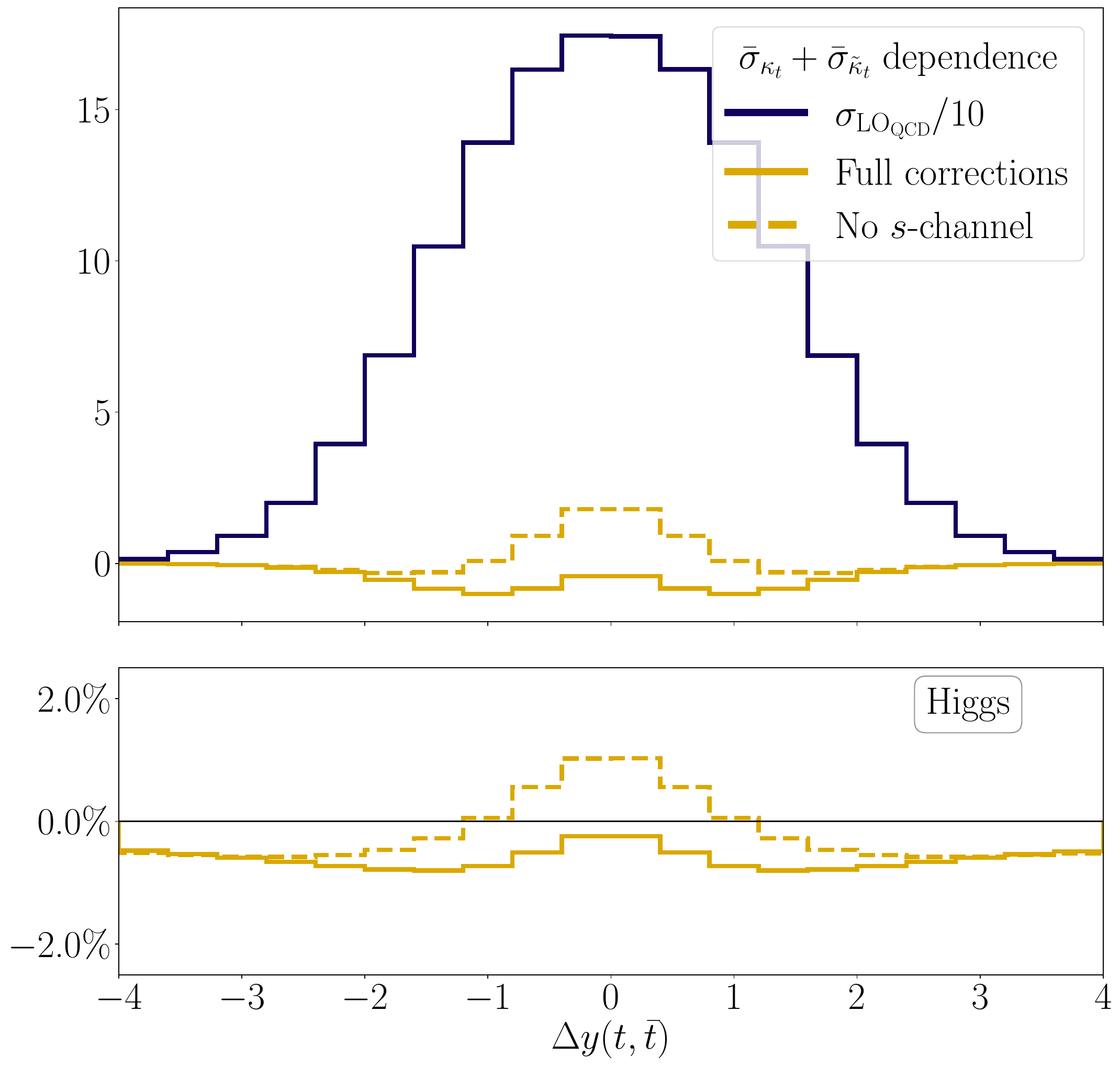}
    \includegraphics[width=0.33\textwidth]{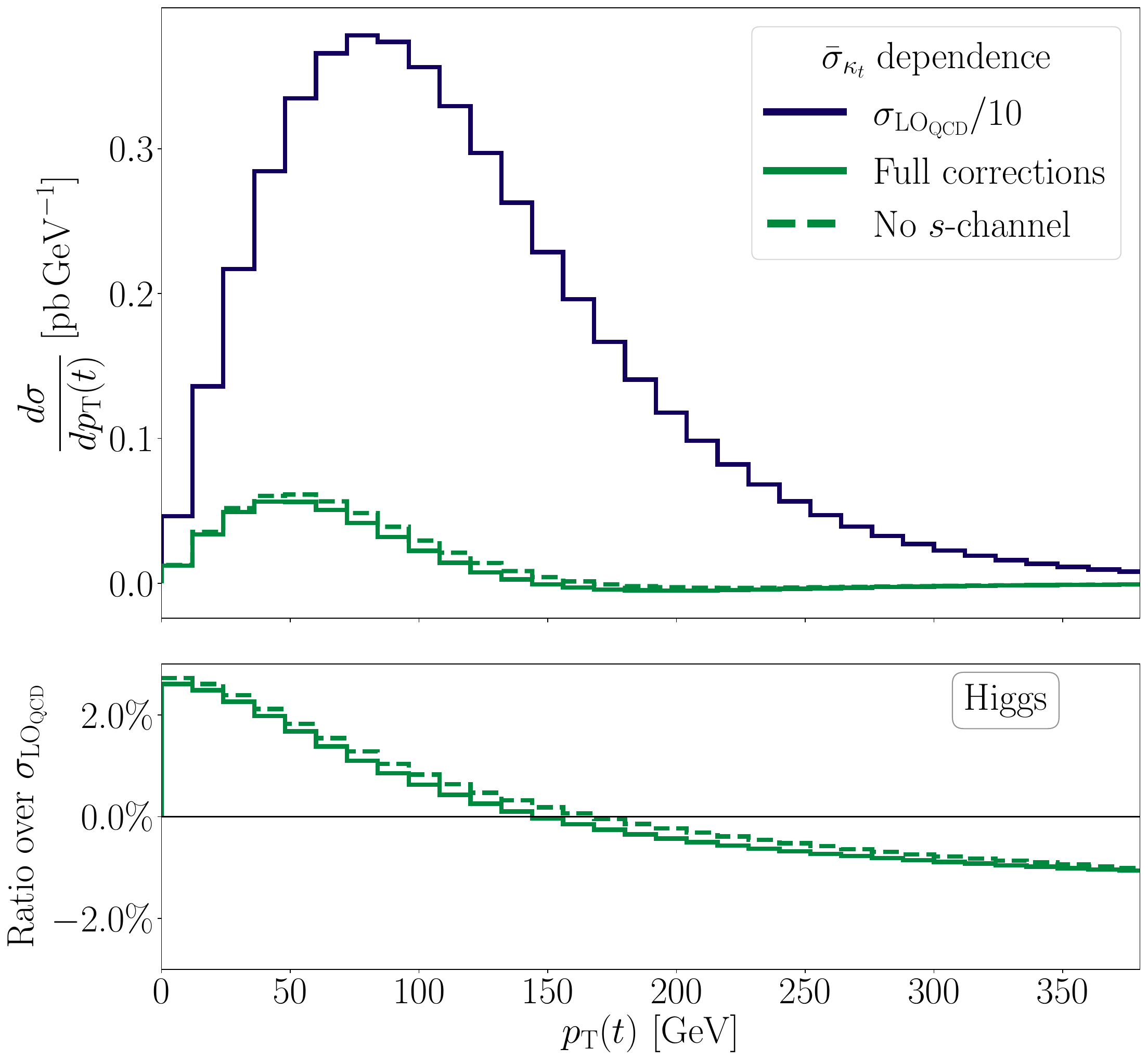} \includegraphics[width=0.32\textwidth]{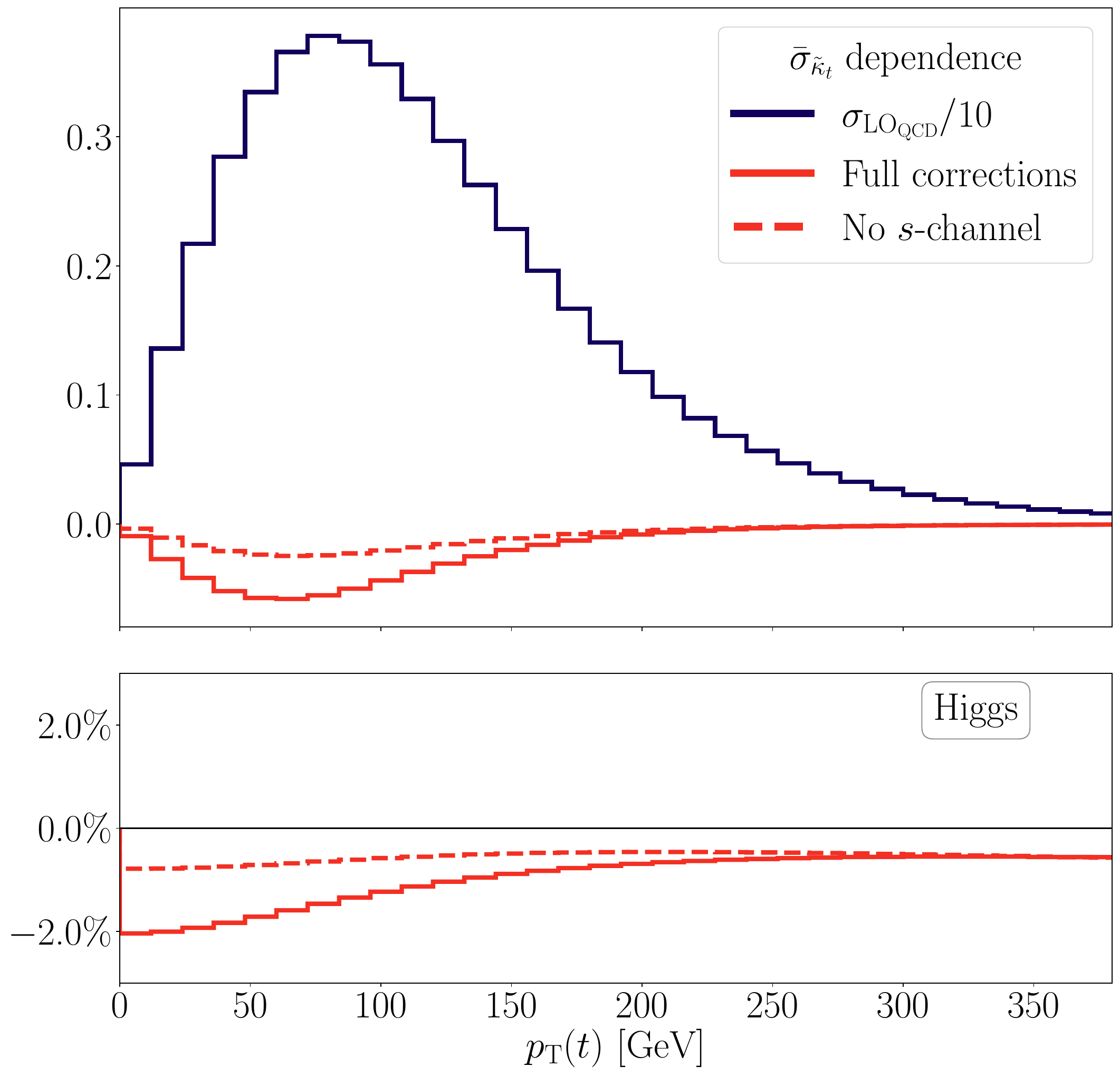} \includegraphics[width=0.32\textwidth]{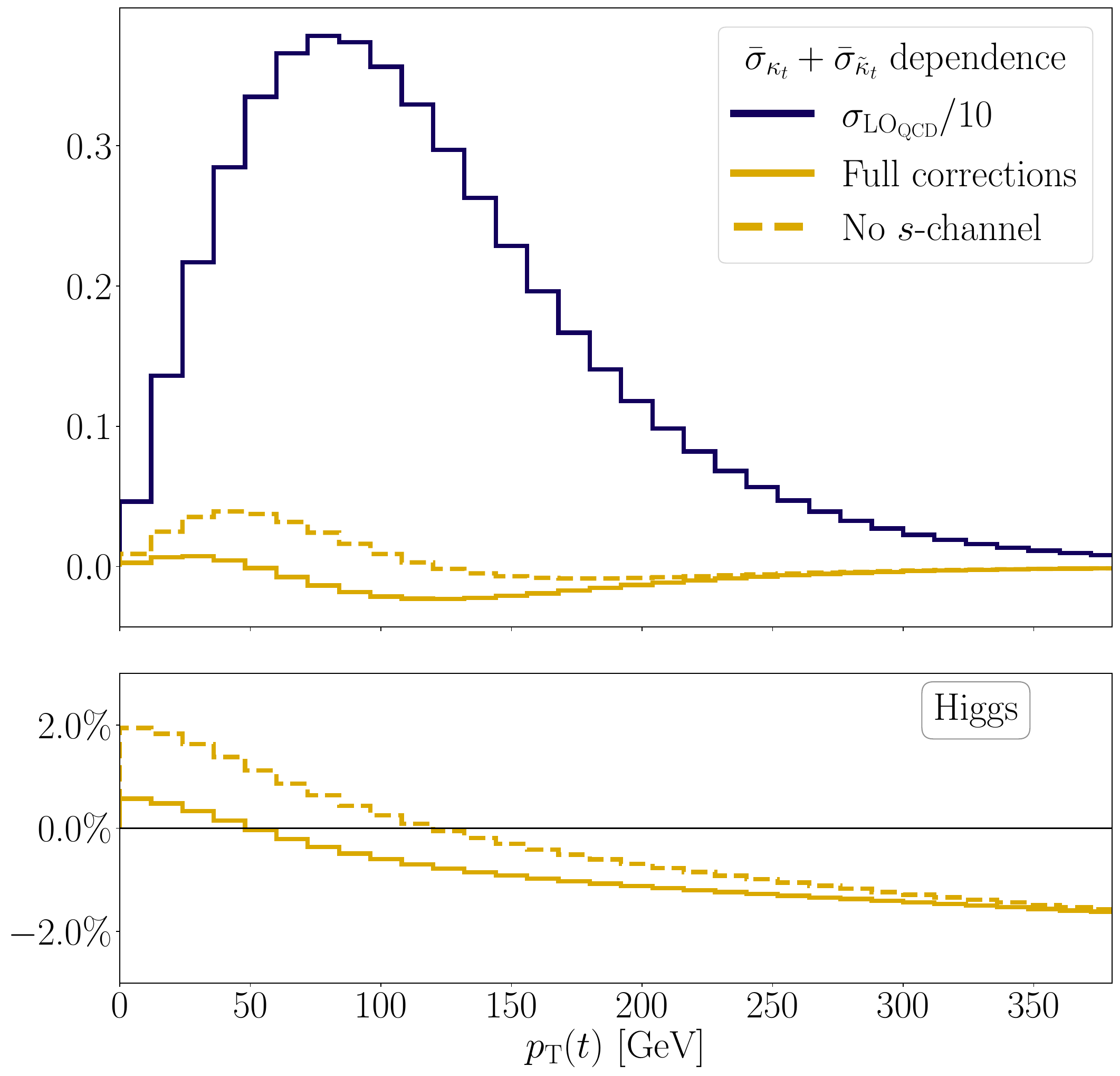}
    \includegraphics[width=0.33\textwidth]{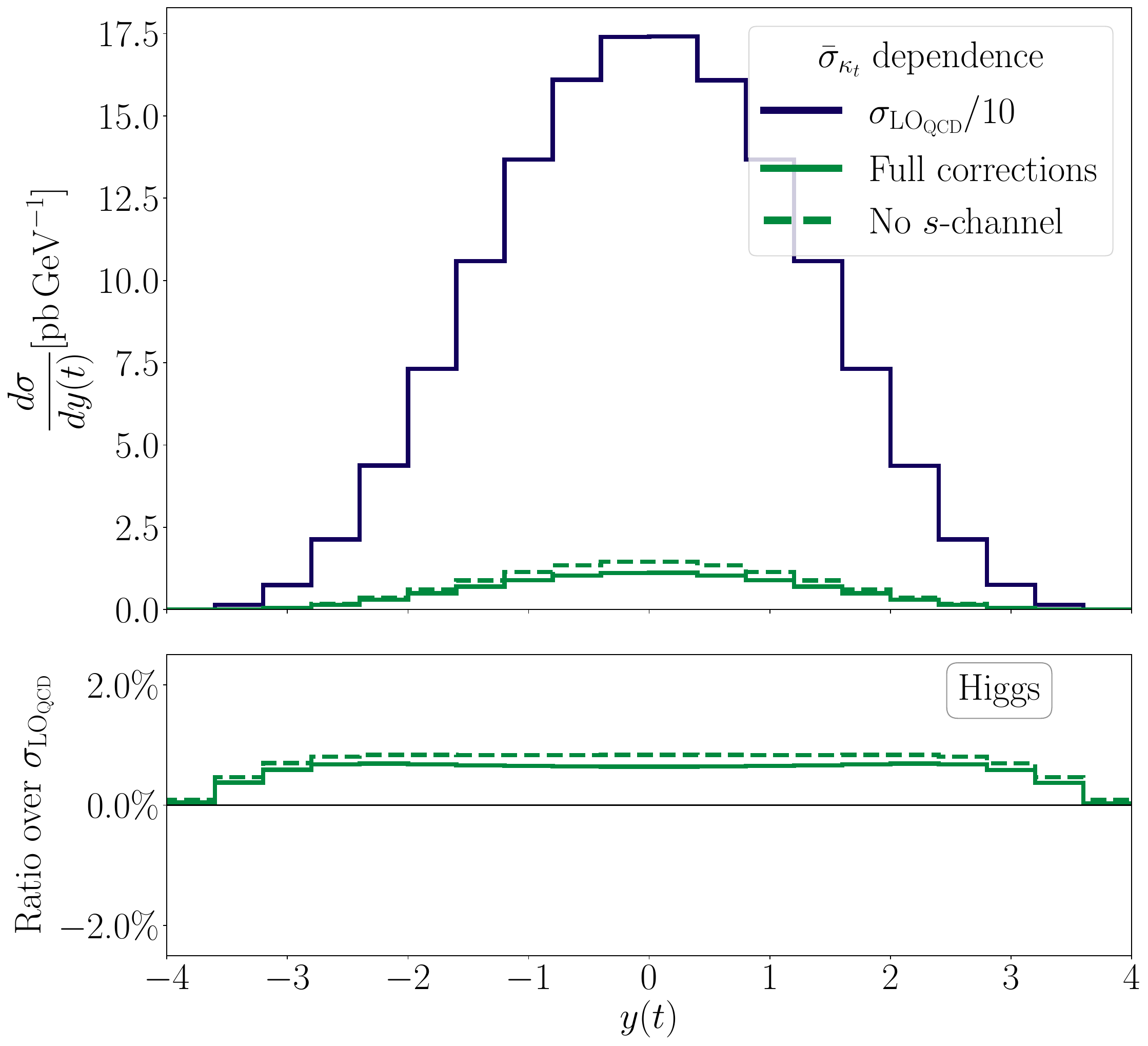} \includegraphics[width=0.32\textwidth]{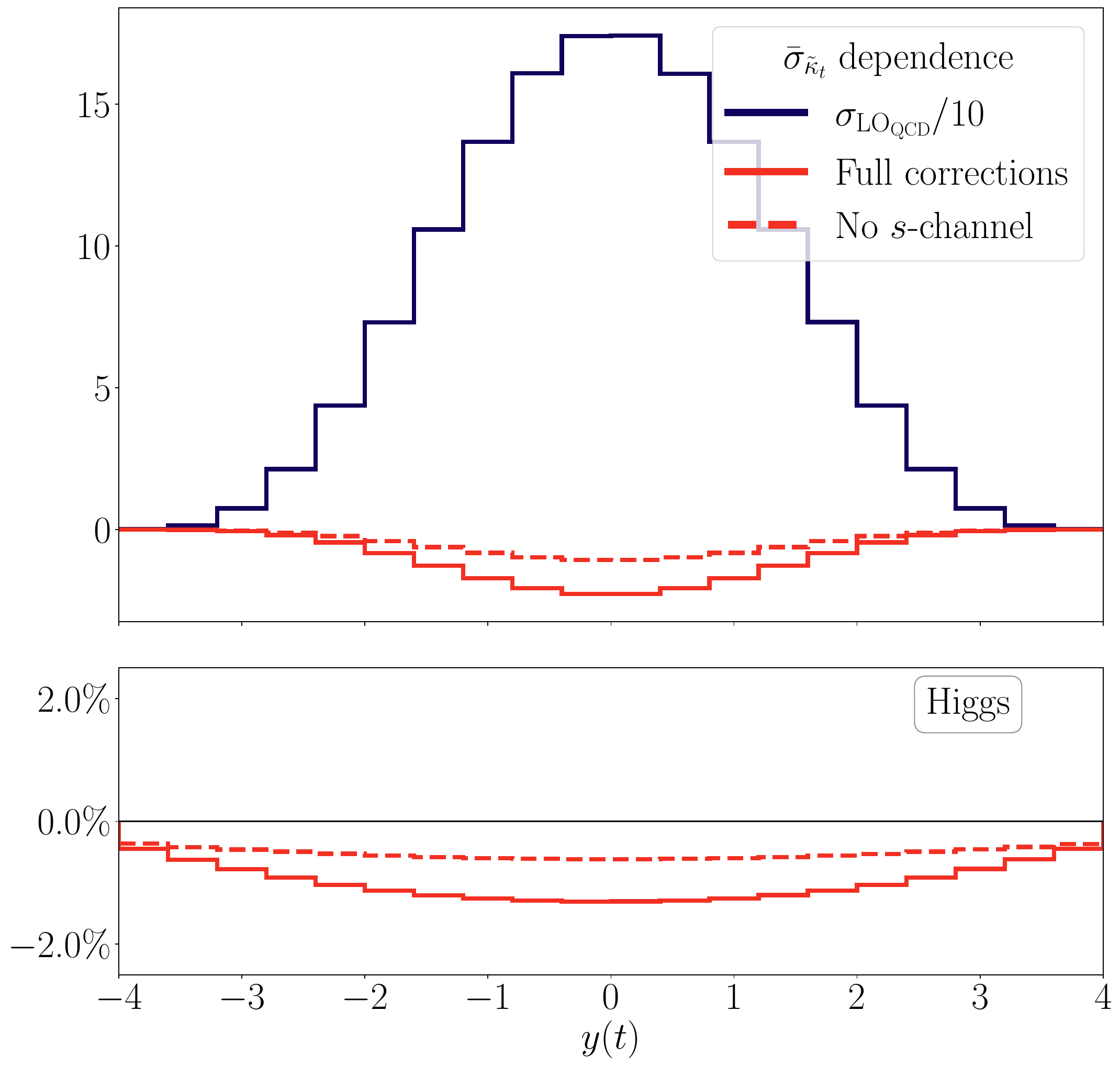} \includegraphics[width=0.32\textwidth]{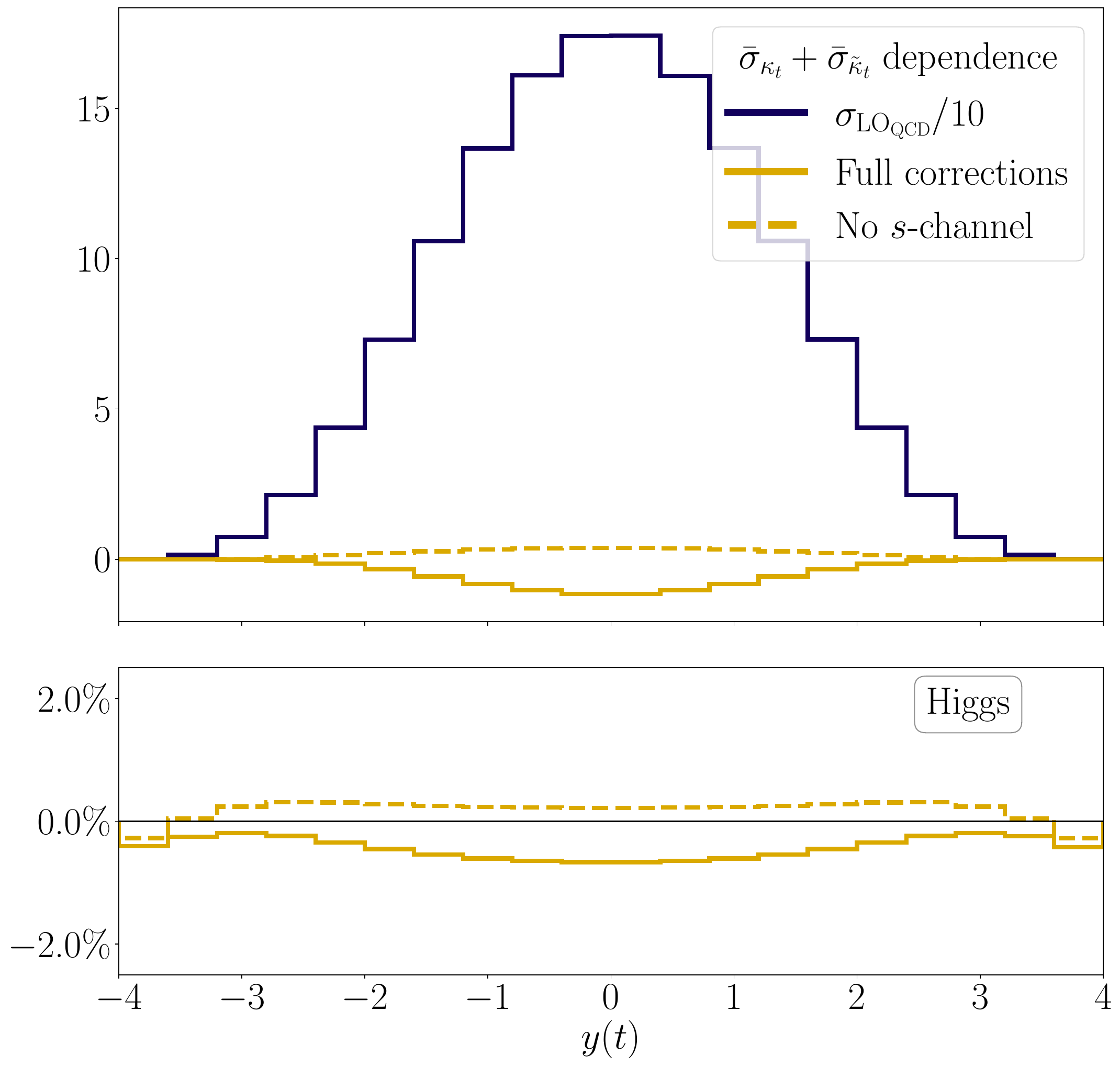}
    \caption{Higgs Boson.  In the main panel of each plot: $\SigmaLOQCD$ divided by ten (blue) and the loop corrections $\SigmaNPH$ with (solid line) and without (dashed line) the contribution from the $s$-channel diagram. In the inset of each plot: ratio of $\SigmaNPH$ over $\SigmaLOQCD$. Left plots:  $(\kt, \ktt)=(1,0)$ green. Central plots: $(\kt, \ktt)=(0,1)$  red. Right plots: $(\kt, \ktt)=(1,1)$ mustard.  }
    \label{fig:sVSnos}
\end{figure}

In the upper plots of Fig.~\ref{fig:sVSnos} we consider the $\mtt$ distribution and we show in the main panel  $\SigmaLOQCD$ divided by ten, as in Fig.~\ref{fig:alldistr}, and $\SigmaNPH$ with (solid) and without (dashed) the contribution from the $s$-channel diagram. The left plot shows the purely scalar case $\SigmaNPH=\Sigmakt$, the central one the purely pseudoscalar one   $\SigmaNPH=\Sigmaktt$ and the right one the aligned mixing case. In the main inset we show the ratio $\SigmaNPH/\SigmaLOQCD$. 

As can be seen,  in the purely scalar benchmark (left plot) the contribution coming from the $s$-channel is almost negligible. The opposite happens for the purely pseudoscalar benchmark;  the inclusion of the $s$-channel diagram plays a dominant role in the corrections as it appears from the central plot of Fig.~\ref{fig:sVSnos}. This behaviour is coherent with what is observed in Sec.~\ref{sec:distrct} for the case of an additional scalar $S$ when $m_S=300~\gev$. In that case, see Fig. \ref{fig:differetmass}, effects can also be larger since the $s$-channel diagram, approaching the resonance region $m_S=2m_t$, becomes dominant and indeed the $m_S=300~\gev$ case can be distinguished from all the others in Fig. \ref{fig:differetmass}, for the purely pseudoscalar case.
Clearly, this effect, although milder, is observed also in the right plot of Fig.~\ref{fig:sVSnos}, since the aligned mixed benchmark is a linear combination of the purely scalar and pseudoscalar ones.

In the second row of plots in Fig.~\ref{fig:sVSnos} we can also observe the similarities with the scalar $S$, when $m_S=300~\gev$, for the $\Dytt$ distributions, {\it cf.}~Fig.~\ref{fig:realemissiondrap}. Also in this case the difference between the scalar and the pseudoscalar benchmark is manifest and similarly this feature can be observed in the plots of the third row, $\ptt$ distributions, and of the fourth row, $\ytt$ distributions. We notice that, excluding the $s$-channel contribution, our predictions for the $\mtt$ and $\Dytt$ distributions are in agreement with those presented in Ref.~\cite{Martini:2021uey}, both for the purely scalar and pseudoscalar case.

 What has been discussed and summarised in Fig.~\ref{fig:mttvspt} for the relation between the $\ptt$ and $\mtt$ distributions explains also the differences between the corresponding distributions in  Figs.~\ref{fig:sVSnos}; the impact of the $s$-channel diagram is large for the $\mtt$ threshold, where the suppression of the off-shell propagator  is not present, and this phase-space region is correlated  with large $\ptt$ values where a mild impact of such diagram, at variance with the case of at large $\mtt$ values, can be observed.

The reason for the different behaviour in the scalar benchmark and pseudoscalar one can be understood by looking at the analytic form of the interference between the QCD SM top-pair production and the $s$-channel virtual mediated diagram (a) computed in Ref.~\cite{Dicus:1994bm}. The SM cross section is proportional to the top-pair velocity $\beta$ that can be written as
\begin{equation}
\beta=\sqrt{1-\frac{4m_t^2}{m(t\bar t)^2}} \,,
\end{equation}
while the scalar benchmark correction due to the $s$-channel is proportional to $\beta^3$ and the pseudoscalar benchmark one to $\beta$. At threshold, where the invariant mass of the top-quark pair approaches the lower bound $2m_t$, the $s$-channel relative contribution with respect to the SM goes to zero as $\beta^2$, while in the pseudoscalar case, it assumes a constant value. A top-quark pair produced via the mediation of a scalar particle  is in a total spin zero state. If the interaction is CP-odd the pair is produced in an $S$-wave configuration and spin singlet ${}^1S_0$ state, while for a CP-even coupling they are produced in a $P$-wave state-spin triplet ${}^3P_0$ state, thus the difference $\beta$ dependence and behaviour, see, {\it e.g.}, Ref.~\cite{Maltoni:2024tul}.
 
\section{Higgs boson $H$: Sensitivity study}
\label{sec:sensitivityH}

In this section, using the same set-up discussed in Sec.~\ref{sec:statistics} for the case of the scalar $S$, we present an explorative study on how our calculation, which consistently takes into account the $s$-channel diagram in Fig.~\ref{fig:diagrams}, can affect the bounds that can be obtained at the LHC on the $\kt$ and $\ktt$, which parametrise respectively  the CP-even and CP-odd interactions of the Higgs boson with the top-quark. Equations in Sec.~\ref{sec:statistics} can be easily extended to the case of the Higgs, {\it e.g.}~replacing $\SigmaNP$ with $\SigmaNPH$ and similarly for other quantities.  The main difference in the set-up w.r.t.~what has been discussed in Sec.~\ref{sec:Ssensitivity} is that we will not include here the contribution from $t\bar t H$ production, since the radiation of a Higgs boson can be experimentally distinguished from $t \bar t$ production. 

In Sec.~\ref{sec:SMkfit} we consider the case of pseudo-data that corresponds to the SM, in Sec.~\ref{sec:pseudofit} instead the case of pseudo-data corresponding {\it not} to the SM but to the purely pseudoscalar configuration $(\kt,\ktt)=(0,1)$. In Sec.~\ref{sec:schannelfit} we discuss for both scenarios the impact of the $s$-channel diagram in the fits and bounds on $\kt$ and $\ktt$.

\subsection{SM-like pseudodata}
\label{sec:SMkfit}

In this section we perform the sensitivity study assuming that the pseudo-data correspond exactly to the SM prediction, as done for the case of the scalar $S$ in Sec.~\ref{sec:boundscts}.

In Tab.~\ref{tab:Higgscalarfit} we  report the result for the one-dimensional parameter fit, {\it i.e.}~separately fitting  $\kt$ and $\ktt$. We  perform the fit by using  for theory predictions both $\SigmaSMNPHmult$ and $\SigmaSMNPHadd$ (see Eqs.~\eqref{eq:multkts} and Eqs.~\eqref{eq:multkts}) quantities. For pseudo data we consistently use in the two cases the prediction for $(\kt, \ktt)=(1,0)$. Thus, by construction, the fit on the scalar coupling must result in $\kappa_t=1$ and indeed we find that. In Tab.~\ref{tab:Higgscalarfit} we report besides central values the $1\sigma$, $2\sigma$ and $3\sigma$ errors. In fact, by looking at the expression of $\SigmaNPH$ in Eq.~\eqref{eq:sigmanpinkts}, it is clear that not only $(\kt, \ktt)=(1,0)$ but also $(\kt, \ktt)=(-1,0)$ is a solution of the $\chi^2$ function minimisation, and it will be manifest when discussing the plots for the simultaneous $\kt$ and $\ktt$ fits. When the errors become larger than 1, the error intervals for the $(\kt, \ktt)=(1,0)$ and $(\kt, \ktt)=(-1,0)$ solutions overlap. This is the origin of the exact $-1$ value in some of the results displayed for the errors. It can also be noted that the $1\sigma$ error in the case of the $\kt$ fit is compatible with the ones of Refs.~\cite{CMS:2019art,CMS:2020djy}. This is precisely due to the strategy that we have described in Sec.~\ref{sec:statistics} and that we have adopted in order to obtain a simplified framework for mimicking the accuracy already achieved at the experimental level.
The difference between results obtained between the fit in the multiplicative and additive approaches can be regarded as theory uncertainties in modelling predictions for both the SM and NP contributions.

\renewcommand{\arraystretch}{1.5}

\begin{table}[!t]
    \centering
    \begin{tabular}{l|c|c}
       &  ${\kappa_t}^{+1\sigma,\,2\sigma,\,3\sigma}_{-1\sigma,\,2\sigma,\,3\sigma}$ & ${\tilde\kappa{_t}}^{+1\sigma,\,2\sigma,\,3\sigma}_{-1\sigma,\,2\sigma,\,3\sigma}$ \\[1.5pt]
              \hline
              $ \SMmult$ LHC & $1.00^{+0.28,\,0.52,\,0.72}_{-0.41,\,1.0,\,1.0}$&$0.0^{+0.59,\,1.05,\,1.43}_{-0.59,\,1.06,\,1.44}$\\
       $\SMadd$ LHC & $1.00^{+0.38,\,0.68,\,0.94}_{-0.72,\,1.0,\,1.0}$ &$0.0^{+0.81,\,1.39,\,1.82}_{-0.82,\,1.39,\,1.84}$
        
    \end{tabular}
    \caption{Best fit and corresponding errors values for $\kappa_t$ and $\tilde\kappa_t$ for the case in which pseudo-data corresponds to the case of $H$ being the SM Higgs, $(\kt,\ktt)=(1,0)$. In the first(second) row numbers refer to the usage of the multiplicative(additive) approach for both theory predictions and pseudo-data. }
    \label{tab:Higgscalarfit}
\end{table}

It is also very interesting to notice that, fitting  $\ktt$, the best result is obtained for $\ktt=0$, which was not {\it a priori} obvious. It means that assuming that the Higgs has only pseudo-scalar interactions, then if the pseudo-data are simulated via the SM  (Higgs purely scalar), the absence of the Higgs, or equivalently no interactions of it with the top-quark, would yield the best fit.
The origin of this effect is the opposite signs of $\Sigmakt$ and $\Sigmaktt$  associated to  the purely scalar and pseudoscalar benchmarks, respectively, for the $\mtt$ distribution, {\it cf.} Fig. \ref{fig:alldistr}. The specific values of these two quantities for the binning we considered  can be read in Tab~\ref{tab:inputforfit} in Appendix \ref{app:table}.

In Fig.~\ref{fig:exclusionHiggs} we show our results for the two-parameter fit in $\kt$ and $\ktt$, where we display the $1\sigma$ (mustard), $2\sigma$ (red) and $3\sigma$ (green) error contours. In the left plot, the multiplicative approach has been used, while in the right plot  the additive approach has been used. In the two plots, the best fits are for $(\pm 1,0)$ and are indicated by light-blue dots. We also display for convenience in the plot, as a dark-blue dot, the SM scenario with a CP-even Higgs, $(\kt,\ktt)=(1,0)$ and as purple one the case where the Higgs is purely pseudo scalar and with the strength of the top-Higgs coupling equal to the one in the SM, {\it i.e.}, $(\kt,\ktt)=(0,1)$.

As expected, in the multiplicative approach bounds are more stringent and, {\it e.g.}, the $1\sigma$ contours around the two minima are separated (the best fit is not compatible with the no Higgs scenario, {\it i.e.}, $(\kt,\ktt)=(0,0)$), at variance with the case of the usage of the additive approach. Moreover, in the multiplicative approach, the purely pseudoscalar benchmark (purple dot) is almost excluded at $2\sigma$, while in the additive approach it is not. Again, differences in the two approaches can be considered as theory uncertainties.

It is manifest from  Fig.~\ref{fig:exclusionHiggs}  that the bound that can be set on $|\Kt|$ strongly depends on the value of $\phiK$ and in particular that around $\phiK\simeq \pi/4$ bounds are much less stringent than those at $\phiK=0$, the purely scalar case, and especially those at $\phiK=\pi/2$, the purely pseudoscalar case.
The origin of this effect is precisely the cancellations taking place between the $\Sigmakt$ and $\Sigmaktt$ quantities discussed in the plots of Sec.~\ref{sec:distrkt} for the aligned mixing benchmark.\footnote{In fact the almost flat direction, which corresponds the case where the contributions proportional to $\Sigmakt$ and $\Sigmaktt$ are similar in absolute value but with opposite sign and therefore leading to large cancellations, rather than being associated to $\phiK\simeq \pi/4$ is in first approximation aligned around the  relation $\kt^2-1=\ktt^2$, which minimises $\SigmaNPH$ in Eq.~\eqref{eq:sigmanpinkts}.  }

\begin{figure}[!t]
    \centering
    \includegraphics[width=0.36\textwidth]{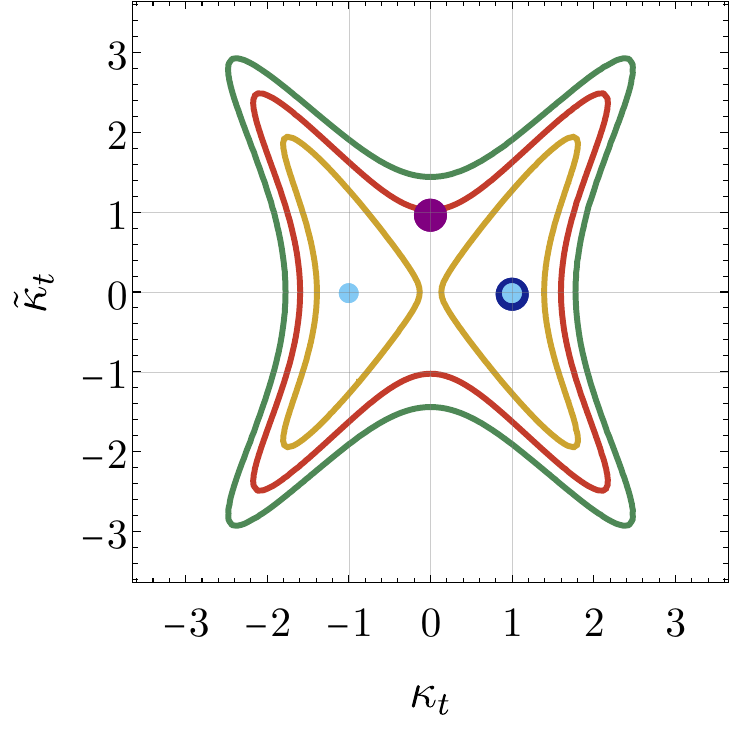}
    \hspace{0.5cm}
\includegraphics[width=0.56\textwidth]{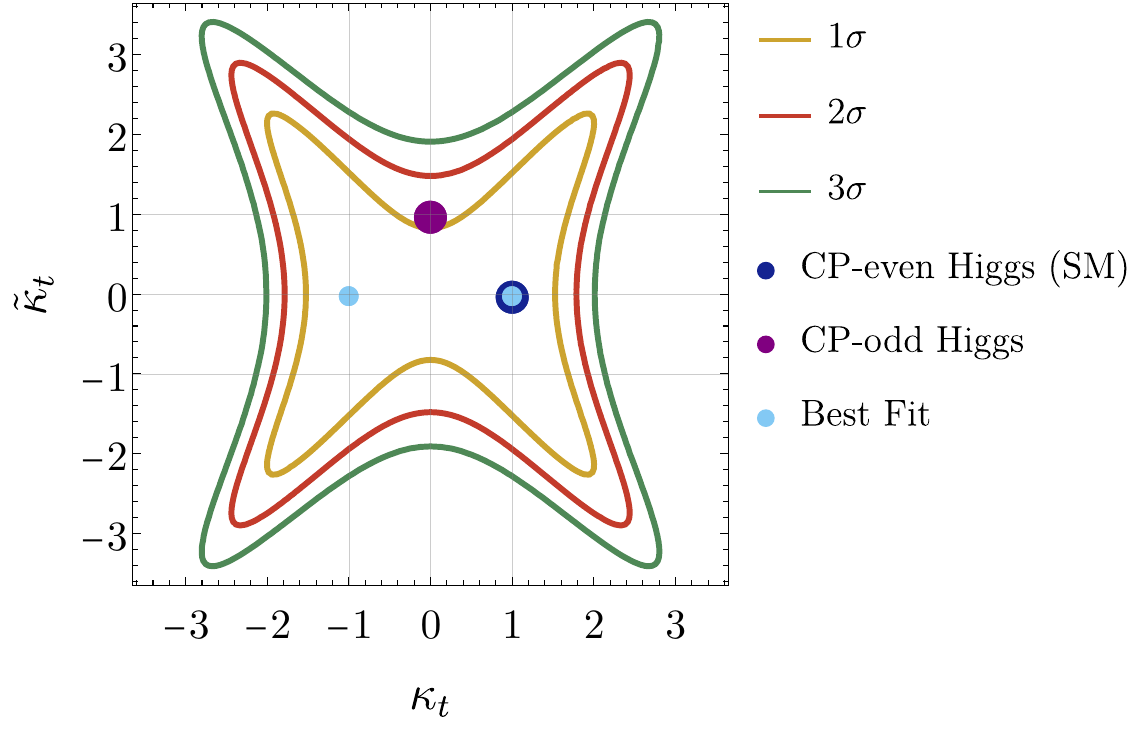}
    \caption{Bounds in the $(\kt, \ktt)$ plane at different confidence levels, assuming $H$ being the SM Higgs, $(\kt,\ktt)=(1,0)$ in the pseudo-data. Left: multiplicative approach. Right: additive approach.}
    \label{fig:exclusionHiggs}
\end{figure}

\subsection{CP-odd Higgs with $|\Kt|=1$}
\label{sec:pseudofit}

In this section we perform the fit with the pseudo-data corresponding {\it not} to the SM but to the configuration $(\kt,\ktt)=(0,1)$, {\it i.e.}~the  purple dot in Fig.~\ref{tab:Higgscalarfit} where the Higgs boson is a purely pseudo scalar and  the strength of the top-Higgs coupling is equal to the one in the SM.
The statistical analysis is the same discussed already in the previous sections for the  scalar $S$ and the Higgs boson, the only difference is that the pseudodata $\Sigmaexp$ entering Eq.~\ref{eq:chi} have been obtained via  $\SigmaSMNPHadd$($\SigmaSMNPHmult$) in the additive(multiplicative) approach evaluated at $(\kt,\ktt)=(0,1)$.

The results of the one-parameter fits, similar to those in Tab. \ref{tab:Higgscalarfit},  can be read in Tab. \ref{tab:Higgspseudoscalarfit}. As expected, the best fit for the parameter $\ktt$ is at $\ktt=1$ and, analogously to Sec.~\ref{sec:SMkfit}, we find that the best fit for the other parameter, in this case $\kt$, is $\kt=0$. The latter result was not obvious a priori and it again originates from the opposite signs of $\Sigmakt$ and $\Sigmaktt$.  
\begin{table}[!t]
    \centering
    \begin{tabular}{l|c|c}
       &  ${\kappa_t}^{+1\sigma,\,2\sigma,\,3\sigma}_{-1\sigma,\,2\sigma,\,3\sigma}$ & ${\tilde\kappa{_t}}^{+1\sigma,\,2\sigma,\,3\sigma}_{-1\sigma,\,2\sigma,\,3\sigma}$ \\[1.5pt]
       \hline
       $ \SMmult$ LHC & $0.00^{+0.55,\,0.93,\,1.22}_{-0.55,\,0.93,\,1.22}$&$1.0^{+0.44,\,0.78,\,1.06}_{-1.00,\,1.00,\,1.00}$\\
              $\SMadd$ LHC & $0.00^{+0.73,\,1.18,\,1.51}_{-0.73,\,1.18,\,1.51}$ &$1.0^{+0.60,\,1.02,\,1.38}_{-1.00,\,1.00,\,1.00}$
         \end{tabular}
    \caption{Same as Tab.~\ref{tab:Higgscalarfit} but with pseudo-data corresponding to $H$ being a Higgs boson with a CP-odd coupling to the top-quark, in particular $(\kt, \ktt)=(0,1)$. }
    \label{tab:Higgspseudoscalarfit}
\end{table}
\begin{figure}[!t]
    \centering
    \includegraphics[width=0.368\textwidth]{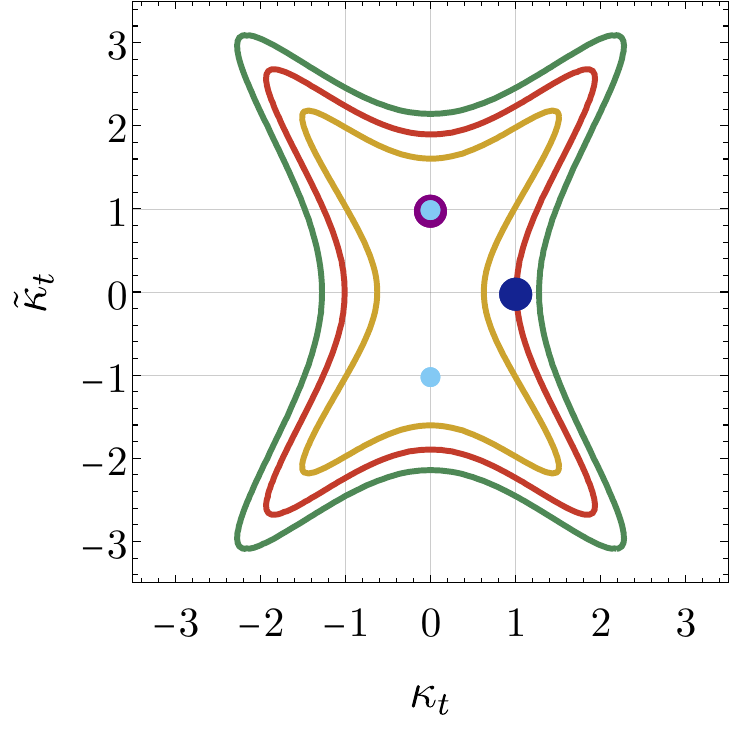} \hspace{0.5cm}
    \includegraphics[width=0.575\textwidth]{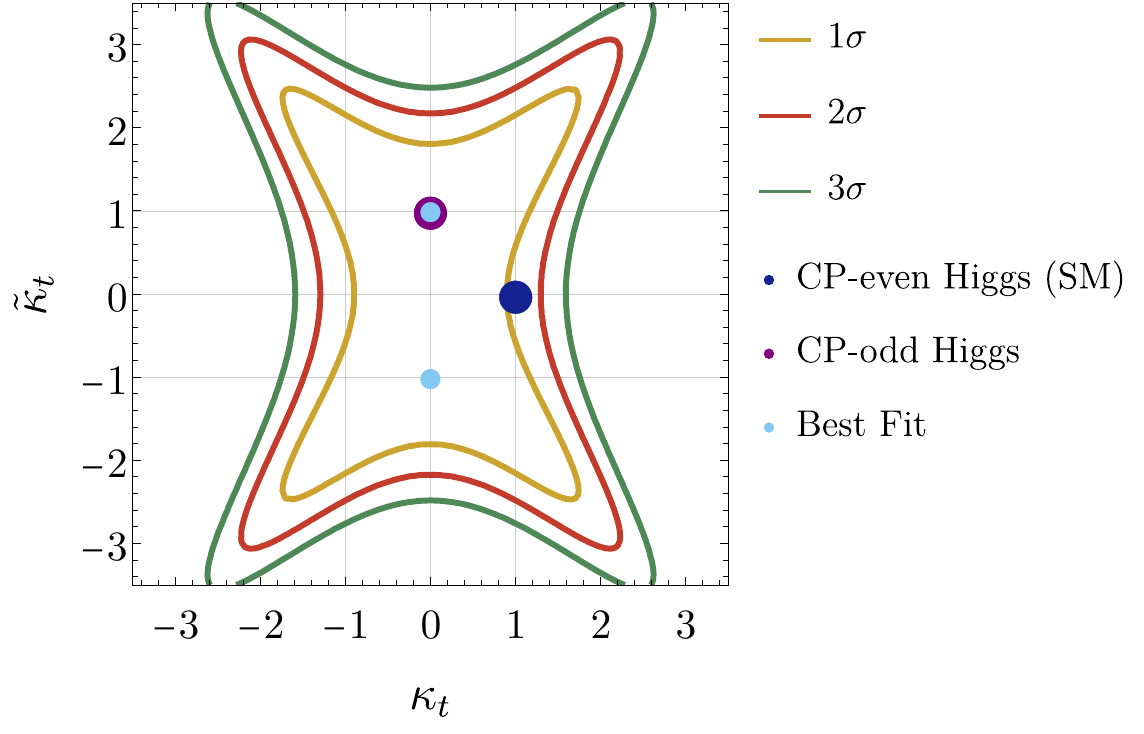}
    \caption{Same as Fig.~\ref{fig:exclusionHiggs} but with pseudo-data corresponding to $H$ being a Higgs boson with a CP-odd coupling to the top-quark, in particular $(\kt, \ktt)=(0,1)$.}
    \label{fig:pseudodatapscalar}
\end{figure}
The two-parameter fit for  $\kt$ and  $\ktt$ is displayed in Fig.~\ref{fig:pseudodatapscalar} and the results we find are  very similar to the one in Fig.~\ref{fig:exclusionHiggs} for the scalar case. The fit performed in the case of the multiplicative approach shows that the SM is compatible at almost $2\sigma$ level.

\subsection{The $s$-channel relevance}
\label{sec:schannelfit}

In this section we study the impact of the $s$-channel diagram in Fig.~\ref{fig:diagrams} on the bounds that have been presented in the two previous sections. We start by presenting  in Tabs.
\ref{tab:higgsnosS} and \ref{tab:higgsnosPS} the same information of respectively Tabs.~\ref{tab:Higgscalarfit} and \ref{tab:Higgspseudoscalarfit}, but having removed the contribution of the $s$-channel diagram from the $\SigmaNPH$ component of $\SigmaSMNPHmult$, which is used for obtaining the predictions of $\Sigmatheo$ entering  the $\chi^2$-function of Eq.~\eqref{eq:chi}. On the contrary, we still simulate the pseudo-data $\Sigmaexp$ entering the $\chi^2$-function {\it with} the contribution of the $s$-channel diagram. In other words, we simulate pseudo-data with the correct expressions, for either the SM or the case of a purely pseudoscalar Higgs boson, and we study how including  (as done in this work) or not including (as done in  Ref.~\cite{Martini:2021uey}) the contribution of the $s$-channel diagram can affect the limits on $\kt$ and/or  $\ktt$.  

Consistently with what has been discussed in Sec.~\ref{sec:schannel}, fits of $\kappa_t$ are almost untouched in Tab.~\ref{tab:higgsnosS} w.r.t.~Tab.~\ref{tab:Higgscalarfit}, while there is a big change for the error associated to the fit for $\tilde\kappa_t$; the $s$-channel diagram has a large impact only for the pseudoscalar. On the other hand, in Tab.~\ref{tab:higgsnosPS} both for the $\kt$ and $\ktt$ fit we observe differences w.r.t.~Tab.~\ref{tab:Higgspseudoscalarfit}. In this case, in the comparison the pseudodata  are different and therefore both fits are affected and also the best result is not centred around $(\kt,\ktt)=(0,1)$.

\begin{figure}[!t]
    \centering
    \includegraphics[width=0.37\textwidth]{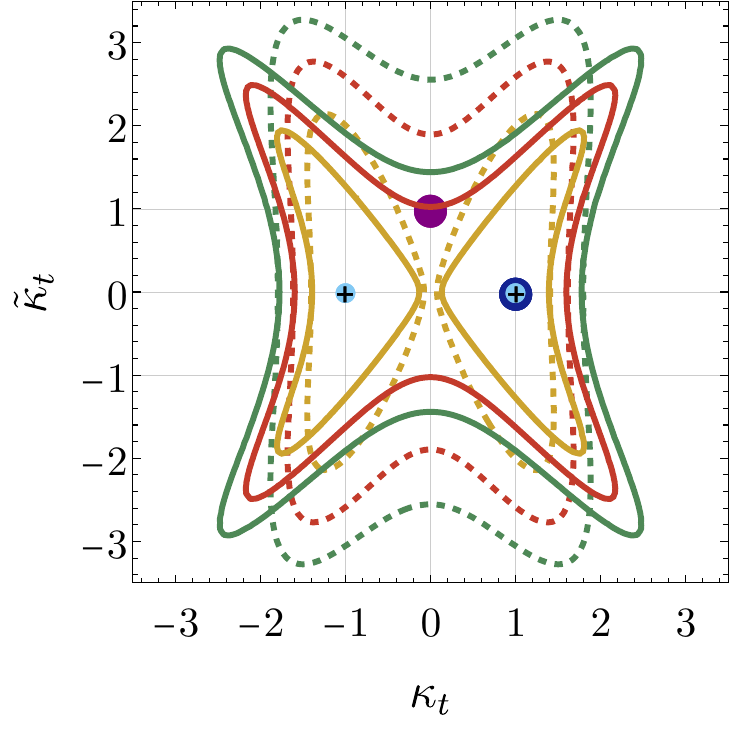} \hspace{0.2cm}
    \includegraphics[width=0.59\textwidth]{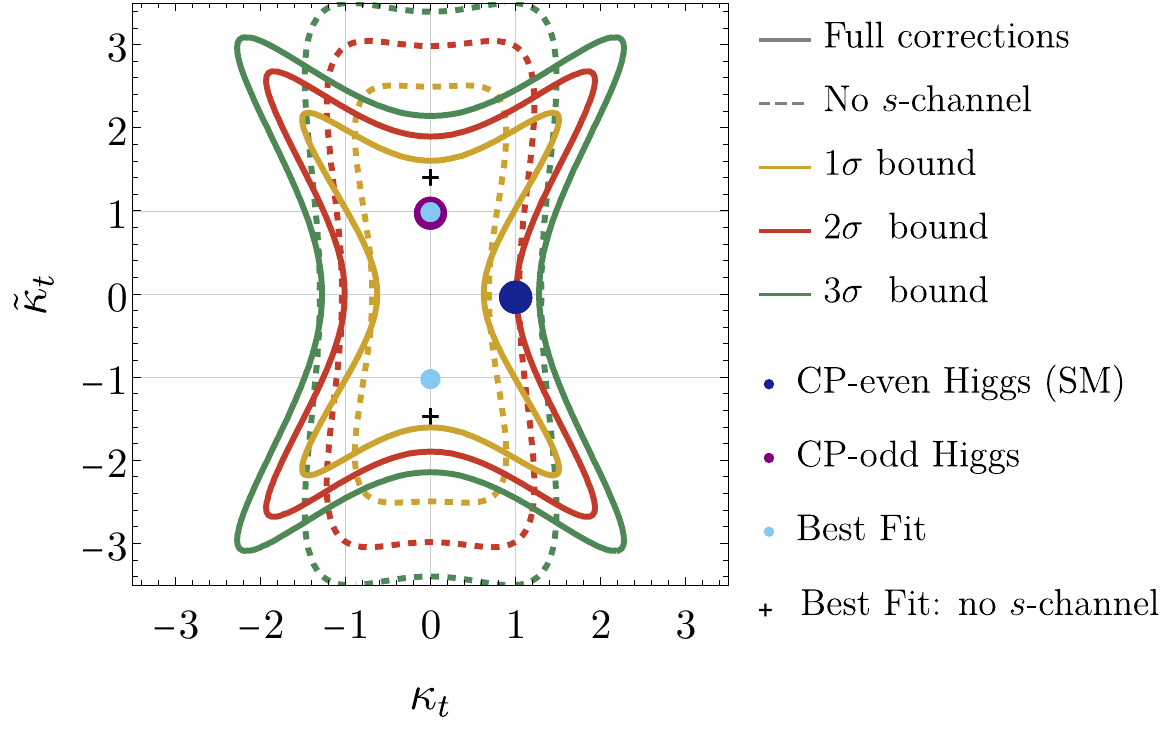}
    \caption{Left plot: Same as the left plot of Fig.~\ref{fig:exclusionHiggs} but we also show the case in which we remove the contribution of the $s$-channel diagram from $\Sigmatheo$ in the fit, but {\it not}  from $\Sigmatheo$, the pseudo-data, as dashed lines and the best fit as a cross.  Right plot: Same as the {\it left} plot of Fig.~\ref{fig:pseudodatapscalar}, but with in addition the information already described for the right plot of this figure.   }    \label{fig:chicomparison}
\end{figure}

In Fig.~\ref{fig:chicomparison} we show the results for the  two-parameter fit using the multiplicative approach,  comparing the complete result (solid lines) and the case where  the $s$-channel contribution has been removed. The left plot refers to SM pseudo-data and the right one to pseudo-data corresponding to $(\kt,\ktt)=(0,1)$.

In the left plot  (SM pseudodata) we can see that for $\kt\simeq0$ the contours with and without the contributions of the $s$-channel almost coincides, again because such contribution is not so relevant around the purely scalar configuration $\phiK\simeq 0$. Instead, around the purely pseudoscalar configuration $\phiK\simeq \pi/2$, the bound that one would obtain without the contribution of the $s$-channel diagram is less stringent because, as explained above, the size of the corrections is sizeably reduced w.r.t.~the complete calculation. This dynamics has also the effect to increase the value of $\phiK$ around which the bounds for $|\Kt|$ are weaker; since the size of $\Sigmaktt$ reduces, the $\ktt/\kt$ ratio leading to large cancellations increases.

\begin{table}[!t]
    \centering
    \begin{tabular}{l|c|c}
              &  ${\kappa_t}^{+1\sigma,\,2\sigma,\,3\sigma}_{-1\sigma,\,2\sigma,\,3\sigma}$ & ${\tilde\kappa{_t}}^{+1\sigma,\,2\sigma,\,3\sigma}_{-1\sigma,\,2\sigma,\,3\sigma}$ \\[1.5pt]
              \hline
       $ \SMmult$ LHC & $1.01^{+0.29,0.53,0.73}_{-0.42,1.01,1.01}$&$0.0^{+1.16,1.95,2.55}_{-1.16,1.95,2.55}$\\
       $\SMadd$ LHC & $1.01^{+0.39,0.70,0.95}_{-0.75,1.01,1.01}$ &$0.0^{+1.56,2.49,3.19}_{-1.56,2.49,3.19}$
        
    \end{tabular}
    \caption{The same information of Tab.~\ref{tab:Higgscalarfit}, but removing the contribution of the $s$-channel diagram from $\Sigmatheo$ in the fit, but {\it not}  from $\Sigmatheo$, the pseudo-data.}
    \label{tab:higgsnosS}
\end{table}
\begin{table}[!t]
    \centering
    \begin{tabular}{l|c|c}
            &  ${\kappa_t}^{+1\sigma,\,2\sigma,\,3\sigma}_{-1\sigma,\,2\sigma,\,3\sigma}$ & ${\tilde\kappa{_t}}^{+1\sigma,\,2\sigma,\,3\sigma}_{-1\sigma,\,2\sigma,\,3\sigma}$ \\[1.5pt]
            \hline
              $ \SMmult$ LHC & $0.00^{+0.56,0.94,1.23}_{-0.56,0.94,1.23}$&$1.44^{+0.78,1.35,1.82}_{-1.44,1.44,1.44}$\\
       $\SMadd$ LHC & $0.00^{+0.74,1.19,1.53}_{-0.74,1.19,1.53}$ &$1.41^{+1.05,1.77,2.35}_{-1.41,1.41,1.41}$
        
    \end{tabular}
    \caption{The same information of Tab.~\ref{tab:Higgspseudoscalarfit}, but removing the contribution of the $s$-channel diagram from $\Sigmatheo$ in the fit, but {\it not}  from $\Sigmatheo$, the pseudo-data. }
    \label{tab:higgsnosPS}
\end{table}

In the right plot, the pseudodata corresponds to the configuration $(\kt,\ktt)=(0,1)$, the case where the Higgs boson is a purely pseudo scalar. We notice that for $\ktt\simeq 0$ the contours without the contribution of the $s$-channel are again similar to the exact calculation. Removing the contribution of the $s$-channel diagram, the size of $\Sigmaktt$ decreases and therefore both the best fit and the contours for the different confidence levels move to larger values of $\ktt$.

 \section{Conclusions}
 \label{sec:conclusions}

In this paper we have studied the effects of unresolvable corrections to the top-quark hadroproduction at the LHC induced by either a light top-philic scalar $S$ or by the Higgs boson (with anomalous Yukawa interactions). We considered in both cases CP-even as well as CP-odd interactions with the top-quark.

In the case of the light top-philic scalar $S$ we have observed that virtual corrections remain finite in the massless limit $m_S \to 0$ when only CP-odd interactions (parametrised in our notation by the quantity $\ctt$) between this particle and the top are present. On the contrary, if   CP-even interactions are present (parametrised in our notation by the quantity $\ct$) then virtual corrections are divergent for  $m_S \to 0$  and they have to be combined with the (soft) real emissions of $S$ in order to obtain IR-safe predictions. 

We have studied the impact on top-quark distributions of such corrections, which depend only on $\ct^2$ and $\ctt^2$ and therefore are insensitive to the relative sign of $\ct$ and $\ctt$. We observe different shapes and signs for the $\ct^2$- and $\ctt^2$-dependent corrections and  large cancellations when  $|\ct|\simeq|\ctt|$.  After that, we have explored the constraints that can be obtained from current top-quark data for $\ct^2$ and/or $\ctt^2$ in the range $m_S<300~\gev$, {\it i.e.}, avoiding the resonant scalar production with subsequent decays into top quarks. In doing so, for the SM predictions we take into account both QCD and EW effects at NLO and also NNLO QCD corrections. 

 When purely CP-even or CP-odd interactions are considered, bounds mildly depend on the value of $m_S$ for $m_S<10~\gev$, while they strongly depend on it for $10~\gev<m_S<300~\gev$, due to the presence of an $s$-channel diagram with the scalar in the propagator. Corrections are larger for the purely CP-even case w.r.t.~the purely CP-odd one and therefore also stronger bounds can be set. When both CP-even and CP-odd interactions are possible, a non trivial pattern of cancellations is present, which depends on the value of $m_S$ and $|\ct|/|\ctt|$.  
 
 The calculation for the scalar $S$ can be recycled for the case of the Higgs boson with both CP-even  (parametrised in our notation by the quantity $\kt$) and CP-odd  (parametrised in our notation by the quantity $\kt$) anomalous interactions, with the SM corresponding to $(\kt, \ktt)=(1,0)$. In this case the calculation is analogous to the one of Ref.~\cite{Martini:2021uey}, where, to the best of our knowledge, the  diagram with the Higgs $s$-channel featuring was not taken into account. We find that while in the case of only purely CP-even interactions ($\kt$) between the top and the Higgs the impact of this diagram is negligible, if CP-odd effects are present ($\ktt$) their contribution cannot be neglected and it is sizeable. We have revisited the bounds that have been set by CMS in $\kt$ and studied the relevance of the inclusion of such diagram for possible analogous analyses that may take into account CP-odd contributions and so the $\ktt$ dependence.
 
 In this paper we have provided technical details and analytical formulas for the more general case of the top-philic scalar, where the mass $m_S$ is not fixed. We also have  explained in depth how the calculation can be recycled for the case of the Higgs boson. Moreover, a few subtleties related to the fact that SM does not include the scalar but does include the Higgs have been discussed in detail and also reinterpreted within the SMEFT framework, which supports the  consistency of the renormalisation procedure we have adopted in our calculation.
 
For what concerns the phenomenological part,  we have performed an exploratory study, focusing on distributions for stable top quarks and in particular on the top-quark invariant mass distribution for analysing the sensitivity from data. Starting from the results presented in this work several other analyses could be performed. First, it is possible to extend this study to the level of fully decayed top-quarks accessing the information form spin correlations and possibly relevant quantities recently studied in the context of quantum tomography, see, {\it e.g.}, Refs.~\cite{Afik:2020onf,Fabbrichesi:2021npl,Severi:2021cnj}. Second, the statistical analyses could be performed on real data and not only pseudodata as in this work and eventually combined with  other processes that allow to set constraints on the couplings with the scalar $S$ or the Higgs boson, such as $t\bar t H$ and four-top production. Last but not least, it would be interesting to consider the case of a lepton collider with an energy not so larger than the top-quark pair production threshold, where sensitivity to this kind of effects is expected to be large.

\section*{Acknowledgments}
 The works of Sonia Delaunay and Kazimir Malevich inspired the choice of the colour palette used in the plots.  We thank Marco Zaro for helping with the code at the initial stage of this project. We also acknowledge Simone Blasi, Alberto Mariotti and Ken Mimasu for discussion on this topic in the context of top-philic ALPs. We acknowledge the use of computing facilities of the Universit\'e catholique de Louvain (CISM/UCL) and the Consortium des \'Equipements de Calcul Intensif en F\'ed\'eration Wallonie Bruxelles (C\'ECI), funded by the Fond de la Recherche Scientifique de Belgique (F.R.S.-FNRS) under convention 2.5020.11 and by the Walloon Region. ST is supported by a FRIA Grant of the Belgian Fund for Research, F.R.S.-FNRS
(Fonds de la Recherche Scientifique-FNRS). This research is partially supported by the IISN-FNRS convention 4.451708,  ``Fundamental
interactions''.
 
  \clearpage
 
\begin{appendices}
\section{Explicit calculation of the UV counterterms}
\label{app:UVcounterterms}

\begin{figure}[!t]
    \centering
    \includegraphics[width=0.4\textwidth]{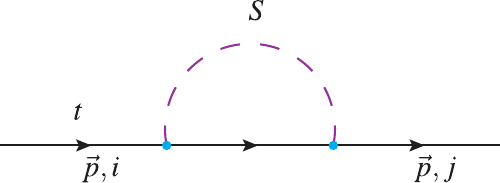}
    \caption{One-loop corrections induced by the scalar $S$ to the top-quark external leg. The $i,j$ indexes refer to the colour.}
    \label{fig:1looplegt}
\end{figure}

The quantity $\widehat \Sigma^{1}_{\rm NP}( p)$ introduced in Eq.~\eqref{eq:Sigma1NP} can be explicitly written as 
\bea
\widehat \Sigma^{1}_{\rm NP}( p)&=&\frac{c_t^2}{16\pi^2}
\left[ m_t B_0\left(p^2;m_t,m_S\right)-\slashed{p} B_1\left(p^2;m_t,m_S\right)\right]\nonumber\\
&-&\frac{\tilde{c_t}^2}{16\pi^2} \left[ m_t B_0\left(p^2;m_t,m_S\right)+ \slashed{p} B_1\left(p^2;m_t,m_S\right)\right] \label{eq:s1npasB}\\
&+&i\frac{c_t\tilde{c_t}}{8\pi^2}  \gamma ^5  m_t  B_0\left(p^2;m_t,m_S\right)\nonumber \,,
\eea
where $B_0$ is the standard scalar one-loop two-point integral \cite{tHooft:1978jhc, Passarino:1978jh} and $B_1$ the analogous tensor coefficient, which we also specify later in Appendix \ref{app:loopintegrals}. The corresponding diagram is depicted in Fig.~\ref{fig:1looplegt}.

From Eqs.~\eqref{eq:s1npasB} and \eqref{eq:Sigma1NP} it is manifest that
\bea
\Sigma_{V}(p^2)&=&-\frac{\ct^2+\ctt^2}{16\pi^2}
B_1\left(p^2;m_t,m_S\right)\, ,\\
\Sigma_{S}(p^2)&=&\frac{\ct^2-\ctt^2}{16\pi^2}
B_0\left(p^2;m_t,m_S\right)\, ,\\
\Sigma_{P}(p^2)&=&\frac{\ct \ctt}{8\pi^2}
B_0\left(p^2;m_t,m_S\right)\, .
\eea
Via Eqs.~\eqref{eq:defdeltamt} and \eqref{eq:defdeltaZt} we get, respectively,
\bea
\delta_{m_t}&=&\frac{\ct^2}{16\pi^2}
\left[B_0\left(m_t^2;m_t,m_S\right)-B_1\left(m_t^2;m_t,m_S\right)\right]\nonumber\\
&-&\frac{\ctt^2}{16\pi^2}
\left[B_0\left(m_t^2;m_t,m_S\right)+B_1\left(m_t^2;m_t,m_S\right)\right]\, ,
\eea
and
\bea
\delta_{\psi_t}&=-&\frac{c_t^2}{16 \pi ^2} \left[2
m_t^2 \left[B'_0\left(m_t^2,m_t,m_S\right)-B'_1\left(m_t^2,m_S,m_t\right)\right]-B_1\left(m_t^2;m_t,m_S\right) \right]\nonumber\\
&+&\frac{\ctt^2}{16 \pi
^2} \left[2 m_t^2 \left[B'_0\left(m_t^2,m_t,m_S\right)+B'_1\left(m_t^2,m_t,m_S\right)\right]+B_1\left(m_t^2;m_S,m_t\right)\right]\,, \label{eq:deltapsiviaB}
\eea
where the prime stands for the derivative $d/dp^2$.

Given the discussion in Sec.~\ref{sec:HversusS}, all the previous equations can be converted for the case $S=H$ via the substitutions 
\bea
m_S&\to& m_H\, , \\
\ct^2&\to& \frac{(\ytSM)^2}{2}(\kt^2-1)\,,\label{eq:link2v2}\\
\ctt^2&\to& \frac{(\ytSM)^2}{2} \ktt^2\,.
\eea
We stress again that a linear $(\kt-1)$ term is present and therefore \eqref{eq:link2v2} is not the same of \eqref{eq:link2}.

In view of what will be discussed in  Appendix \ref{app:qqttanal} it is also useful to show the limit $m_S\to 0$ for the UV counterterms,
\begin{align}\label{eq:countertermsexpanded1}
 \delta_{\psi_t}&=\frac{1}{32\pi^2}\Biggl[c_t^2\left(-\frac{1}{\epsilon}-\log\frac{\tilde \mu^2}{m_t^2}+4\log\frac{m_{S}^2}{m_t^2
 }+7\right)+\tilde c_t^2\left(-\frac{1}{\epsilon}-\log\frac{\tilde \mu^2}{m_t^2}-1\right)\Biggr]\,,\\
\delta_{m_t}&=\frac{1}{32\pi^2}\Biggl[c_t^2\left(\frac{3}{\epsilon}+3\log\frac{\tilde \mu^2}{m_t^2}+7\right)+\tilde c_t^2\left(-\frac{1}{\epsilon}-\log\frac{\tilde \mu^2}{m_t^2}-1\right)\Biggr]\label{eq:countertermsexpanded2}\,.
\end{align}

The quantity $\tilde \mu$ is defined as $\tilde \mu^2=4\pi e^{-\gamma_{\rm E}}\mu^2$, where $\mu$ is the regularisation scale introduced via dimensional regularisation in $d=4-2\epsilon$ dimensions and $\gamma_{\rm E}$ is the Mascheroni constant. 
As can be noted, consistently with what is discussed in details in the main text, the part depending on the CP-odd coupling $\ctt$ does not involve any IR divergency while the   part depending on the CP-even coupling $\ct$ diverges for $m_S\to 0$.

\section{Renormalised $t \bar t g$ vertex and one-loop $qq\to  t \bar t $ amplitude}
\label{app:qqttanal}

The purpose of this appendix is twofold. First we want to explicitly show that via the counterterms of Appendix \ref{app:UVcounterterms} UV-finite one-loop corrections can be obtained for the $q\bar q \to t \bar t$ process. We provide the explicit results for case of the $t \bar t g$ vertex with the top quarks on-shell and a generic timelike $q^2$ value. Second, we can explicitly show the IR sensitivity on $m_S$ for the corrections proportional to $\ct$ that parameterises the CP-even interactions. 

\begin{figure}[!t]
    \centering
    \includegraphics[width=0.25\textwidth]{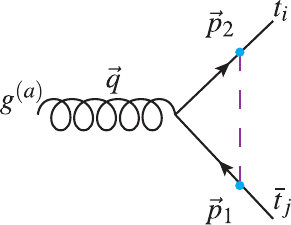}
    \caption{One-loop corrections induced by the scalar $S$ to the $t\bar t g$ vertex. The $i,j$ indexes refer to the colour}
    \label{fig:leg}
\end{figure}

\begin{figure}[!t]
    \centering
\includegraphics[width=0.78\textwidth]{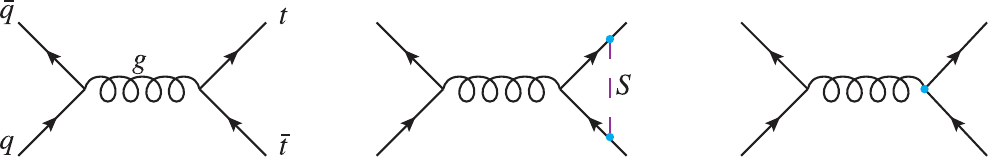}
    \caption{Feynman diagrams entering the calculation of one-loop virtual corrections for the process $q\bar q \rightarrow t\bar t$. From left to right: the tree-level in QCD ($\MzSM$),  the UV-divergent virtual corrections ($\MoNPunren$) and the amplitude with UV counterterms ($\MoNPCT$).}
    \label{fig:qqtt}
\end{figure}

We start by writing the general structure for the  UV-divergent $t\bar t g$ vertex at one-loop accuracy, contracted with the helicity of the top antiquark with outgoing momentum $p_1$, $v(p_1)$ and the one of the top quark  with outgoing momentum $p_2$, $v(p_2)$. The  diagram corresponding to the one-loop correction is depicted in Fig.~\ref{fig:leg} and the aforementioned expression reads
\begin{equation}\label{eq:generalloop}
\bar u_i(p_2)\widehat{\Gamma}_{ij}^{\mu,a}v_j(p_1)=(-ig_s )t^{a}_{ij}~ \bar u_i(p_2)\left[\widehat F_1(q^2)\gamma^\mu+i\frac{\sigma^{\mu\nu}}{2m_t}q_\nu  \left( F_2(q^2)+i F_3(q^2)\gamma_5\right)\right]v_j (p_1),
\end{equation}
where $i,j$ are the colour indexes and $\sigma^{\mu \nu}=\frac{i}{2}[\gamma^\mu,\gamma^\nu]$. The ${F}_i$ functions depend only on $q^2=(p_1+p_2)^2$. While $F_2$ and $F_3$ are UV-finite,  $\widehat F_1$ is UV-divergent and it is also convenient to rewrite it as
\begin{equation}
\widehat F_1(q^2)=1+\widehat f_1(q^2),
\end{equation}
such that for $\ct$ and $\ctt$ vanishing the tree-level expression is recovered.
We write in the following  the expressions of $\widehat f_1$, $F_2$ and $F_3$ in terms of the standard $B$ and $C$ loop-integral functions, where in the case of the $C$ functions is always understood that the invariant entering is $q^2$ and the internal masses are always $m_t$ and $m_S$.
\begin{align}
\label{eq:f1hat}
 \widehat f_1(q^2)=-&\frac{c_t^2}{16\pi^2} \left[2 \left(C_{00}+2 m_t^2 \left(C_{12}+C_{11}\right)\right)-B_0\left(q^2;m_t,m_t\right)+(4 m_t^2-m_S^2) C_0+8 m_t^2 C_1\right]\nonumber\\
-&\frac{\ctt^2}{16\pi^2} \left[2 \left(C_{00}+2 m_t^2 \left(C_{12}+C_{11}\right)\right)-B_0\left(q^2;m_t,m_t\right)-m_S^2 C_0\right]  \,,
\end{align}
\begin{align}
 F_2(q^2)=\frac{c_t^2 }{4\pi^2}m_t^2
\left(C_{12}+2 C_1+C_{11}\right)+\frac{\ctt^2}{4\pi^2} m_t^2  \left(C_{12}+C_{11}\right)\,,
\end{align}
\begin{equation}
 F_3(q^2)=m_t^2\frac{\ct\ctt}{2\pi^2} C_1 \label{eq:F3}\,.
\end{equation}
Having these expressions, it is possible to compute for the $q \bar q \to t \bar t$ process the quantity  $2\Re[\MzSM (\MoNP)^*]$ entering $\SigmaNP$ (see also Eq.~\eqref{eq:proprtoRE} and the notation in that section) induced by the diagrams in Fig.~\ref{fig:qqtt} and expressing it as a function of $|\MzSM|^2$.  

In particular, summing(averaging) over the colours and polarisations of the  final(initial) state, 
\be
\overline \sum 2\Re[\MzSM (\MoNP)^*]= \overline \sum |\MzSM|^2 2\Re\left[f_1(q^2)\right] +(4\pi\alpha_S)^2 \left(2+4\frac{m_q^2}{s}\right) \frac{4}{9}\Re\left[F_2(q^2)\right]\,. \label{eq:qqttexplicit}
\ee
with 
\be
f_1(q^2)\equiv \widehat f_1(q^2)+\delta_{\psi_t}\,. \label{eq:f1ren}
\ee
and
\be
\overline \sum |\MzSM|^2 = \frac{4}{9}(4\pi)^2\alpha_s^2 \frac{s^2-2tu+2(m_q^2+m_t^2)^2}{s^2}
\ee

Taking the expressions of the $C$ and $B$ integrals in Appendix \ref{app:loopintegrals}, noticing that the only $C$ integral that is UV divergent is the $C_{00}$, and plugging them into  Eqs.~\eqref{eq:deltapsiviaB} and \eqref{eq:f1hat}, one easily see that $f_1(q^2)$ is UV finite. One can  in principle obtain in this way also the full expression for $f_1(q^2)$, which we do not provide here and that involve also $C_0$, which is not given in Appendix \ref{app:loopintegrals},  but it that can be found in the literature, {\it e.g.}, in Ref.~\cite{Denner:1991kt}. The same procedure can be used for obtain the expressions for $F_2$ and $F_3$, however, the latter is proportional to $\ct\ctt$ and disappears in the final results consistently to what is written in the main text.

We do provide instead the explicit expression of the IR limit of $m_S\rightarrow 0$ for $f_1(s)$ and $F_2(s)$. In the case of $f_1(s)$ we separately show the case with $\ctt=0$ and $\ct=0$, which are therefore its components proportional to $\ct^2$ and $\ctt^2$, respectively.
In doing so we exploit the fact that in our kinematic configuration,
\be
C_0\xrightarrow{m_S\rightarrow 0}\frac{{\rm DB}(s,m_t,m_t)}{s-4m_t^2}\log\frac{m_S^2}{m_t^2}+\mathcal{O}(1) \label{eq:C0expanded}\,,
\ee
obtaining
\bea
f_1(q^2)\bigg|_{\ctt=0}&\xrightarrow{m_S\rightarrow 0}& \frac{\ct^2}{32\pi^2}\left[4\log \frac{m_S^2}{m_t^2}\left(1-\frac{2m_t^2}{s-4m_t^2}{\rm DB}(s,m_t,m_t)\right)+\mathcal{O}(1)\right]\,, \label{eq:IRf1explicit}\\
f_1(q^2)\bigg|_{\ct=0}&\xrightarrow{m_S\rightarrow 0}& \frac{\ctt^2}{32\pi^2} \left[\frac{s}{s-4m_t^2}{\rm DB}(s,m_t,m_t)+\mathcal{O}(m_S/Q)\right]\,, \label{eq:noIRf1explicit}\\
F_2(q^2)&\xrightarrow{m_S\rightarrow 0}& \frac{\ct^2}{8\pi^2}\left[\frac{3m_t^2}{s-4m_t^2}{\rm DB}(s,m_t,m_t)+\mathcal{O}(m_S^2/Q^2)\right]\nonumber\\
&&-\frac{\ctt^2}{8\pi^2}\left[\frac{m_t^2}{s-4m_t^2}{\rm DB}(s,m_t,m_t)+\mathcal{O}(m_S^2/Q^2)\right]\label{eq:noIRF2}\,.
\eea
In Eq.~\eqref{eq:IRf1explicit} we clearly see double Sudakov  logarithm of IR origin, with a coefficient that scale as $m_t^2/s$ for large $s$,  a  single one with a $s$-independent coefficient.\footnote{We stress that these expressions are obtained in the limit $m_S \to 0$ and not $s\to\infty$. These are not high-energy Sudakov logarithms; we treat $m_t^2,s\gg m_S^2$, which is different than considering $s \gg m_t^2, m_S^2$}. We stress that taking a generic scale $Q^2=m_t^2,s$ and expanding in terms of $(m_S^2/Q^2)$ the leading terms of Eq.~\eqref{eq:IRf1explicit}, which is IR divergent, are of different order than the one of Eqs.~\eqref{eq:noIRf1explicit} and \eqref{eq:noIRF2}, which are not IR divergent. The $\mathcal{O}(1)$ terms of Eq.~\eqref{eq:IRf1explicit} are of the same order of the explicit terms written for Eqs.~\eqref{eq:noIRf1explicit} and \eqref{eq:noIRF2}. The explicit expressions involve dilogarithms and can be found in Ref.~\cite{Dittmaier:2003bc}, from where also Eq.~\eqref{eq:C0expanded} has been derived.

As final remark, we notice the absence of $F_3$ in Eq.~\eqref{eq:qqttexplicit}, consistently to the fact that any term proportional to  $\ct\ctt$ vanishes.

\section{Loop integrals}
\label{app:loopintegrals}
In this Appendix, we provide for the convenience of the reader  the explicit result of the  $B$ and $C$ integrals entering our calculation. To this purpose, it is convenient to define the following $\rm DB$ function 
\begin{equation}
  \text{DB}(a^2,b,c)=\frac{\sqrt{\lambda \left(a^2,b^2,c^2\right)} \log \left(\dfrac{\sqrt{\lambda \left(a^2,b^2,c^2\right)}-a^2+b^2+c^2}{2 b c}\right)}{a^2}\,,
\end{equation}
which as explicitly shown depends on $a^2,b$ and $c$. The $\lambda$ function depends only on $a^2,b^2$ and $c^2$ and it is defined as
    \begin{equation}
  \lambda(a^2,b^2,c^2)=-2 a^2 b^2-2 a^2 c^2+a^4-2 b^2 c^2+b^4+c^4\,.
\end{equation}
In order to obtains the IR limits for $m_S\to0$ is useful to notice that
\begin{equation}
 \text{DB}(a^2,0,a)= 0\, .
\end{equation}

Defining the quantities 
\bea
r&\equiv& \frac{m_S}{m_t}\,,\\
\Delta m^2&\equiv& m_t^2-m_S^2\,,
\eea
the $B$ functions can be expressed  as
{\small\begin{align}
B_0(p^2,m_t,m_S)=&\text{DB}\left(p^2,m_S,m_t\right)+\frac{1}{2}\log \left(\frac{\tilde\mu ^2}{m_S^2}\right)+\frac{1}{2}\left(\frac{\tilde\mu ^2}{m_t^2}\right)+\frac{\Delta m^2 \log \left(r^2\right)}{2 p^2}+\frac{1}{\epsilon }+2\,,\\
B_1(p^2,m_t,m_S)=&-\frac{\left(\Delta m^2+p^2\right)}{2 p^2} (\text{DB}\left(p^2,m_S,m_t\right)+2)-\frac{\log \left(r^2\right) \left(\lambda(p^2,m_S^2,m_t^2)+2p^2m_t^2\right)}{4 p^4}+\nonumber\\
&-\frac{1}{2} \left(\log \left(\frac{\tilde\mu ^2}{m_S^2}\right)+\frac{1}{\epsilon }\right)\,,\\
B_0^1(m_t^2,m_t,m_S)=&\frac{2 \left(r^2-3\right) \text{DB}\left(m_t^2,m_S,m_t\right)-\left(r^2-4 \right) \left(\log \left(r^2\right) \left(r^2-1\right)-2 \right)}{8m_t^2-2m_S^2 }\,,\\
B_1^1(m_t^2,m_t,m_S)=&\frac{2  \left(r^4-5 r^2 +5 \right) \text{DB}\left(m_t^2,m_S,m_t\right)-\left(r^2-4 \right) \left(\log \left(r^2\right) \left(r^4-3 r^2 +1\right)-2 r^2 +3 \right)}{8 m_t^2-2 m_S^2}\,,
\end{align}}
where $\tilde \mu$ has been defined at the end of Appendix \ref{app:UVcounterterms}.

We also list the (combinations of) $C$ functions that enter Eqs.~\eqref{eq:f1hat}--\eqref{eq:F3}, besides as said the $C_0$. In order
to shorten the expression, we understood that the momenta configurations are $C_{n(m)}=C_{n(m)}\left(m_t^2,p^2,m_t^2;m_S,m_t,m_t\right)$, obtaining
{\small\begin{align}
C_1=&-\frac{m_S^2 C_0}{4 m_t^2-p^2}+\frac{\text{DB}\left(m_t^2,m_S,m_t\right)}{4 m_t^2-p^2}-\frac{\text{DB}\left(p^2,m_t,m_t\right)}{4 m_t^2-p^2}-\frac{r^2 \log \left(r^2\right)}{2 \left(4 m_t^2-p^2\right)}\,,\\
C_{00}=&-\frac{m_S^2 \left(m_S^2-4 m_t^2+p^2\right) C_0}{2 \left(4 m_t^2-p^2\right)}+\frac{m_S^2 \text{DB}\left(m_t^2,m_S,m_t\right)}{2 \left(4 m_t^2-p^2\right)}\nonumber\\
&+\frac{\left(4 m_t^2-2 m_S^2-p^2\right) \text{DB}\left(p^2,m_t,m_t\right)}{4 \left(4 m_t^2-p^2\right)}\nonumber\\
&-\frac{m_S^2 r^2 \log \left(r^2\right)}{4\left(4 m_t^2-p^2\right)}+\frac{1}{4} \left(\log \left(\frac{\tilde \mu ^2}{m_t^2}\right)+\frac{1}{\epsilon }\right)+\frac{3}{4}\,,\\
C_{11}+C_{12}=&\frac{m_S^2 \left(3 m_S^2-4 m_t^2+p^2\right) C_0}{\left(4 m_t^2-p^2\right){}^2}-\frac{r^2 \left(10 m_t^2-p^2\right) \text{DB}\left(m_t^2,m_S,m_t\right)}{2 \left(4 m_t^2-p^2\right){}^2}-\frac{r^2}{2 \left(4 m_t^2-p^2\right)}\nonumber\\
+&\frac{r^2 \log \left(r^2\right) \left(2 p^2-p^2 r^2 +10 m_S^2 -8 m_t^2\right)}{4  \left(4 m_t^2-p^2\right)^2}+\frac{\left(6 m_S^2+4 m_t^2-p^2\right) \text{DB}\left(p^2,m_t,m_t\right)}{2 \left(4 m_t^2-p^2\right)^2}\,.
\end{align}}

\section{Numerical inputs for the fits}
\label{app:table}

In this Appendix we explicitly show the relevant quantities leading to  $\Sigmaexp$ (the pseudo-data) for the bins of the $m(t\bar t)$ distribution measured in \cite{CMS:2018htd}, which is entering the fits discussed both in Sec.~\ref{sec:Ssensitivity} for the scalar $S$ and in Sec.~\ref{sec:sensitivityH} for the Higgs boson $H$. 
For the latter case, where the mass $m_H$ is fixed unlike $m_S$ for the scalar $S$, we provide the relevant quantities leading to  $\Sigmatheo$.

All the information is reported in Tab.~\ref{tab:inputforfit}, where all the quantities, besides $\SigmaSMadd$ and $\SigmaLOQCD$, are normalised w.r.t.~$\SigmaLOQCD$.

\begin{table}[!h]\scriptsize
\centering
\begin{tabular}{c|c|c|c|c|c|c|c}
\toprule
\hline
$\mtt~[\gev]$&$\SigmaSMadd [
\frac{\rm pb}{\rm GeV}]$&$\SigmaLOQCD [\frac{\rm pb}{\rm GeV}]$ &$\SigmaNLOQCD~[\%]$&$\SigmaNNLOQCD~[\%]$&$\SigmaNLOEW~[\%]$&$\Sigmakt~[\%]$&$\Sigmaktt~[\%]$\\
\midrule
300-360&0.173&0.101&51.8&20.2&2.77&9.61&-6.01\\

360-430&1.07&0.731&38.9&7.64&-0.188&3.67&-3.85\\

430-500&0.84&0.592&35.2&6.63&-1.42&0.646&-2.23\\

500-580&0.519&0.368&34.9&6.05&-1.93&-0.334&-1.55\\

580-680&0.286&0.2&34.0&8.81&-2.21&-0.747&-1.22\\

680-800&0.141&0.0977&33.9&10.1&-2.54&-0.923&-1.06\\

800-1000&0.0563&0.0385&35.4&10.7&-3.06&-0.998&-1.01\\

1000-1200&0.0192&0.013&36.2&11.3&-3.56&-1.04&-1.03\\

1200-1500&0.00614&0.00416&34.8&12.7&-4.27&-1.07&-1.07\\

1500-2500&0.000772&0.000514&35.8&14.4&-5.22&-1.17&-1.18\\
\bottomrule
\end{tabular}
\caption{ All the quantities are defined in Secs.~\ref{sec:SMcombination} and \ref{sec:LagrangiansH}.}
\label{tab:inputforfit}
\end{table}

\end{appendices}
\newpage\bibliographystyle{JHEP}

\bibliography{bibfr.bib}
\end{document}